\documentclass[preprint]{aastex}

\newcommand{\kms}{\, {\rm km\, s}^{-1}}
\newcommand{\ikms}{\, {\rm s\, km}^{-1}}
\newcommand{\mpc}{\, {\rm Mpc}}
\newcommand{\kpc}{\, {\rm kpc}}
\newcommand{\hmpc}{\, h^{-1} \mpc}
\newcommand{\ihmpc}{(\hmpc)^{-1}}
\newcommand{\hkpc}{\, h^{-1} \kpc}
\newcommand{\lya}{Ly$\alpha$}
\newcommand{\lyaf}{Ly$\alpha$ forest}
\newcommand{\bF}{\bar{F}}
\newcommand{\xrei}{x_{rei}}
\newcommand{\lr}{\lambda_{{\rm rest}}}
\newcommand{\hi}{\mbox{H\,{\scriptsize I}\ }}

\newcommand{\PF}{P_F(k,z)}
\newcommand{\PL}{P_L(k_p,z_p)}
\newcommand{\gprime}{g^\prime}
\newcommand{\sprime}{s^\prime}
\newcommand{\tgamma}{\tilde{\gamma}}
\newcommand{\gmo}{\gamma-1}
\newcommand{\Tp}{T_{1.4}}
\newcommand{\DL}{\Delta_L^2(k_p,z_p)}
\newcommand{\neff}{n_{\rm eff}(k_p,z_p)}
\newcommand{\aleff}{\alpha_{\rm eff}(k_p,z_p)}

\begin{document}

\title{The Linear Theory Power Spectrum from the 
Lyman-$\alpha$ Forest in the Sloan Digital Sky Survey}

\author{Patrick McDonald\altaffilmark{1,2},  
Uro\v s Seljak\altaffilmark{2},
Renyue Cen\altaffilmark{3}, 
David Shih\altaffilmark{2},
David~H.~Weinberg\altaffilmark{4}, 
Scott Burles\altaffilmark{5}, 
Donald~P.~Schneider\altaffilmark{6},
David~J.~Schlegel\altaffilmark{3,7}, 
Neta A. Bahcall\altaffilmark{3},
John W. Briggs\altaffilmark{8},
J. Brinkmann\altaffilmark{9},
Masataka Fukugita\altaffilmark{10},
\v{Z}eljko Ivezi\'{c}\altaffilmark{3,11}, 
Stephen Kent\altaffilmark{12}, and
Daniel E. Vanden Berk\altaffilmark{6}
}

\altaffiltext{1}
{Canadian Institute for Theoretical
Astrophysics, University of Toronto, Toronto, ON M5S 3H8, Canada;
pmcdonal@cita.utoronto.ca}

\altaffiltext{2}{Physics Department, Princeton University, Princeton NJ 08544,
USA}

\altaffiltext{3}{Princeton University Observatory, Princeton, NJ 08544, USA}

\altaffiltext{4}
{Department of Astronomy, Ohio State University, Columbus, OH 43210, USA}

\altaffiltext{5}{Physics Department, MIT, 77 Massachusetts Av., 
Cambridge MA 02139, USA}

\altaffiltext{6}
{Department of Astronomy and Astrophysics,
The Pennsylvania State University, University Park, PA 16802, USA}

\altaffiltext{7}
{Lawrence Berkeley National Laboratory, One Cyclotron Road, Mailstop 50R232,
Berkeley, CA 94720, USA}

\altaffiltext{8}
{National Solar Observatory, Sunspot, NM 88349, USA}

\altaffiltext{9}
{Apache Point Observatory, 2001 Apache Point Rd, Sunspot, NM 88349-0059, USA}

\altaffiltext{10}
{Inst. for Cosmic Ray Research, Univ. of Tokyo, Kashiwa 277-8582, Japan}

\altaffiltext{11}
{Department of Astronomy, University of Washington, Seattle, WA 98195, USA}

\altaffiltext{12}
{Fermi National Accelerator Laboratory, P.O. Box 500, Batavia, IL 60510, USA}

\begin{abstract}

We analyze the SDSS \lyaf\ $\PF$ measurement to determine the
linear theory power spectrum.
Our analysis is based on fully hydrodynamic simulations, 
extended using hydro-PM simulations.
We account for the effect of absorbers with damping
wings, which leads to an increase in the slope of 
the linear power spectrum.
We break the degeneracy between the mean level of absorption 
and the linear power spectrum 
without significant use of external constraints.
We infer linear theory power spectrum amplitude 
$\Delta^2_L(k_p=0.009~{\rm s/km},z_p=3.0)=
0.452_{-0.057~-0.116}^{+0.069~+0.141}$ 
and slope $\neff=-2.321_{-0.047~-0.102}^{+0.055~+0.131}$
(possible systematic 
errors are included through nuisance parameters in the fit --- a
factor $\gtrsim 5$ smaller errors would be obtained on both
parameters if we ignored modeling uncertainties). 
The errors are correlated and not perfectly Gaussian,
so we provide a $\chi^2$ table to accurately describe the 
results.
The result corresponds to $\sigma_8=0.85$,
$n=0.94$, for a  $\Lambda$CDM model with $\Omega_m=0.3$,
$\Omega_b=0.04$, and $h=0.7$, but is most useful in a
combined fit with the CMB.
The inferred curvature of the
linear power spectrum and the evolution of its amplitude
and slope with redshift are consistent with expectations for
$\Lambda$CDM models, with the evolution of the slope, in particular,
being tightly constrained. 
We use this information to constrain systematic 
contamination, e.g., fluctuations in the UV background. 
This paper should serve as a starting point for more work
to refine the analysis, including
technical improvements such as
increasing the size and number of the hydrodynamic simulations, 
and improvements in the treatment of the various forms of 
feedback from galaxies and quasars. 

\end{abstract}


\section{Introduction}

While the \lyaf\ was discovered long ago \citep{1971ApJ...164L..73L},
a clear physical picture was not settled on until relatively 
recently.
Observations of absorption in pairs of spectra showing coherence 
of \lyaf\ absorption over hundreds of kpc 
\citep{1994ApJ...437L..83B,1994ApJ...437L..87D}
demonstrated the key result that came
from numerical simulations of the intergalactic medium (IGM), 
that the absorption features arose in low density
structures that must contain a large fraction of the baryons and 
merge
continuously with the background, instead of being dense, discrete 
systems.
This was confirmed when the Keck HIRES 
spectrograph \citep{1994SPIE.2198..362V}
produced fully resolved spectra that were qualitatively explained
by hydrodynamic simulations and semi-analytic models 
\citep{1992A&A...266....1B,1994ApJ...437L...9C,1995ApJ...453L..57Z,
1996ApJ...457L..51H,1996ApJ...471..582M,
1997MNRAS.292...27H,1997ApJ...477...21D,
1998MNRAS.301..478T,1998MNRAS.296...44G}. 
The \lyaf\ absorption appears to arise from continuously 
fluctuating photoionized gas in the IGM,
with density near the universal mean and temperatures around $10^4$ K.
The structure
of the absorption field can be derived from the primordial
density field with reasonable accuracy using 
numerical simulations, smoothed on scales smaller than
a few hundred comoving $\hkpc$ by gas pressure and thermal 
broadening in redshift space.

Starting with \cite{1998ApJ...495...44C}, the statistic of choice
for comparing observations of the \lyaf\ to predictions of different
cosmologies has been the power spectrum, $\PF$, of the transmitted
flux fraction, $F(\lambda)=\exp[-\tau(\lambda)]$.    
Observational measurements of $\PF$ have been presented in several 
recent papers
\citep{1998ApJ...495...44C,2000ApJ...543....1M,
2002ApJ...581...20C,2004MNRAS.347..355K,2004MNRAS.351.1471K,
2004astro.ph..5013M}.
In parallel with the observational efforts there has been considerable
effort to interpret these measurements using 
numerical simulations
\citep{1998ApJ...495...44C,
2000ApJ...543....1M,2001ApJ...557..519Z,2002ApJ...581...20C,
2002MNRAS.334..107G,2003ApJ...590....1Z,2003MNRAS.342L..79S,
2004MNRAS.354..684V}.
Other statistics of the fluctuations in transmitted flux are also 
useful, with recent papers studying the bispectrum 
\citep{2003MNRAS.344..776M,2004MNRAS.347L..26V,2004ApJ...606L...9F} and
very large scale fluctuations \citep{2004ApJ...617....1T}.

In the standard picture of the Ly-$\alpha$ forest
the gas in the IGM is in ionization equilibrium.  
The rate of ionization by the UV background balances the rate of
recombination of protons and electrons.
The recombination rate depends on the temperature of the gas, which
is a function of the gas density.
The temperature-density relation can be parameterized by an amplitude,
$T_0$, and a slope $\gamma-1=d\ln T/d\ln \rho$.
The uncertainties in the intensity of the UV background, the mean
baryon density, and other
parameters that set the normalization of the relation between optical
depth and density can be combined into one parameter: the mean transmitted
flux, $\bF$.  We always
treat $\Tp$ (we follow \cite{2001ApJ...562...52M} in 
specifying the temperature-density relation at density 1.4 times the mean), 
$\gmo$, and $\bF$ as the independent (adjustable) variables
in our analysis.  For example, when we perform a convergence test comparing
two simulations with different resolution we compare at fixed $\bF$, even
though this may require us to use different strengths of the ionizing 
background when constructing the simulated spectra. 

In general the flux power spectrum, $\PF$, is a function of 
the linear matter power spectrum, $P_L(k)$, cosmological parameters
such as the matter density, $\Omega_m$, which we 
denote collectively as $\mathbf{p}_{\rm cosmology}$, and
parameters of the \lyaf\ model,
which we denote as
$\mathbf{p}_{\rm forest}$ (parameters in addition to $\Tp$, $\gmo$, and 
$\bF$ are introduced 
later). 
In observationally favored $\Lambda$CDM models 
the universe is Einstein-de Sitter at $z>2$, so 
if velocity units are used for $k$ 
we can drop the dependence on cosmological parameters, 
which determine the relation between velocity and comoving 
coordinates. (This relation must of course be reinstated when 
comparisons to explicit cosmological models are performed, but 
this is not a subject of this paper.)  

We do not attempt to
invert the flux power spectrum to a band-power 
description of $P_L(k)$.  The linear power spectrum
$P_L(k')$ contributes to $P_F(k)$ at all $k$ (we 
use $k'$ and $k$ here to make it clear that the 
two are not fundamentally connected), and
the transformation is generally nonlinear.  
As a result inversion requires a
large number of simulations in which the power in the 
bands is varied, in principle in combination and by
varying amounts. 
This does not mean that such an inversion is 
impossible, but simple attempts we tried to devise have failed 
and current inversion treatments that exist in the literature 
are not sufficiently reliable for this purpose  
\citep{2003ApJ...590....1Z,2003MNRAS.342L..79S}. 

Instead we parametrize the information we wish to extract in terms 
of $\Delta^2_L(k,z)\equiv k^3 P_L(k,z)/2 \pi^2$,
$n_{\rm eff}(k,z)\equiv d\ln P_L/d\ln k$, and
$\alpha_{\rm eff}(k,z)\equiv d n_{\rm eff} /d\ln k$, 
the amplitude, logarithmic slope, and curvature of
$P_L$, all evaluated 
at a pivot redshift $z_p$ and pivot wavenumber $k_p$, 
at which the information is near maximum. 
We adjust these variables in simulations, 
covering a broad range of 
values to obtain predictions of the flux power spectrum over the 
whole range of interest.

Our analysis is based on the $\PF$ measurement of
\cite{2004astro.ph..5013M}, which used 3300
Sloan Digital Sky Survey spectra from data releases one
and two 
\citep{1996AJ....111.1748F,1998AJ....116.3040G,2000AJ....120.1579Y,
2001AJ....122.2129H,2002AJ....123..485S,2002AJ....123.2121S,
2002AJ....123.2945R, 2003AJ....125.1559P,
2003AJ....125.2276B,2003AJ....126.2081A,2004AJ....128..502A}. 
The SDSS sample is nearly two orders of magnitude larger than the 
samples available previously. Because the spectra are of lower 
resolution than HIRES spectra the small scale information 
is erased, so we supplement our study with 
the HIRES-based 
$\PF$ measurement of \cite{2000ApJ...543....1M}.
We do not include in our standard analysis the more recent measurements by
\cite{2002ApJ...581...20C} and 
\cite{2004MNRAS.347..355K,2004MNRAS.351.1471K}, 
because these show 
signs of a systematic discrepancy and/or underestimation of errors
when compared to SDSS \lyaf\ data \citep{2004astro.ph..5013M};
however, we do present an alternative analysis using these results with
some allowance for systematic errors, which gives results 
consistent with our standard analysis.

This paper is part of a closely intertwined set of four papers,
including \cite{2004astro.ph..5013M}, \cite{2005MNRAS.360.1471M},
and \cite{2005PhRvD..71j3515S}. 
The observational measurement of $\PF$ was presented 
in \cite{2004astro.ph..5013M}, 
which stands alone independent of theory,
and makes a strong case that the systematic errors in the measured
flux power spectrum are for practical purposes 
smaller than the statistical errors.
The present paper transforms the flux power spectrum measurement into a
constraint on the amplitude, slope, and curvature of the linear theory
matter power spectrum at z=3 and comoving scale of a few Mpc.  This
constraint should apply to a wide range of cosmological models with
linear power spectra similar to those favored by current observations,
though it should not be applied to models with sharp breaks in the power
spectrum on the scales of the measurement or to warm dark matter models
(see more discussion below).  For this class of models, we believe that the
systematic errors in our inferred linear $P(k)$ constraints are also below
the statistical errors (after several effects that would otherwise lead
to systematic errors are included in the fit through nuisance parameters), 
though more testing with hydrodynamic simulations
is desirable as discussed below.  The results of this paper allow the
SDSS flux power spectrum measurement to be incorporated in a straightforward
way into cosmological parameter constraints drawing on multiple cosmological
observables.  We defer this task to a separate paper,
\cite{2005PhRvD..71j3515S}, since it requires
discussion of the other data sets to be used and the methodology for
combining them.  However, we note that the additional leverage provided
by the \lyaf\ power spectrum at small scales allows much improved
constraints on the inflationary spectral index, $n$, the running of that
index with scale, and neutrino masses.  Also, some of the details on
how we treat high column density systems and UV background fluctuations, and
an investigation of galactic winds are described in another paper,
\cite{2005MNRAS.360.1471M}.

The layout of this paper is as follows:
Section \ref{secbasesims} gives a detailed 
description of how we make our
prediction of $\PF$ given $P_L(k)$, $\mathbf{p}_{\rm cosmology}$, and
$\mathbf{p}_{\rm forest}$.  Section 3 describes how we perform 
$\chi^2$ fits to the observations to estimate $\DL$ and $\neff$
and their errors.  Finally, \S 4 contains our conclusions.

\section{Numerical Simulations of $\PF$ \label{secbasesims}}

In this section we explain how we translate any given set of model 
parameters into a prediction of $\PF$.  
We assume that any winds from galaxies do not effect $\PF$ beyond
the modest effect of the local energy injection in our hydrodynamic
simulations (we do allow for some uncertainty in this effect by 
marginalizing over the differences
between three versions of the feedback in the simulations). 
Winds are explored in more detail in a companion paper
\citep{2005MNRAS.360.1471M}. 
We also assume that the density-temperature-neutral density relation 
is not made inhomogeneous by inhomogeneous reionization and heating
(i.e., patchy reionization of either hydrogen or helium).
We expect to investigate these issues in the future. 

\subsection{Background}

In the redshift range of interest, $2\lesssim z \lesssim 4$, the
Universe is expected to be nearly Einstein-de Sitter (EdS) in 
typical $\Lambda$CDM models.
The growth factor for 
linear perturbations, $D(z)$, is nearly  
proportional to $a=1/(1+z)$, e.g.,  
$[(1+2) D(z=2)]/[D(z=4) (1+4)]=(1.0, 0.992, 0.981)$
for flat models with $\Omega_m=(1.0, 0.4, 0.2)$.
Similarly, to a good approximation $H(z)=\dot{a}/a$ evolves 
like $(1+z)^{3/2}$, e.g.,
$[H(z=2)/(1+2)^{3/2}]/[H(z=4)/(1+4)^{3/2}]=(1.0, 1.021, 1.055)$
for the same three models.  
This means that when analyzing the \lyaf\ alone, we generally
do not need to specify a model, as long as we measure distances
in $\kms$.
Conversion to comoving $\hmpc$ for comparison of the power spectrum 
to measurements at other redshifts of course requires
a model.  
We only display our results in $\kms$.  
Conversions factors for flat 
$\Lambda$CDM models range from
$83 (\kms)/(\hmpc)$ at $z=2$ for $\Omega_m=0.2$ to $142 (\kms)/(\hmpc)$
at $z=4$ for $\Omega_m=0.4$, so one can get a 
qualitative idea of the comoving $\hmpc$ 
scale of a figure by dividing $\kms$
by 100.

As stated in the introduction, our goal is to generate a grid 
of simulations covering the range of interest. 
When this project started, it was impractical to run hydrodynamic 
simulations for every model needed, 
because of the CPU requirements for these simulations combined with
the large range of parameter space allowed by 
existing constraints. 
For this reason in this paper we use 
hydro-particle-mesh (HPM) simulations \citep{1998MNRAS.296...44G},
calibrated by a limited 
number of fully hydrodynamic simulations.  
For the next generation analysis, it should be possible 
to employ hydrodynamic simulations only, 
both because of 
increasing computer power, but also because we can now 
focus on a smaller volume in parameter space 
(note, however,
that freedom in the temperature-density relation 
will inevitably be cumbersome to implement within hydro 
simulations and  
approximations similar to those made in HPM simulations may still 
be required).  

Our standard set of HPM simulations were normalized to 
$\Delta^2_L(k_s,z_p)=0.29$, 
with $k_s=0.0078~\ikms$ at $z_p=3.0$ (note
that this pivot point is slightly different from the 
one at which we report the final inferred power, because
the simulations were performed before the observational
pivot point was known).
We generally use outputs at different redshifts (labeled
by expansion factor) in 
place of explicit changes in the power spectrum 
amplitude, although we also have some simulations 
with alternative normalizations (our final measured power 
corresponds to a $\sim 20$\% higher expansion 
factor in the most common simulations than the real Universe). 
Throughout this section on numerical details
we will usually show three examples, $a=0.24~(z=3.17)$, 
$\bF=0.67$, 
which is near the center of weight of our 
data, $a=0.32~(z=2.12)$, $\bF=0.85$, which is near the low redshift
end of our data, and $a=0.2~(z=4)$, $\bF=0.4$, which is near the
high redshift end of our data.  Unless otherwise noted, we
show simulations with $n_{\rm eff}(k_s,z_p)=-2.3$ and 
$\alpha_{\rm eff}(k_s,z_p)=-0.2$,
values near the best fit to the data.

Our basic simulation strategy is as follows,
with the details explained in the rest
of the section:
We use $L=40\hmpc$ simulations
for our main grid for three reasons:  
we need to predict $\PF$ to this scale, we 
expect that there is a small systematic error
related to finite box size for smaller 
simulations, and use of these larger simulations produces
smaller statistical errors on $\PF$.  
We do not have the capability to run large 
numbers of $N\geq 1024^3$ simulations 
($N$ is the number of particles and cells, which 
are always equal in number in this paper), which
are needed to compute $\PF$ to the
accuracy we require, so we use $N=512^3$, 
with a correction for the limited resolution.
The correction is made by comparing (20,512)
simulations to (20,256) simulations,
where we describe simulation size and resolution
using the shorthand notation
($L$,$N^{1/3}$), where $L$ is the box size in 
$\hmpc$ (we always use an equal number of particles
and cells in this paper).
Finally, we calibrate the approximate HPM
method by comparing (10,256) simulations 
to fully hydrodynamic simulations with 
identical initial conditions.
We now describe this procedure in detail, 
building up from the hydrodynamic 
simulations.

\subsection{Hydrodynamic Simulations \label{sechydrosims}}

Our hydrodynamic simulations use the code described 
in \cite{2003ApJ...598..741C}.  We use an $L=10\hmpc$ 
box, with $N=256^3$ cells.  To the limited extent that
it matters, the cosmological model is 
flat $\Lambda$CDM with $\Omega_m=0.3$, $\Omega_b=0.04$,
and $h=0.7$.  The power spectrum has 
$\Delta^2_L(k_s=0.0078~{\rm s/km},z_p=3.0)=0.29$, 
$n_{\rm eff}(k_s,z_p)=-2.41$, 
and $\alpha_{\rm eff}(k_s,z_p)=-0.2$.  
The main simulation, which we will 
call FULL (full physics), has feedback in the form of 
localized energy injection by supernovae. 
The winds that are produced do 
not have a large effect on $\PF$.  We explore 
the effects of winds in more detail in a companion paper 
\citep{2005MNRAS.360.1471M}. 
The supernovae also inject
metals which are followed dynamically and influence cooling.

\begin{figure}
\plotone{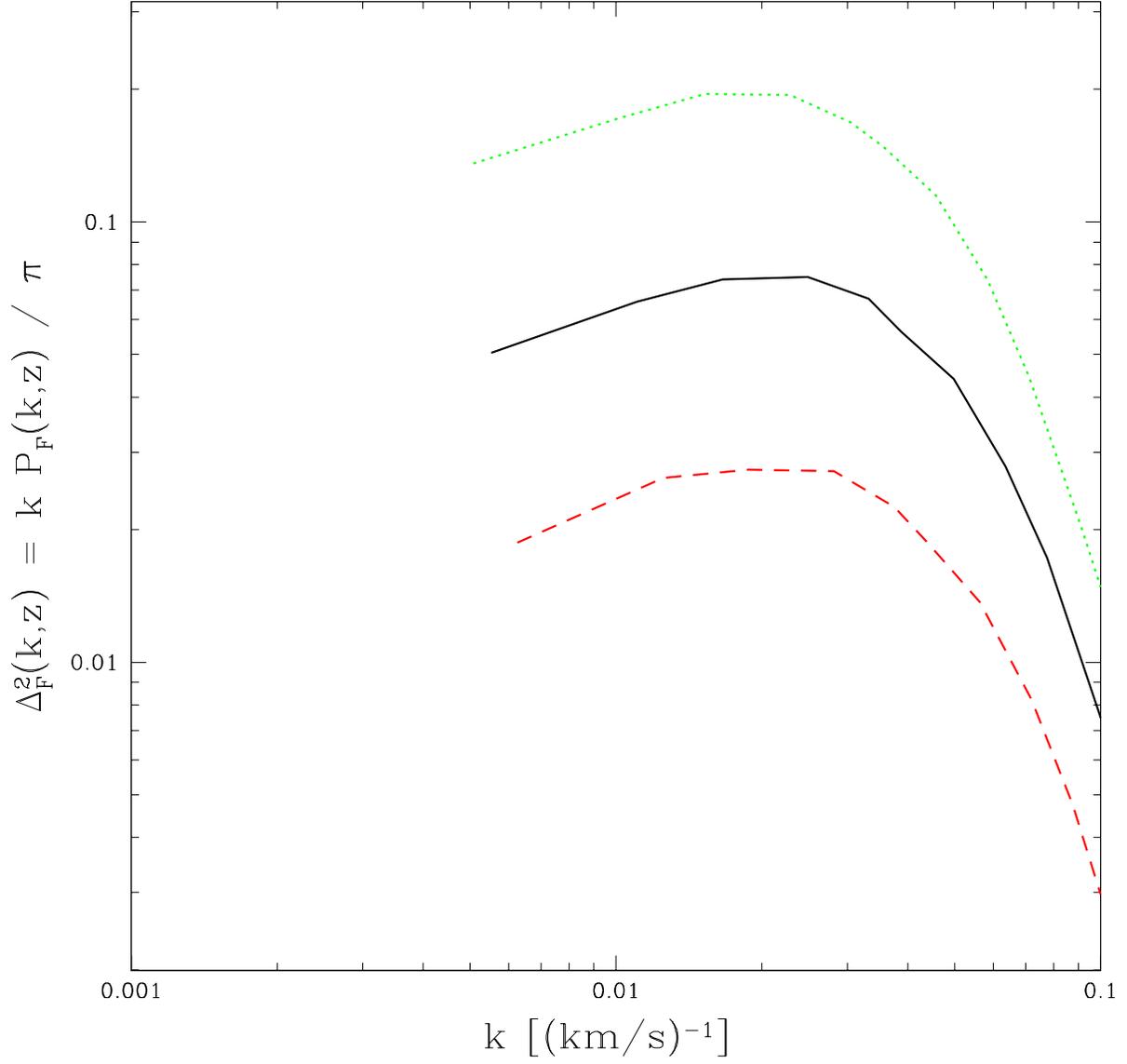}
\caption{
$\PF$ prediction from our basic hydrodynamic simulation (FULL).
The lines show, from bottom to top, $a=0.32$, 0.24, and 0.20, 
with $\bF=0.85$, 0.67, and 0.4.
}
\label{basichydropower}
\end{figure}

For the rest of this section, we will generally show ratios
of $\PF$ calculations, but, for reference,
Figure \ref{basichydropower} shows $\PF$ results from our
main hydro simulation, for outputs representing, roughly,
the central redshift of our data, $z\sim 3$, and the 
low and high redshift extremes, $z\sim 2$ and $z\sim 4$. 
We see the expected increase in power with increasing 
redshift, due to the increase in mean absorption.
This simulation box is too small to compare directly to 
the data,
and we need simulations of many more models, but this is
the base on which the analysis rests.

\begin{figure}
\plotone{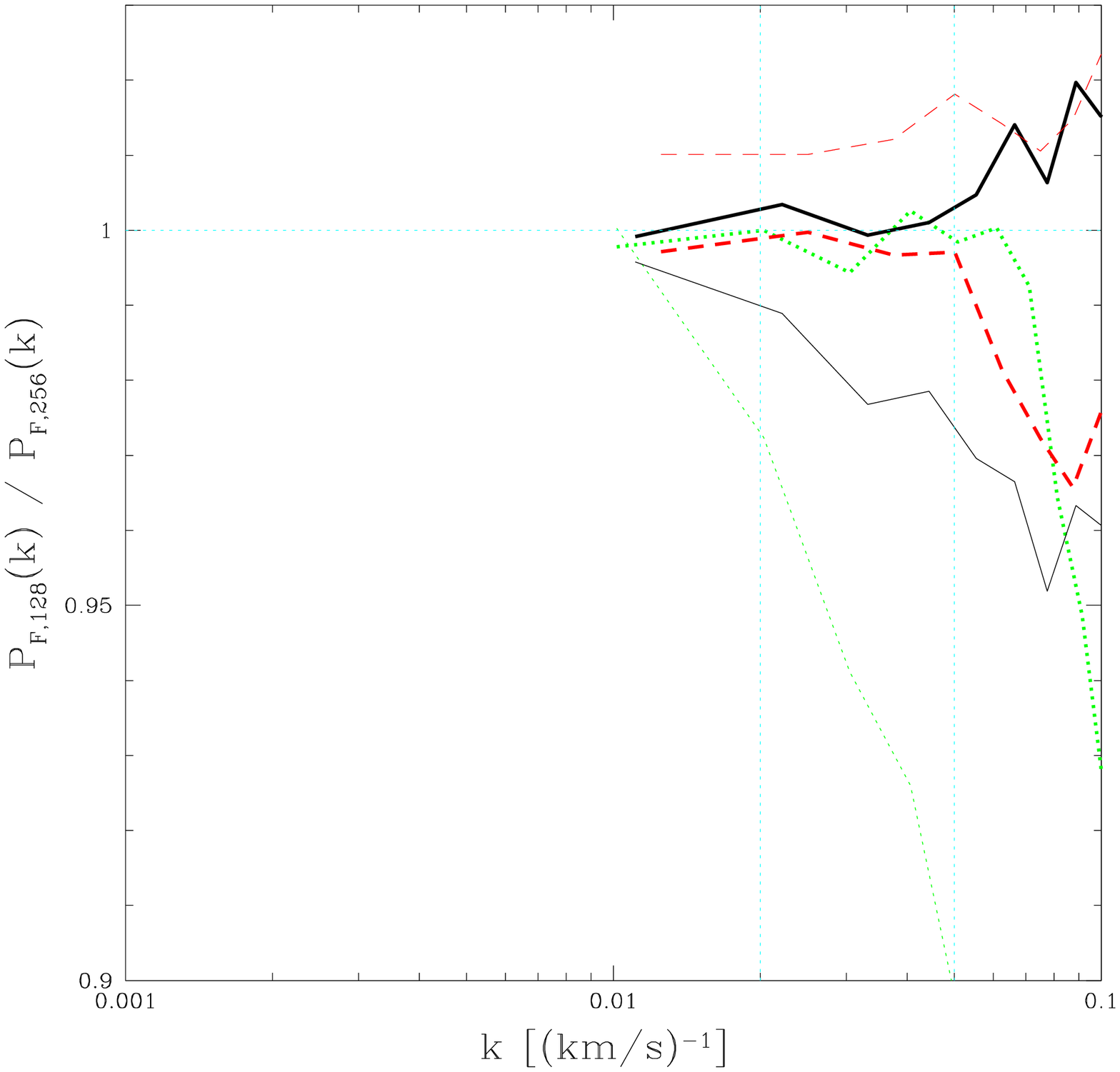}
\caption{
Resolution test for the hydrodynamic simulation, showing the 
ratio of $\PF$ in a (5,128) run to $\PF$ in a (5,256) run.
The (red/dashed, black/solid, green/dotted) lines show
$a=$(0.32, 0.24, 0.20), with $\bF=$(0.85, 0.67, 0.4).
Thin and thick lines, respectively, show before and after
the redshift of reionization adjustment.  The vertical
cyan (dotted) lines mark the upper limits on $k$ used
for SDSS and HIRES $\PF$ measurements, while the 
horizontal dotted line guides the eye to 1.  We use
these same cyan/dotted lines in many figures 
(and occasionally another at $k=0.0013\ikms$,
which marks the lower limit on $k$ used in our fits).
}
\label{hydrorestest}
\end{figure}

We show a resolution convergence test in Figure \ref{hydrorestest}.
For this test we compared fully hydrodynamic runs of (5,256) 
and (5,128) (the latter has the same resolution as our base 
simulations).  

Interpreting a resolution test of our hydrodynamic simulations 
requires some subtlety.  
Because of the detailed small-scale physics in the simulations,
the time of reionization and the amount of heating during it 
are somewhat sensitive to resolution, even when we use the 
same ionizing background in both simulations (as we did for
this resolution test -- usually the homogeneous radiation background is 
computed from stars and AGN generated within the running simulation).  
For example: while the simulations do not include realistic 
radiative transfer, 
we do use a rough self-shielding approximation to attenuate
the radiation background seen by high neutral density cells.
In this resolution test the lower resolution simulation is 
$\sim 5000$ K hotter between reionization at $z\sim 10$ and
$z\sim 7$, with the difference decreasing at lower redshift.
Simple differences in the thermal history do not concern us 
in practice. 
In our power spectrum analysis we marginalize over the 
temperature-density relation and the
small-scale smoothing level (which is sensitive to the full
thermal history back to reionization), so changes of this 
kind will be automatically accounted for.
In Figure \ref{hydrorestest} we first show (thin lines) the
comparison when we correct only for the difference in 
temperature-density relation at the time of observation,
i.e., differences in $\Tp$ and $\gmo$.  
We see that, while the two resolutions agree to a few 
percent at $a=0.32$ and $a=0.24$, the disagreement 
at $a=0.2$ (and, probably more importantly, $\bF=0.4$) 
is relatively large. 
We next allow for an adjustment in the filtering scale, 
equivalent to a change in the redshift of reionization. 
We implement this, as described in more detail below, by
interpolating between HPM runs with reionization at $z=7$
heating the gas to 25000 K and reionization at $z=17$ with 
heating to 50000 K (in other contexts we have spot-checked
that this interpolation is accurate).  We require 27\% of
the difference between these two cases to produce the 
thick lines in Figure \ref{hydrorestest} (we also adjusted
$\bF$ in the two lower $z$ bins by 0.002 -- a tiny
amount relative to the uncertainties in $\bF$).  The 
agreement is excellent, indicating that any effect of 
limited resolution is degenerate with the nuisance
parameters we are already marginalizing over.  Some 
further investigation using HPM simulations with thermal
histories matching those in the different resolution 
hydrodynamic simulations suggests that only about 1/3
of the effect is simply differences in thermal history.
The other 2/3 must be an early-time smoothing of the gas 
by limited resolution.  

The reader may at this point 
wonder why we believe that 5 $\hmpc$ simulations are
sufficient for this resolution test.  They would not 
be adequate if we needed to make any kind of correction using
them directly, because the extrapolation to large
scales would be very uncertain; however, we use them
only to motivate a physical interpretation of the 
effect of limited resolution as a modification of the
early-time thermal history (i.e., the 
reionization history).  Since this seems to work
so well, we believe the freedom we allow in the 
fits (see below) is sufficient to absorb any 
resolution-related error.  This will be checked in the 
future with larger simulations.

\begin{figure}
\plotone{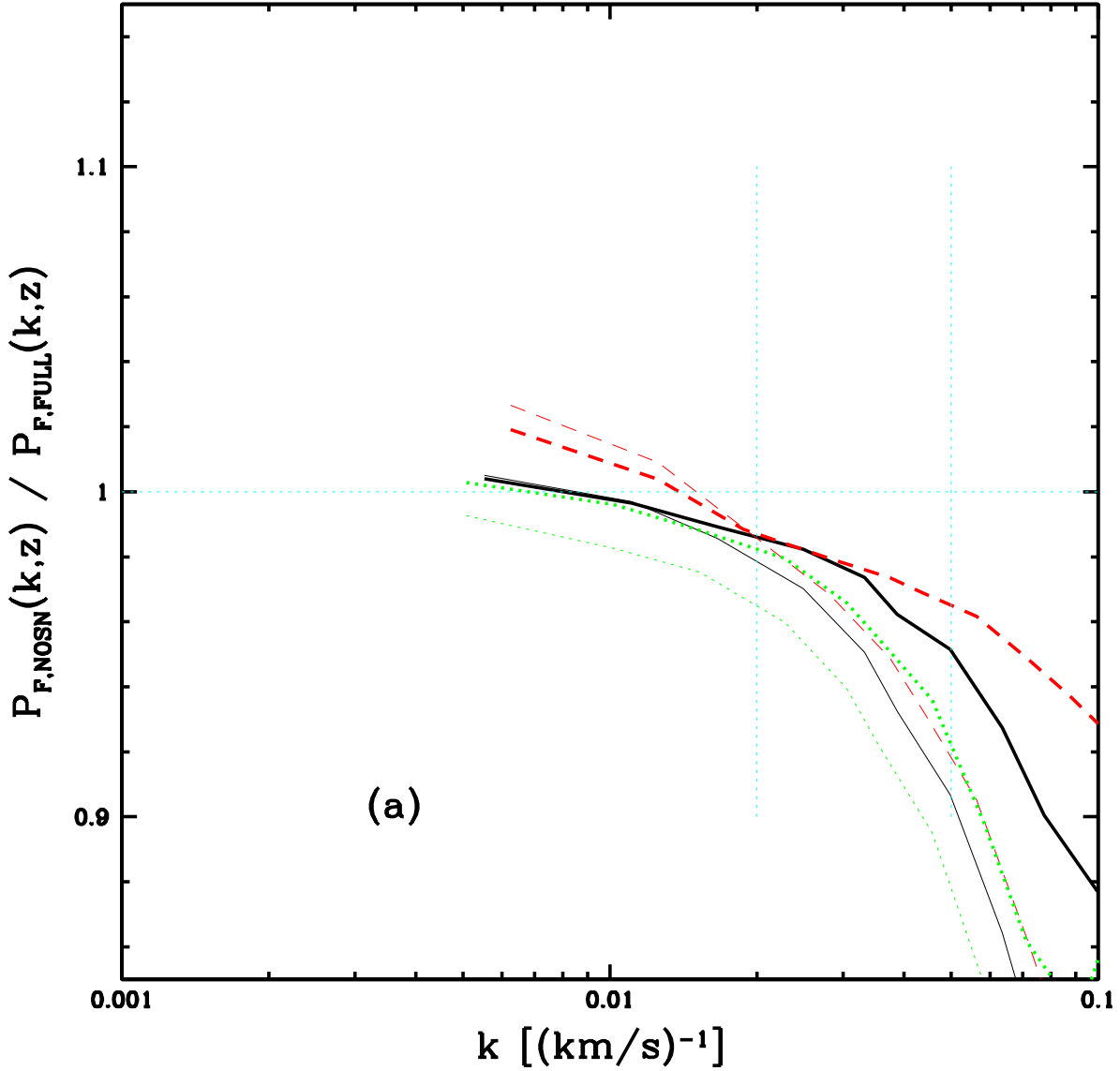}
\caption{
Comparison of hydro simulations including different physics.
(a) shows the ratio of $\PF$ in the NOSN (no energy feedback
from supernovae) simulation to the FULL simulation.
(b) shows $P_{\rm NOMETAL} / P_{\rm FULL}$ (NOMETAL means
no metal cooling).  
The thick lines show the power after we correct for differences in the 
bulk temperature-density relations in the simulations, 
while the thin lines show the uncorrected power.
The (red/dashed, black/solid, green/dotted) lines show 
$a=$(0.32, 0.24, 0.20), with $\bF=$(0.85, 0.67, 0.4).
The horizontal dotted line guides the eye to 1, while the
vertical dotted lines mark the $k$ to which we use SDSS
and HIRES data [$k<0.02~\ikms$ and $k<0.05~\ikms$, 
respectively]. 
}
\label{hydrophysicscomp}
\end{figure}

We have two additional alternative-physics hydrodynamic runs. 
The first one does not have metal cooling 
and we call it NOMETAL, the second one 
does not have energy feedback from supernovae
and we call it NOSN
(the metals in the NOSN
simulation still come from supernovae, i.e., they
are not evenly distributed).
Figures \ref{hydrophysicscomp}(a) and (b) show the 
ratio of $\PF$ from NOSN and NOMETAL, respectively,
to $\PF$ from FULL.  
\begin{figure}
\plotone{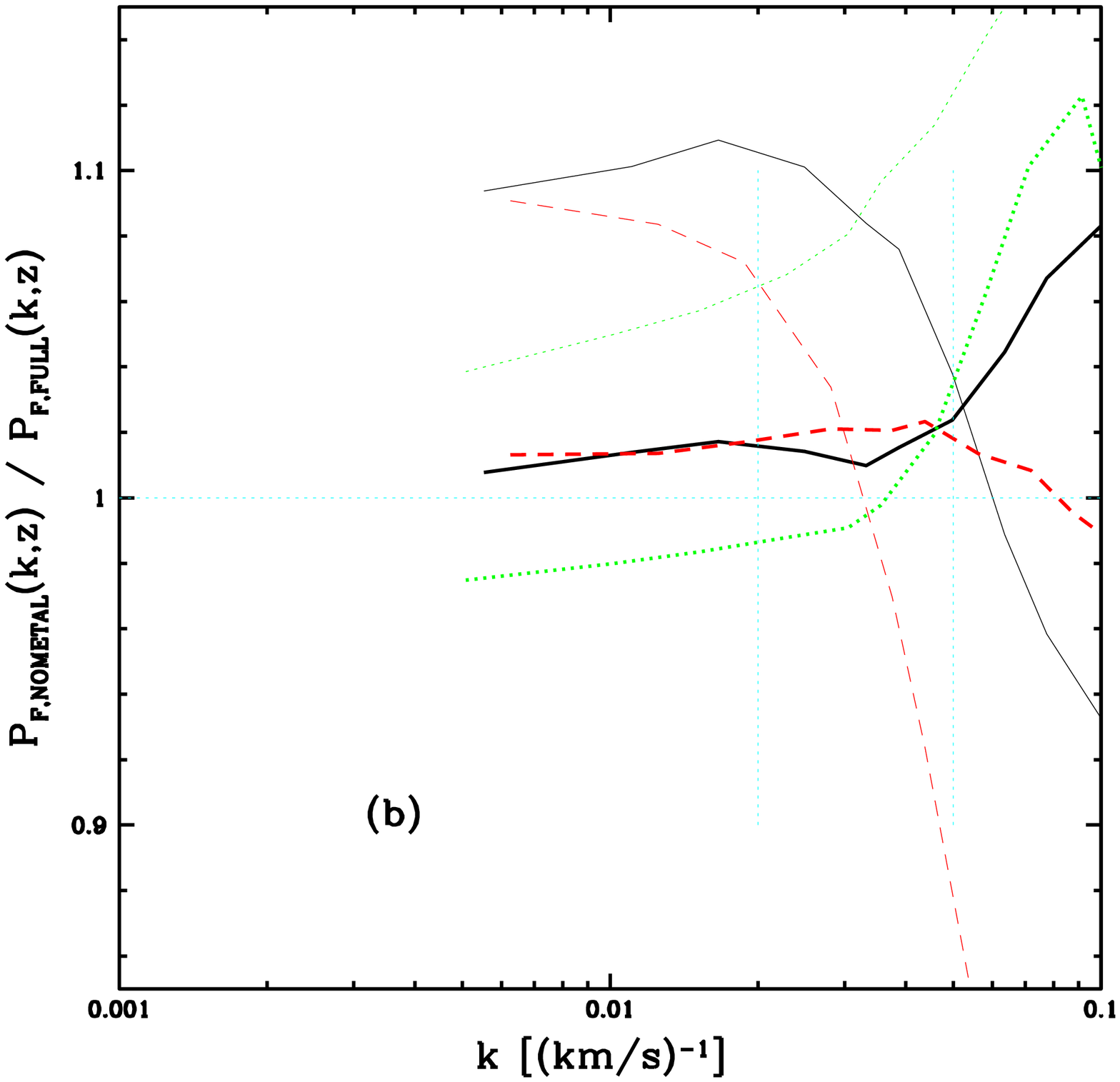}
\end{figure}

The results from these simulations
are not the same, at a level that, we will see later,
does matter to us at the $\sim 1\sigma$ level.  
Some of the difference is simply 
a difference in the temperature-density relation 
within these simulations, which will be automatically
accounted for when we use them to calibrate our HPM
simulations. For example, the NOMETAL simulation 
is typically hotter, with
smaller $\gmo$ --- the $\sim 10$\% disagreements 
seen in figure \ref{hydrophysicscomp}
are reduced to below $5$\% when this 
is accounted for, as one can see by comparing  
Figure \ref{cencomp}(a) and (c). 
These differences are thus not necessarily worrisome and 
only a full fit to the data can reveal their impact 
on the cosmological conclusions.  
When we perform our final fit to determine the 
mass power spectrum, we will include the differences
between these simulations as an uncertainty in the
fit by defining $P_{\rm hydro} = a~P_{\rm NOMETAL} +
b~P_{\rm NOSN} + c~P_{\rm FULL}$, with 
$a$, $b$ and $c$ free parameters subject to the constraints
$0< a, b, c < 1$ and $a+b+c=1$.  
This procedure thus includes the systematic uncertainties that arise
from these simulations, but also allows the possibility 
that simulations which better fit the data receive more weight. 

\subsection{Calibrating the HPM Simulations}

Our hydro-particle-mesh (HPM) simulations model the IGM as simply 
particles evolving under gravity plus a pseudo-pressure term 
computed from an arbitrarily imposed temperature-density relation 
\citep{1998MNRAS.296...44G}.   
They are not expected to simulate high density regions accurately
because they do not contain shocked or cooled gas, but these
regions occupy very little of the volume of 
the IGM and typically produce saturated absorption, 
and for both of these 
reasons have minimal influence on the \lyaf\ power spectrum.
The other approximation in the code we use (kindly provided
by N. Gnedin) is the treatment of gas and dark matter with a
single set of particles.  Ultimately, the accuracy of the
simulations must be verified by direct comparison with fully
hydrodynamic simulations.  As we will see, the agreement on 
$\PF$ is very good.  In fact, the HPM simulations agree with 
the hydrodynamic simulations as well as hydrodynamic simulations
with different forms of galaxy feedback agree with each other.

We use the approximate HPM simulations 
for our main grid of models for 
two reasons:  The obvious and most important one is that they 
are less costly to run -- while we do not have a direct 
comparison with the fully hydrodynamic
code that we use for the simulations in this paper, simulations 
using the 
publicly available ENZO code \citep{2004astro.ph..3044O} 
require a factor of $\sim 50$ more  
CPU time than similar HPM simulations.
Another useful advantage of the HPM simulations
is that we can control the thermal history in them very easily.  It is 
unlikely that we will ever be able to predict the thermal
history from first principles using a hydrodynamic simulation, 
because of uncertainty in the simulation of radiation sources.
Therefore, any proper analysis of the \lyaf\ observations must 
marginalize 
over all plausible thermal histories.  While it will certainly 
be possible in the future to manipulate fully hydrodynamic 
simulations to achieve this marginalization (this has been
done on a small scale before, e.g., \cite{2000MNRAS.318..817S}), 
for now we do
it using HPM simulations.  We do not, however, assume that
the HPM simulations are perfectly accurate.  In this subsection
we explain how we use a limited set of hydrodynamic simulations
to calibrate the HPM simulations, i.e., to correct for any 
error in the HPM simulations.

\begin{figure}
\plotone{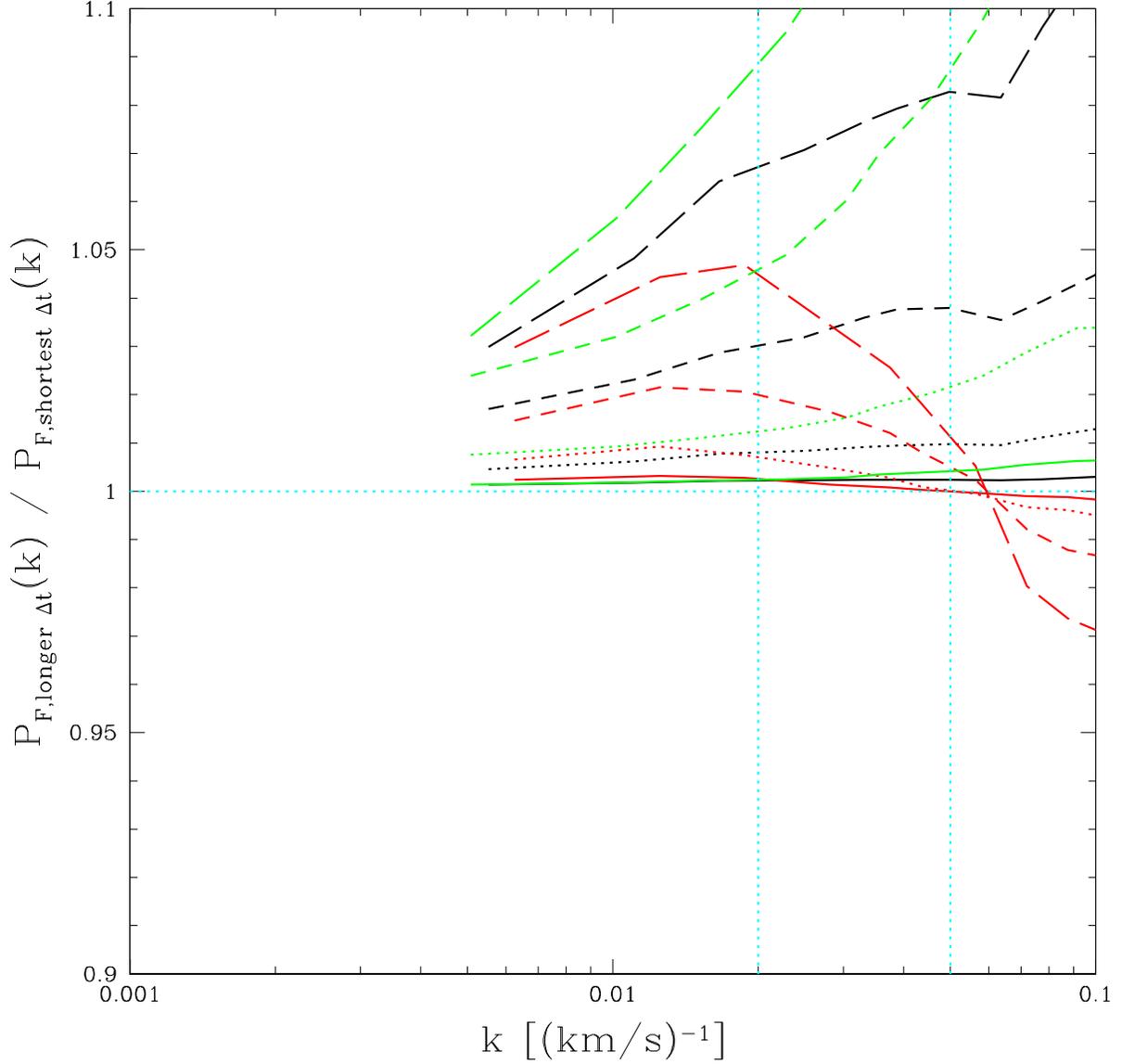}
\caption{
Convergence of $\PF$ with decreasing time step size for
the (10,256) HPM simulations that we compare to hydro
simulations.  Note that, of the simulations of this size, only the  
ones with the smallest
timestep (most total steps) are used in our analysis.
The denominator is the result for 876 times steps down to
$z=1.5$, while solid, dotted, dashed, and long-dashed 
lines show, respectively, 429, 205, 89, and 42 steps.
Red, black, and green indicate $\PF$ at, respectively, 
$a=0.32$, 0.24, and 0.20, with $\bF=0.85$, 0.67, and 0.4
(these run from bottom to top in each case when looking 
at $k=0.05\ikms$).
}
\label{timecencomp}
\end{figure}

We compare the hydrodynamic simulations discussed in
\S \ref{sechydrosims} to 
a (10,256) HPM simulation with identical initial 
conditions.
We use $N=256^3$ to match the resolution of our
(20,512) simulations.
We show the convergence with the time step in 
Figure \ref{timecencomp}. 
We use 876 steps down to $z=1.5$, although we have 
checked explicitly that 205 would have been sufficient
to produce the same final $P_L$ result (note that we 
use the fully converged HPM simulation for 
our comparison with full-hydro simulations,
not the corrected long-timestep HPM simulations discussed 
below).

\begin{figure}
\plotone{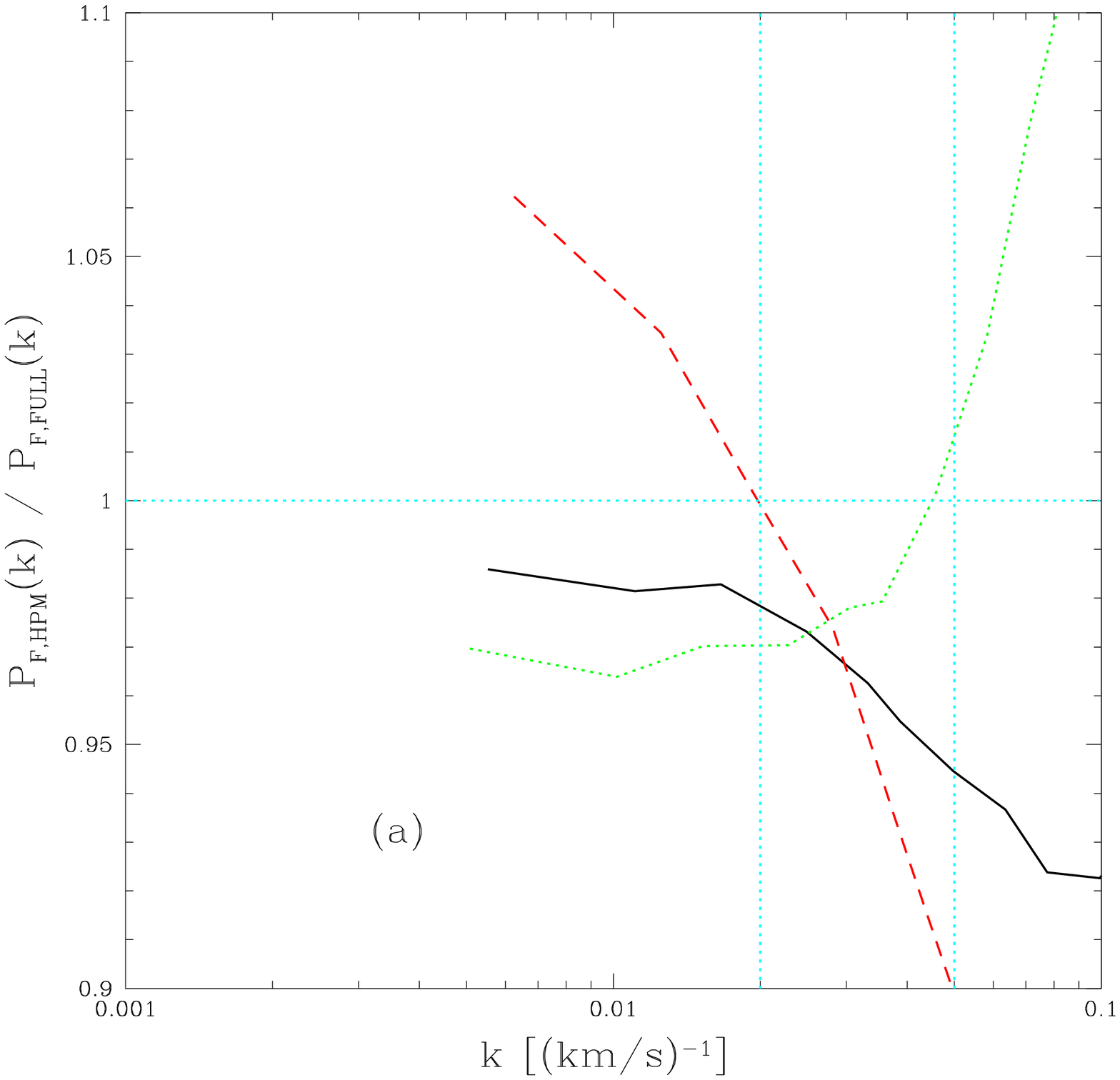}
\caption{
Comparison of the hydrodynamic results to the HPM 
results, for the same initial conditions and 
temperature-density relation.  (a), (b), and (c)
show, respectively, the comparison for the FULL,
NOSN, and NOMETAL hydro simulations.
The (red/dashed, black/solid, green/dotted) lines show 
$a=$(0.32, 0.24, 0.20), with $\bF=$(0.85, 0.67, 0.4).
}
\label{cencomp}
\end{figure}

Figures \ref{cencomp}(a,b,c) show the HPM simulation
compared to the three hydrodynamic simulation versions
(FULL, NOSN, NOMETAL).  
\begin{figure}
\plotone{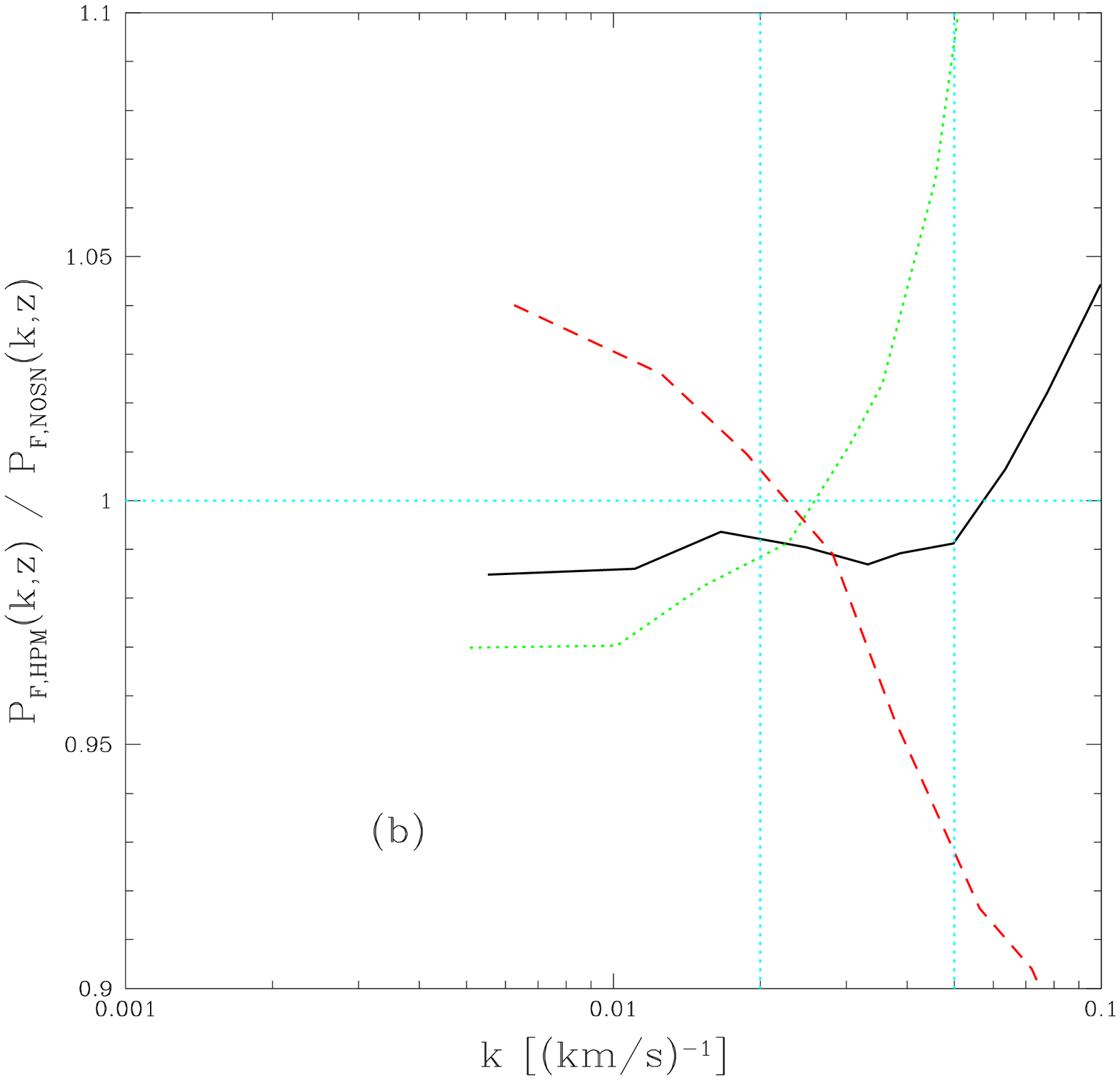}
\end{figure}
\begin{figure}
\plotone{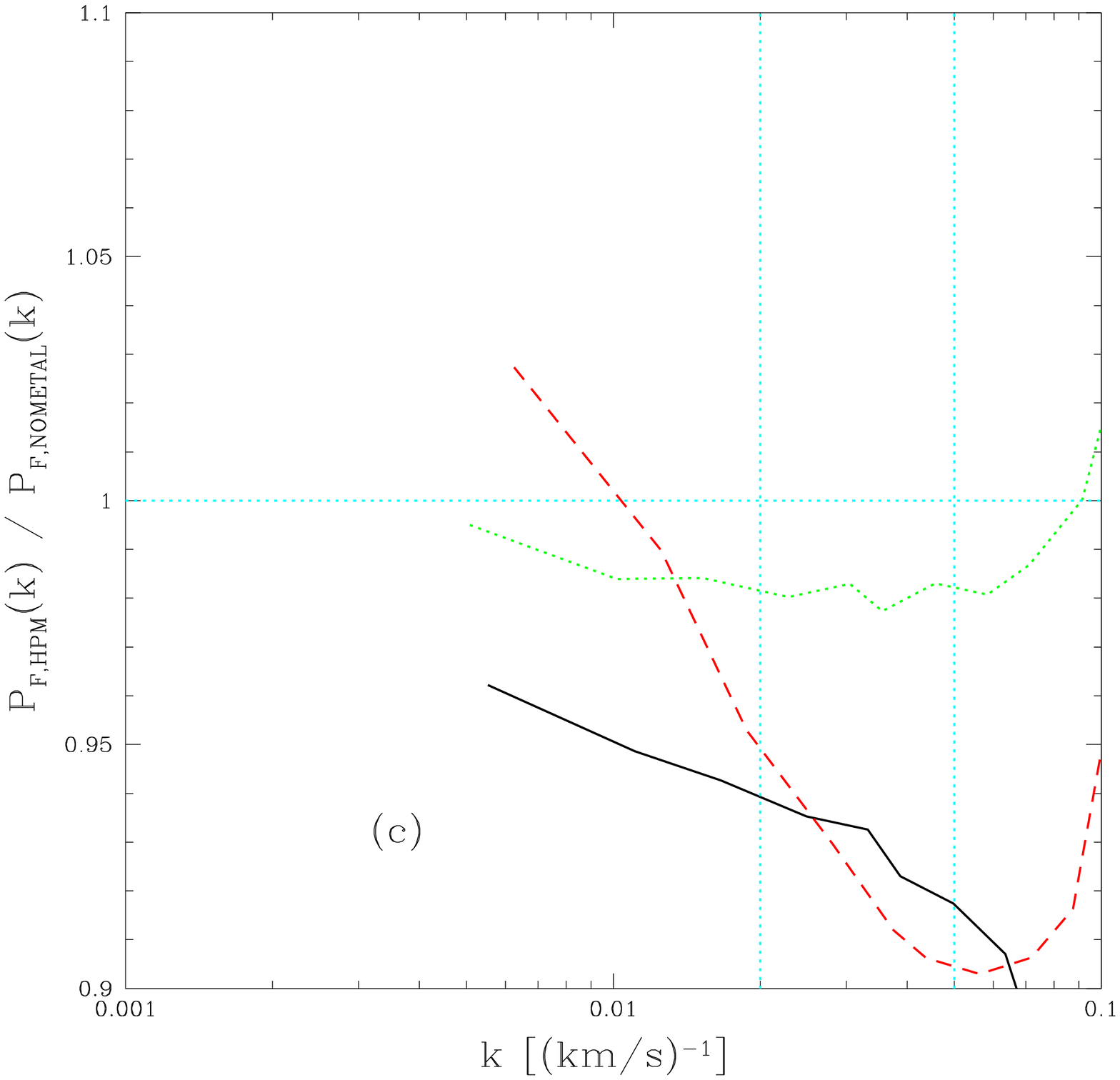}
\end{figure}
In each case we have 
used the temperature-density relation computed from
the hydro simulation when creating the HPM spectra.
Operationally, we estimate $T_{1.4}$ and $\gmo$ for
the hydro simulation by a least absolute deviation 
fit \citep{1992nrca.book.....P} to $\ln T$ vs. $\ln (\rho/\bar{\rho})$, 
limited to the range $1<\rho/\bar{\rho}<2$ (there is no
unambiguously best way to make this estimate).
We see that the agreement is generally quite good in the $k$ range that we use,
although this is less true at $z=4$, $\bF=0.4$.  The general increase in 
disagreement at $k\sim 0.03\ikms$ can be understood qualitatively by looking 
forward
to Figure \ref{pardep}, which shows the parameter dependence of $\PF$.  We see
that this scale corresponds to the scale where thermal broadening suppression
of the power is rapidly becoming significant.  Furthermore, changes in pressure
history (i.e., reionization) are becoming more important, and the ``fingers of 
god''-like suppression of small-scale power by non-linear peculiar velocities 
becomes so 
significant that increasing the linear power actually begins to reduce the 
flux power.  In other words:  all of the details that make HPM an approximation
are becoming significant for $k\gtrsim 0.03\ikms$.  It is not entirely clear
why the disagreement generally becomes substantially worse at $z=4$, but it 
seems likely that the hydrodynamic simulation has better effective resolution
here, where resolution is most important (e.g., see Figure \ref{resL20}).

\begin{figure}
\plotone{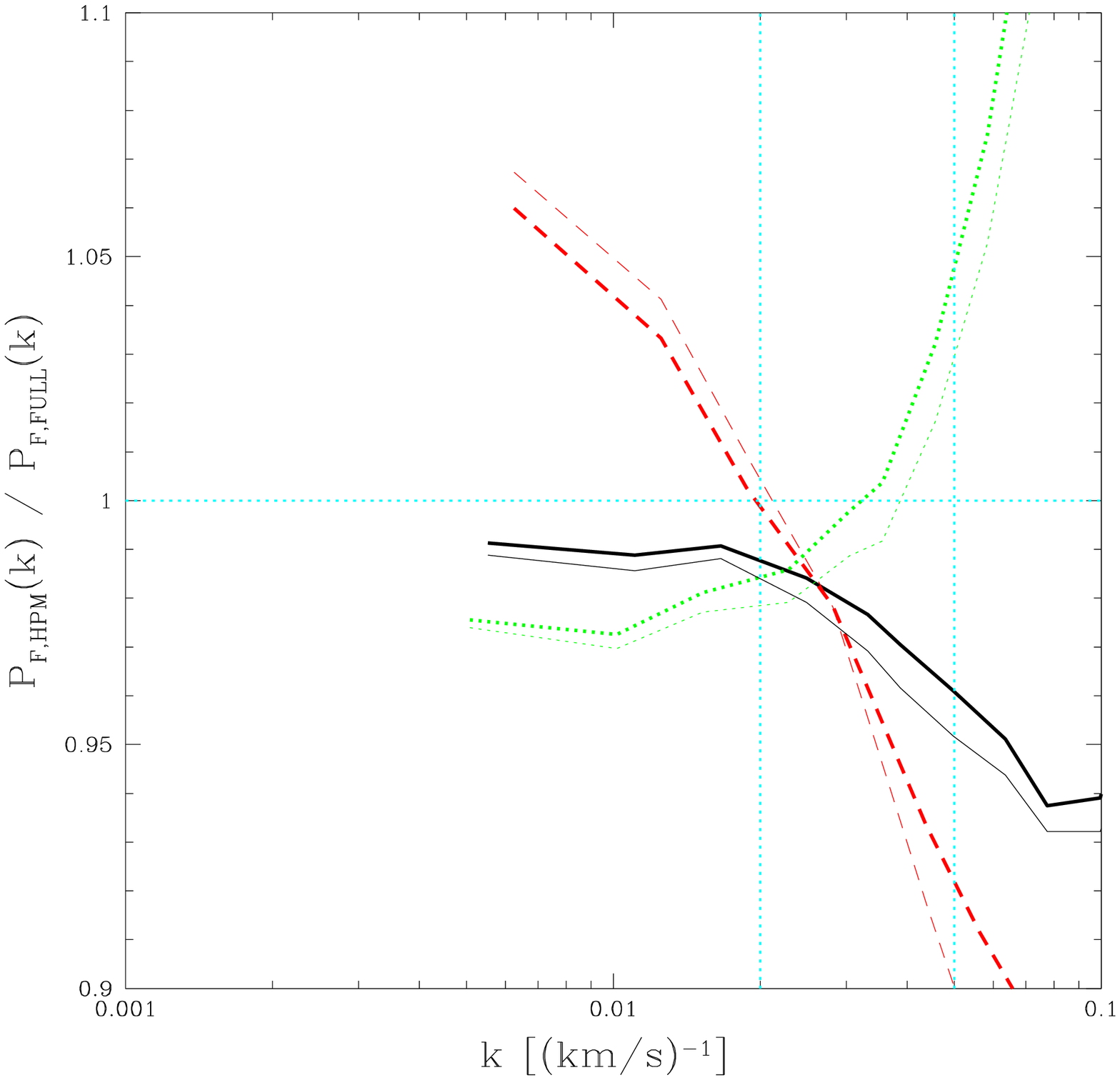}
\caption{
Unimportance of the late-time thermal history
assumed in the HPM simulations.
The thick lines show the $P_{\rm HPM} / P_{\rm hydro}$ 
ratio when the pressure in the HPM simulation is 
computed using the true thermal history in the 
hydrodynamic simulation, while the thin lines show
the same for our default (closer to observed) 
thermal history (see text for numbers).
The (red/dashed, black/solid, green/dotted) lines show 
$a=$(0.32, 0.24, 0.20), with $\bF=$(0.85, 0.67, 0.4).
}
\label{matchcenthermhist}
\end{figure}

When running our standard HPM simulations, we usually 
use the same thermal history to compute the pressure
term.  We turn the pressure on at $z=7$, and then 
use linear interpolation in $\ln T_{1.4}$, $\gmo$,
and $\ln (1+z)$ to connect the points 
$(T_{1.4},~\gmo,~z)=$(24511 K, 0.0, 7.0), (19939 K, 0.2, 3.9), 
(19542 K, 0.3, 3.0), 
and (20071 K, 0.55, 2.4), with the temperature decreasing 
like $a^{-1}$ and constant $\gmo$ at lower $z$. 
This does not exactly match the hydro simulations,
e.g., FULL has $(T_{1.4},~\gmo,~z)=$(15527 K, 0.0, 7.33), 
(21180 K, 0.23, 5.25), 
(18754 K, 0.47, 4.00), (16618 K, 0.55, 3.17), 
(14910 K, 0.58, 2.57), (13561 K, 0.6, 2.12).
To gauge the effect of the difference, we ran an HPM 
simulation using these points for the interpolation.
Figure \ref{matchcenthermhist} shows that the 
results barely change, i.e., changes in the thermal
history at relatively low redshift do not have much
effect on $\PF$. 
This is not to say that the thermal
history is irrelevant -- we will show below that
early reionization can substantially smooth the gas,
and we will allow for this in our fits. 

We use these results as a correction to our larger HPM
simulation results by multiplying the $\PF$ prediction
from the larger simulations by the ratio  
$P_{\rm hydro}/P_{\rm HPM}$.  We account for the 
dependence of this correction on power spectrum 
amplitude and $\bF$, but not temperature-density
relation or power spectrum shape, since this would
require more hydro simulations.  Since the corrections
are small, and the allowed variations in these 
parameters are also small, the change in the 
correction should be negligible.  
The procedure used to extrapolate the simulations to 
large scales not covered by the simulation is described 
below in section \ref{seccombinterpsim}.

\begin{figure}
\plotone{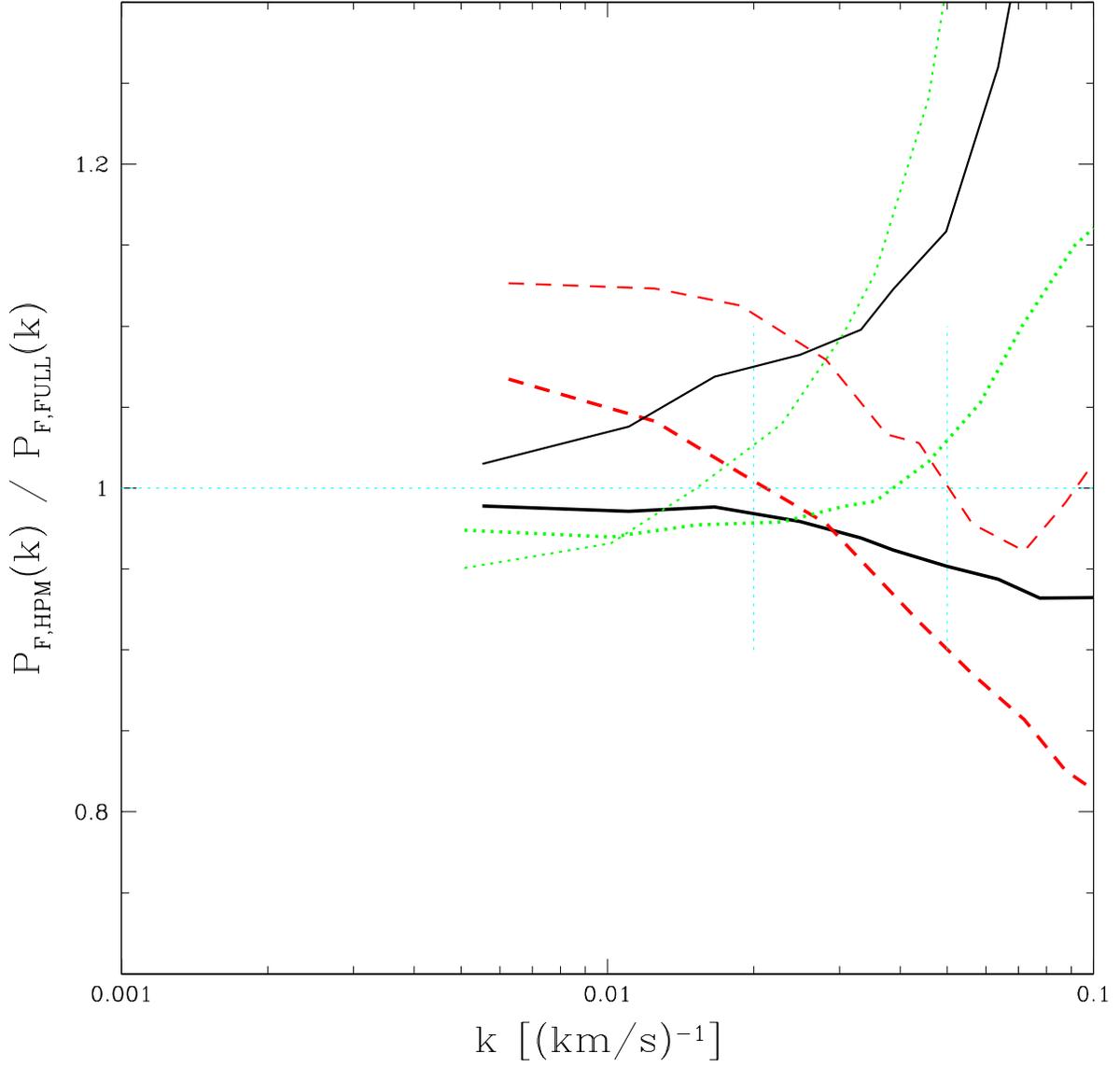}
\caption{
Comparison of PM (thin lines) vs. HPM (thick lines) 
relative to hydrodynamic 
results, for the same initial conditions and 
temperature-density relation.
The (red/dashed, black/solid, green/dotted) lines show 
$a=$(0.32, 0.24, 0.20), with $\bF=$(0.85, 0.67, 0.4).
}
\label{cencomppmhpm}
\end{figure}

Incidentally, we have also compared the results of a
pure PM run to the hydrodynamic simulation (using the
same code as for HPM, just without the pressure term).  
Figure \ref{cencomppmhpm} shows this along with the 
HPM comparison.  
In contrast to the findings of \cite{2001MNRAS.324..141M}, 
we find that the 
pressure component in the HPM simulation substantially
improves the agreement with the hydrodynamic simulation,
in the way that one would intuitively expect, i.e., the 
PM result has too much small scale power (it 
is less obvious what is happening at the 
lowest redshift).  While corrections need to be made
in either case, our tests suggest HPM is as good or 
better than PM. However, given the recent computational 
advances in the development of fast fixed grid hydrodynamic codes
\citep{2004NewA....9..443T}, there may be no 
need to use these approximate methods in the future. 

\subsection{$20 \hmpc$ and $40 \hmpc$ HPM Simulations}

\begin{figure}
\plotone{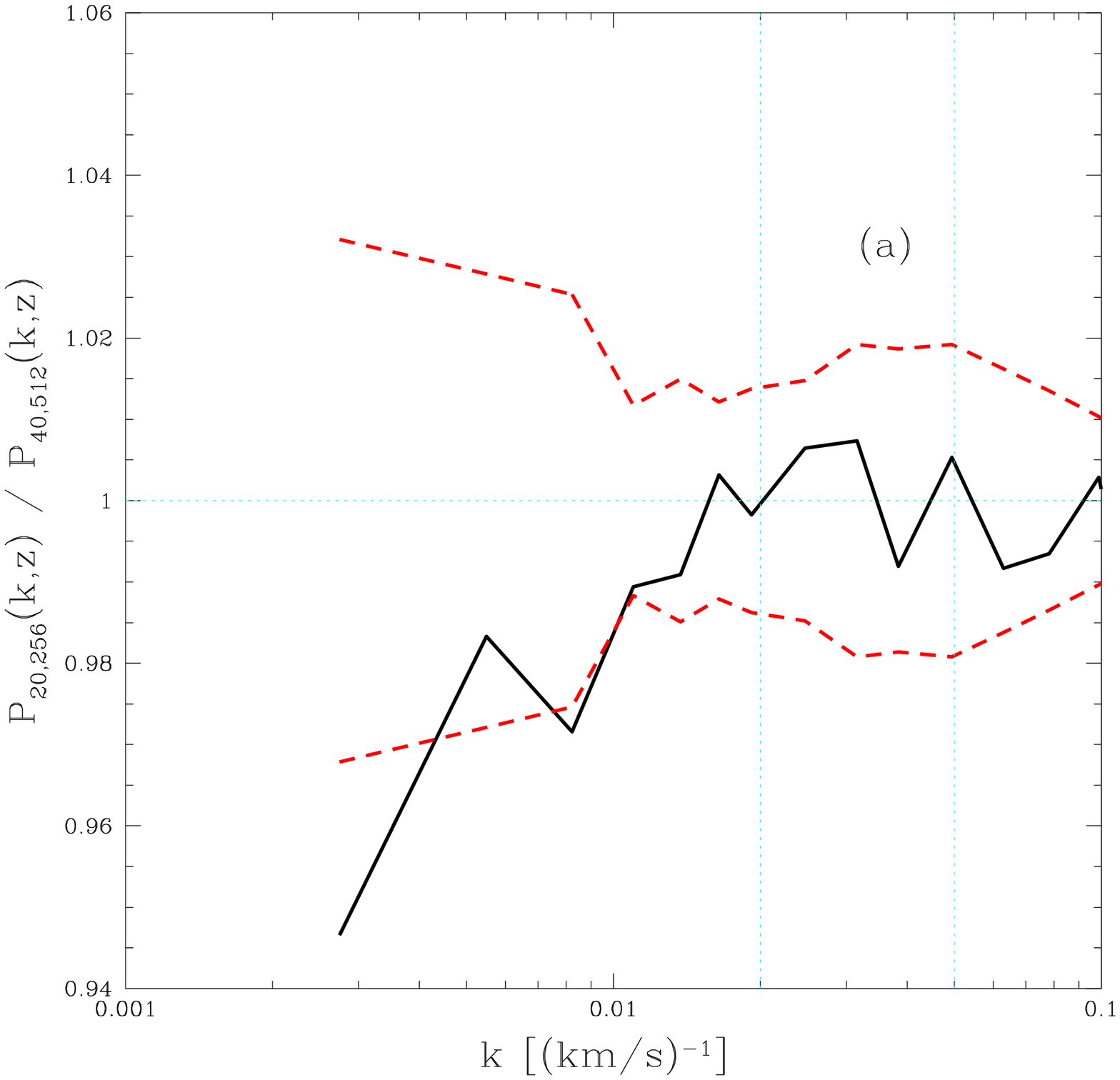}
\caption{
Box size test (note small range on the vertical axis).  
Black, solid line:  ratio of $P_{20,256}$ to $P_{40,512}$, 
averaged over 8 and 6 simulations, respectively, 
with different seeds.
Red, dashed line:  plus and minus the rms error on the
mean for each bin (estimated from the variance between
the eight $N=256^3$ runs).  (a, b, c) show 
$a=$(0.32, 0.24, 0.20), with $\bF=$(0.85, 0.67, 0.4).
}
\label{box2040}
\end{figure}

Figure \ref{box2040} shows a test for systematic error in
$\PF$ related to finite box size, comparing $P_{20,256}$
to $P_{40,512}$.
\begin{figure}
\plotone{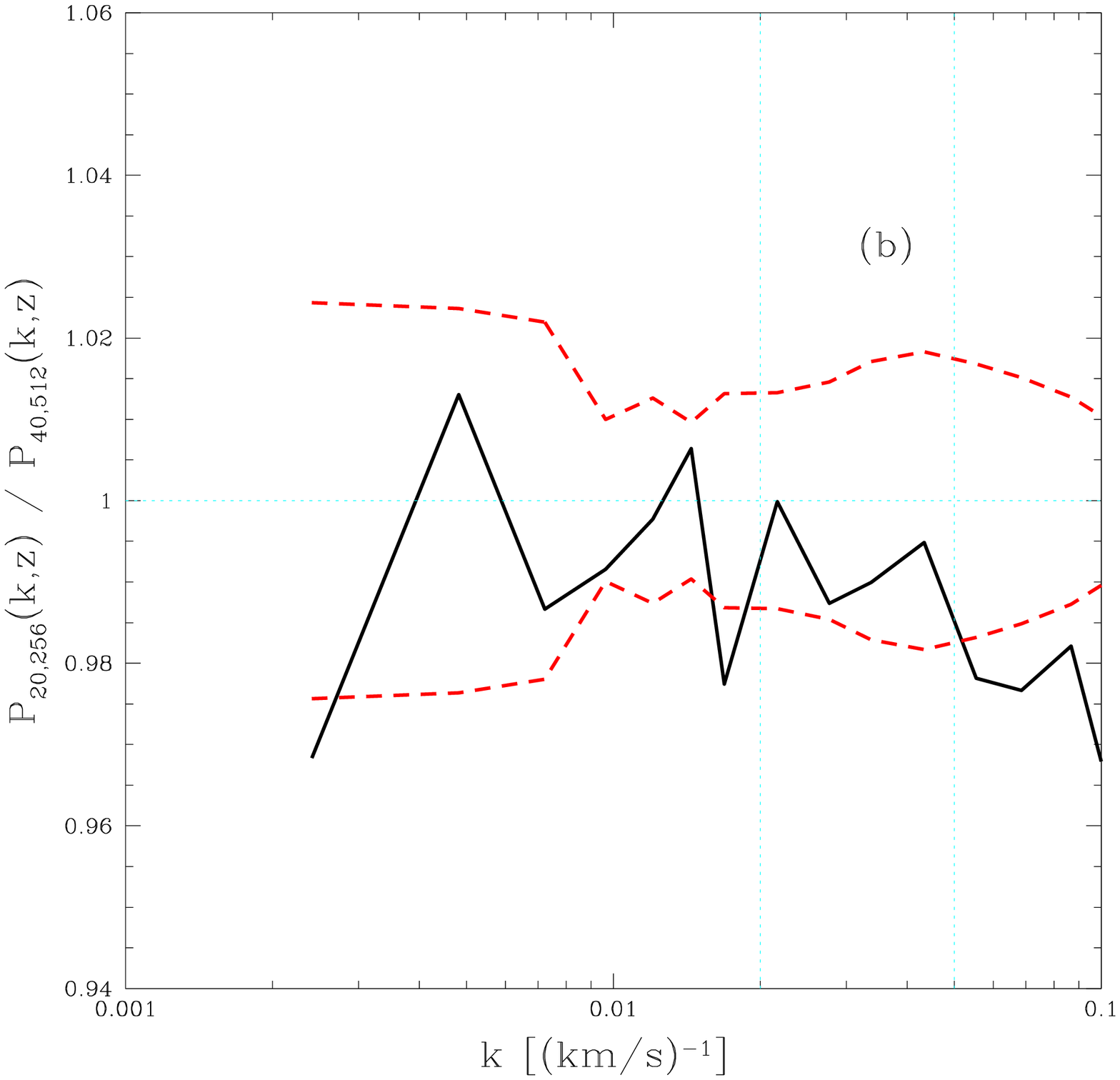}
\end{figure}
\begin{figure}
\plotone{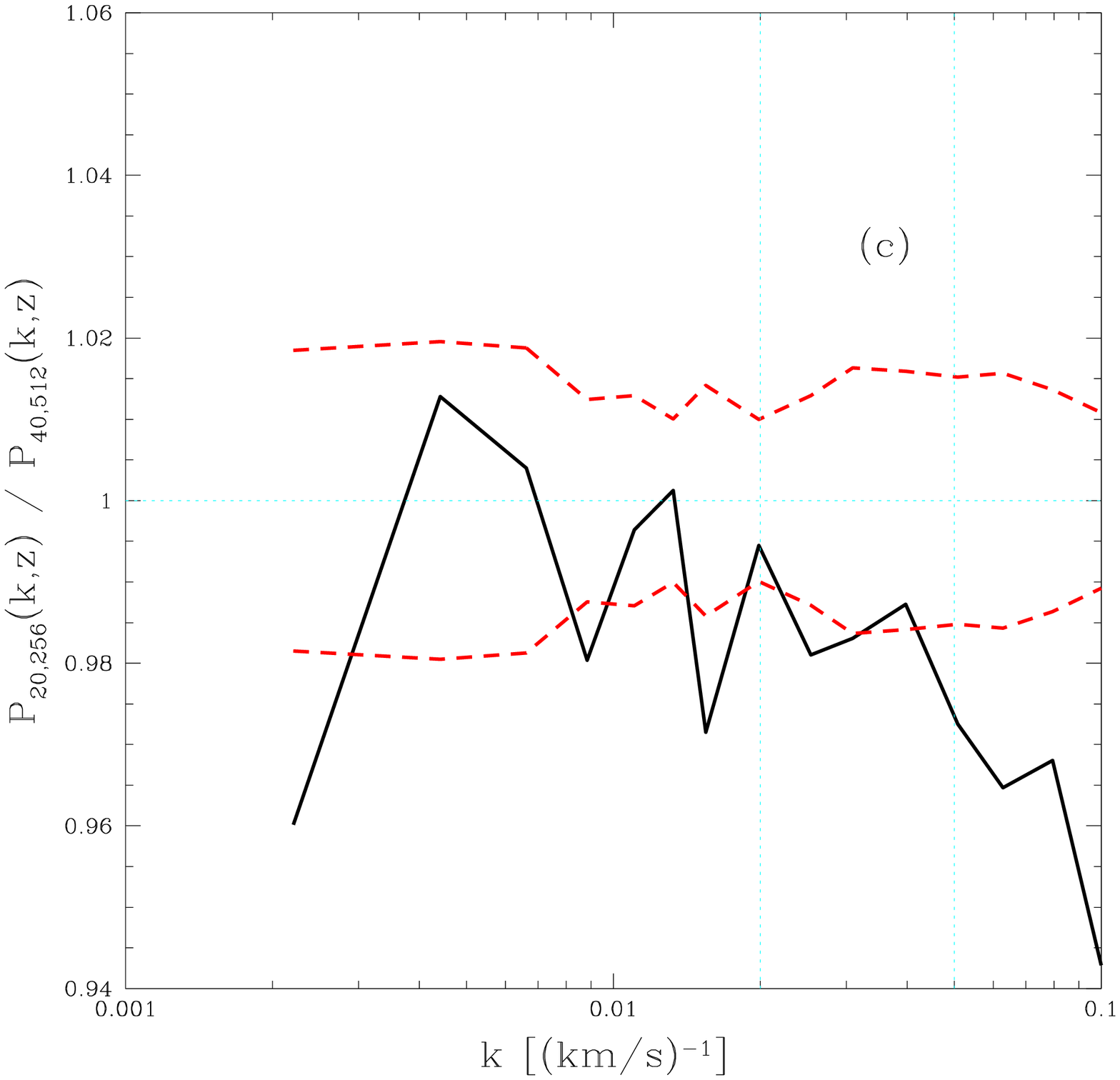}
\end{figure}
Note that, unlike most of our tests, we cannot perform 
a box size test with identical initial conditions in
each simulation.  
To suppress the resulting larger statistical fluctuations, 
we averaged $P_{20,256}$ over eight runs with different 
seeds, and $P_{40,512}$ over six runs.  
We see that
any systematic error in the $20 \hmpc$ boxes is for the
most part limited to be $\lesssim 2$\%, although there
probably is some error at that level.  This error alone 
might not compel us to go to $L=40\hmpc$ simulations,
but we need to predict the power spectrum on somewhat
larger than $20\hmpc$ scales anyway, and the larger boxes 
give much smaller statistical errors per box, at fixed $k$.

\begin{figure}
\plotone{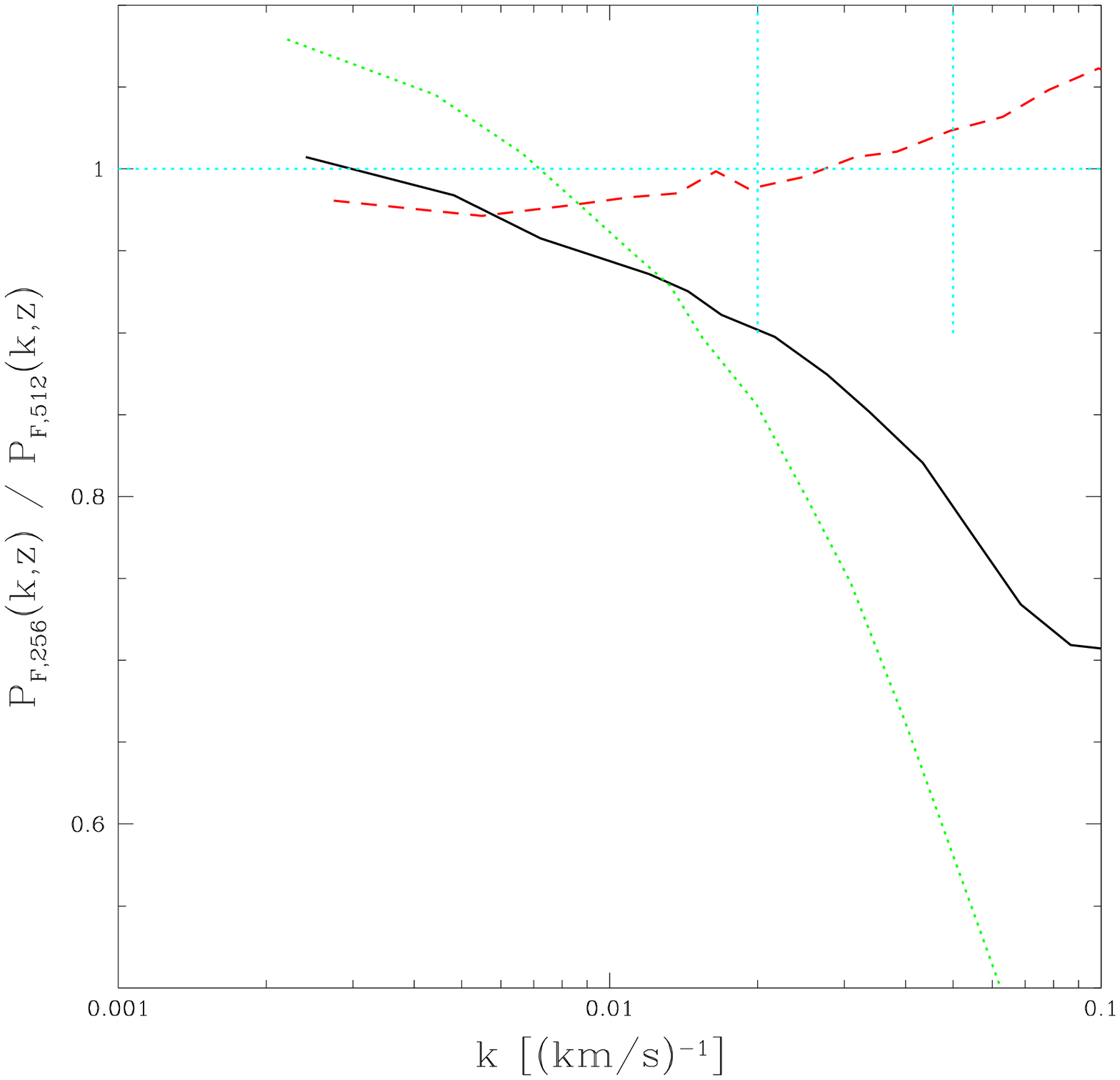}
\caption{
Resolution correction factor for the (40,512) HPM simulations.
Plotted is the ratio of $P_{20,256}$ to $P_{20,512}$.
Red/dashed, black/solid, and green/dotted indicate $\PF$ 
at, respectively, 
$a=0.32$, 0.24, and 0.20, with $\bF=0.85$, 0.67, and 0.4.
}
\label{resL20}
\end{figure}

Figure \ref{resL20} shows the ratio 
$P_{20,256}/P_{20,512}$, demonstrating clearly that 
(40,512) simulations do not have 
sufficient resolution.   
Note that, while the eye is drawn to the very large 
difference at high $k$ and $z$, for the scales probed
by SDSS data [$k<0.02~\ikms$] the errors are no more
than 15\%, and usually less. 
The counter-intuitive small increase in 
small-scale power with decreasing resolution at $a=0.32$
is probably a case of limited resolution reducing the 
small-scale 
smoothing by peculiar velocities more than it reduces the
real-space power.
This prompts us to use
$L=40\hmpc$ simulations, but correct them for the 
resolution error.  We do this by dividing by the 
correction factor given by Figure \ref{resL20}.  
Including the hydro correction, the
formula for our predicted $\PF$ is then:
\begin{equation}
\PF = P_{40,512}~\frac{P_{20,512}}{P_{20,256}}~
\frac{P_{\rm hydro}}{P_{\rm HPM}}.
\end{equation}
As we discuss below, we also tried fitting to observations 
using predictions based 
simply on (20,512) simulations,
i.e., $\PF=P_{20,512}~P_{\rm hydro}/P_{\rm HPM}$,
and get essentially the same result, suggesting that
several potential problems (limited box size, statistical
errors, accuracy of the resolution correction) are 
not significant.  Note that the convergence of the hydrodynamic
simulations is a separate issue. 

\begin{figure}
\plotone{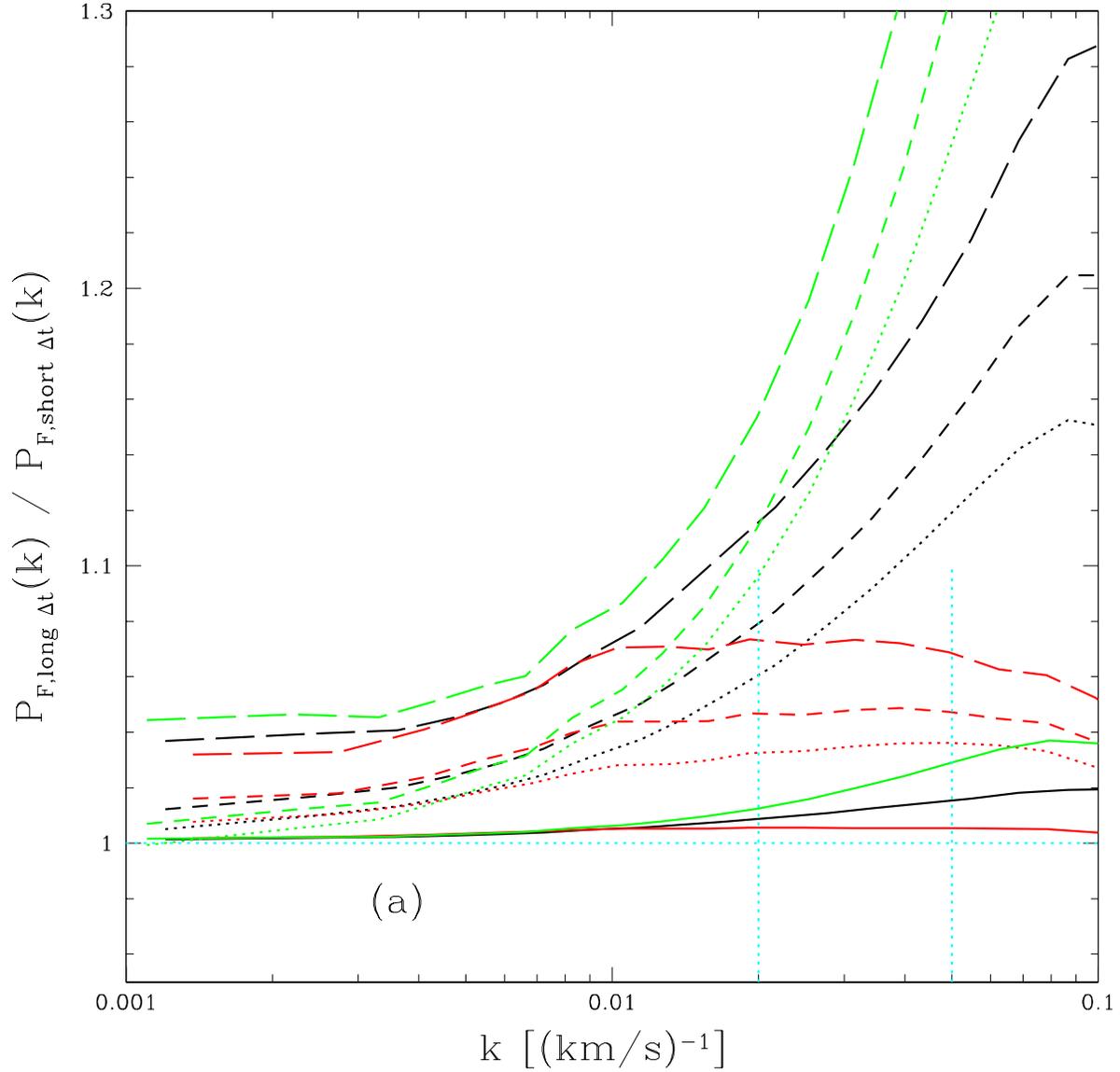}
\caption{
Time-step convergence test.
(a-c) show $\PF$ from (40,512), (20,512), and (20,256) simulations, 
respectively, for different numbers of time steps relative to $\PF$ for 
the largest
number we tried.  The timesteps down to $z=1.5$ used for the (denominators,
solid lines, dotted, dashed, long-dashed) are (479, 227, 94, 41, 20), 
(589, 256, 116, 50, 20), and (702, 336, 158, 67, 33) for (a-c).
In each case
$a=$(0.32, 0.24, 0.20), with $\bF=$(0.85, 0.67, 0.4)
run from bottom to top at $k=0.05\ikms$.
}
\label{time}
\end{figure}

Finally, we need to check the timestep convergence of these HPM simulations.
Because we wanted literally hundreds of simulations to cover the 
pre-SDSS 
allowed range of parameter space, and to make sure we did not have statistical
errors, we intentionally ran the main grid with
rather large time steps ($\sim 20-30$ steps to reach $z=1.5$).  
We will have to make a small correction for the
error this causes.
Figures \ref{time} (a-c) show the tests for each relevant simulation size.
\begin{figure}
\plotone{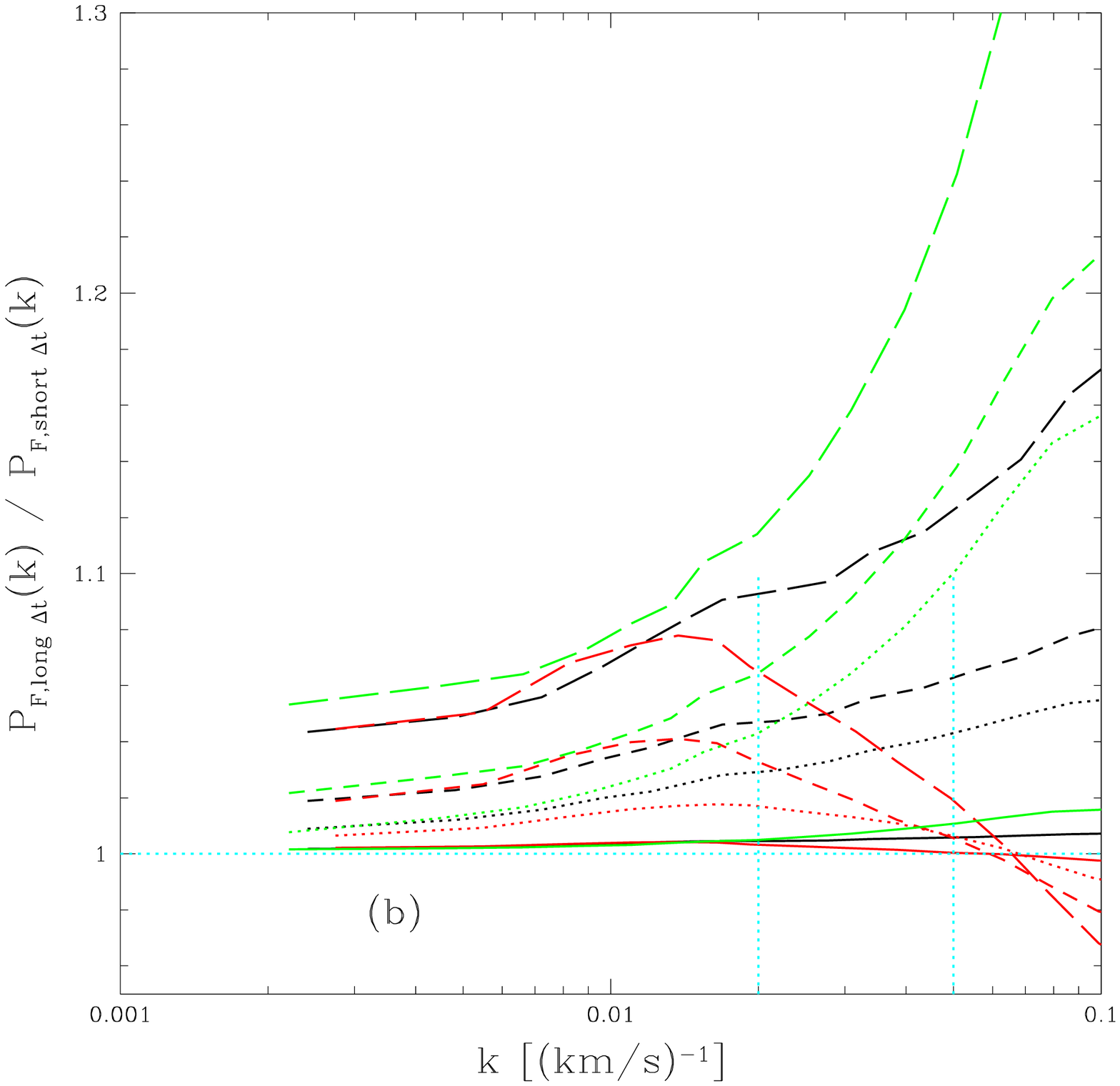}
\end{figure}
\begin{figure}
\plotone{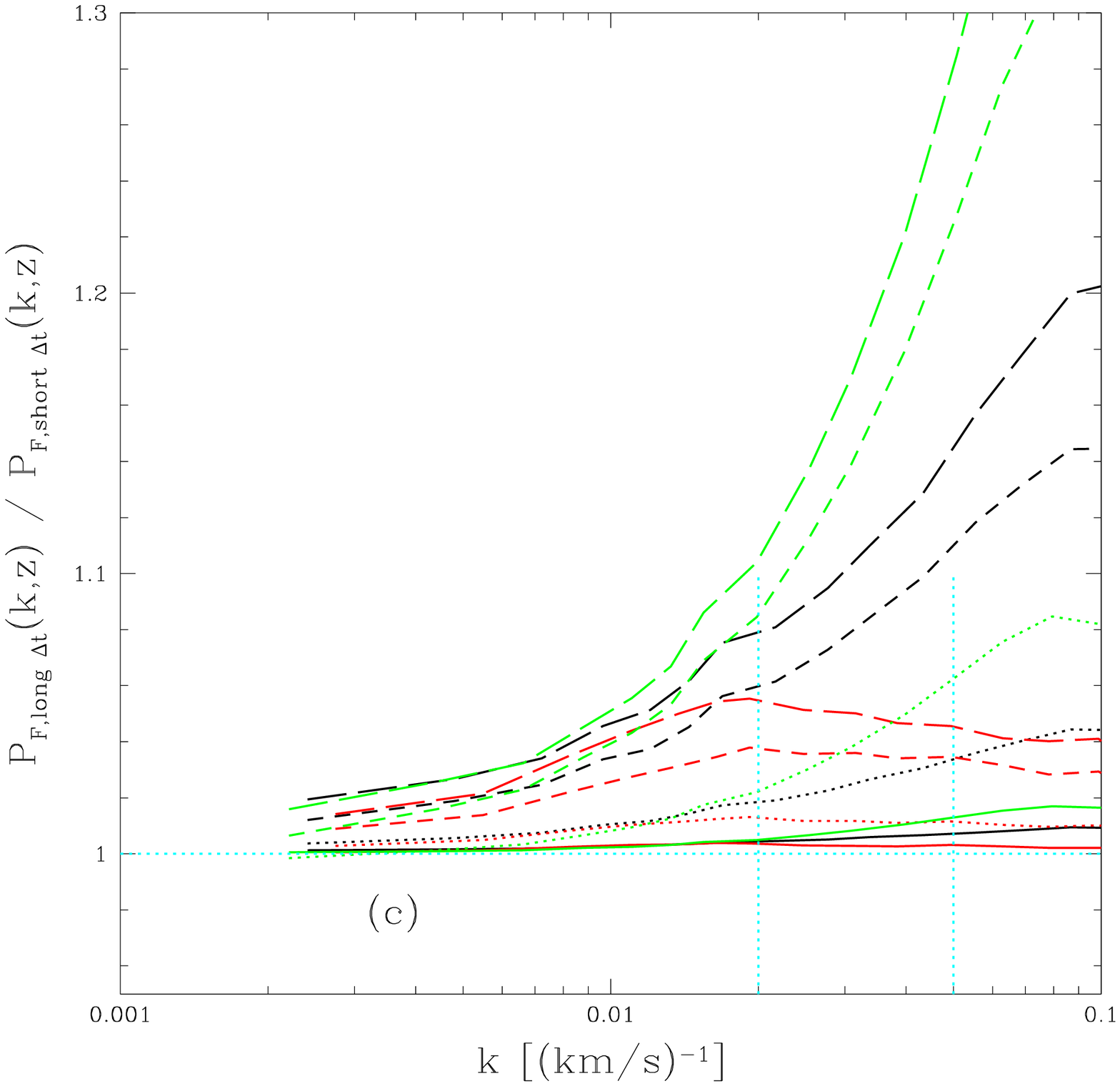}
\end{figure}
We make the corrections in the usual way, i.e., 
multiplying the main grid $\PF$ by 
$P_{{\rm short} \Delta t} / P_{{\rm standard} \Delta t}$
(separately for each box size and resolution).
The time savings comes about because we do not allow the
correction to depend on power spectrum shape, or compute
it for more than one random seed for the initial conditions
(we do include dependence on power spectrum amplitude, 
$\bF$, $T_{1.4}$, and $\gmo$, because these do not require
extra simulations).  We see that a huge savings in time 
can be obtained at a small price in accuracy.

\subsection{Summary of Numerical Simulation Error Control}

We emphasize that our analysis attempts to fully account for all of the 
possible numerical error sources discussed above.  Any residual systematic 
error can
only enter through imperfections in the corrections we make.  
The finite resolution of the 
hydrodynamic simulations is allowed for by introducing extra freedom 
in the
filtering scale of the gas \citep{2003ApJ...583..525G}, which we showed in 
Figure \ref{hydrorestest} has an effect practically equivalent to a change in 
resolution. 
The sensitivity to physics details in the hydrodynamic simulations,
shown in Figure \ref{hydrophysicscomp}, is allowed for by making the 
hydrodynamic simulation prediction an average over the three physics versions, 
with the relative weightings of the average as free parameters.  
The error in the HPM approximation is corrected by comparison to the 
hydrodynamic simulations, with the uncertainty in extrapolation from $10\hmpc$
box size to larger scales accounted for by a free parameter that allows 
anything between a constant value of $\PF$ and a constant slope of $\PF$.
Limited box size in our $40\hmpc$ simulations is not a significant source of 
statistical or systematic error, as shown by Figure \ref{box2040} and the 
fact that our results using only $20\hmpc$ simulations are consistent 
(see below, Table \ref{modtab}).  Limited resolution in the (40,512) 
simulations is corrected for using full grids of (20,512) and (20,256)  
simulations, with any remaining resolution error in the (20,512) 
simulations incorporated into the hydrodynamic correction.  Finally, error
from insufficiently small timesteps in the main grids of HPM simulations is
corrected by comparison to fully converged simulations.  

\subsection{High Density Absorbers and UV Background Fluctuations}

\begin{figure}
\plotone{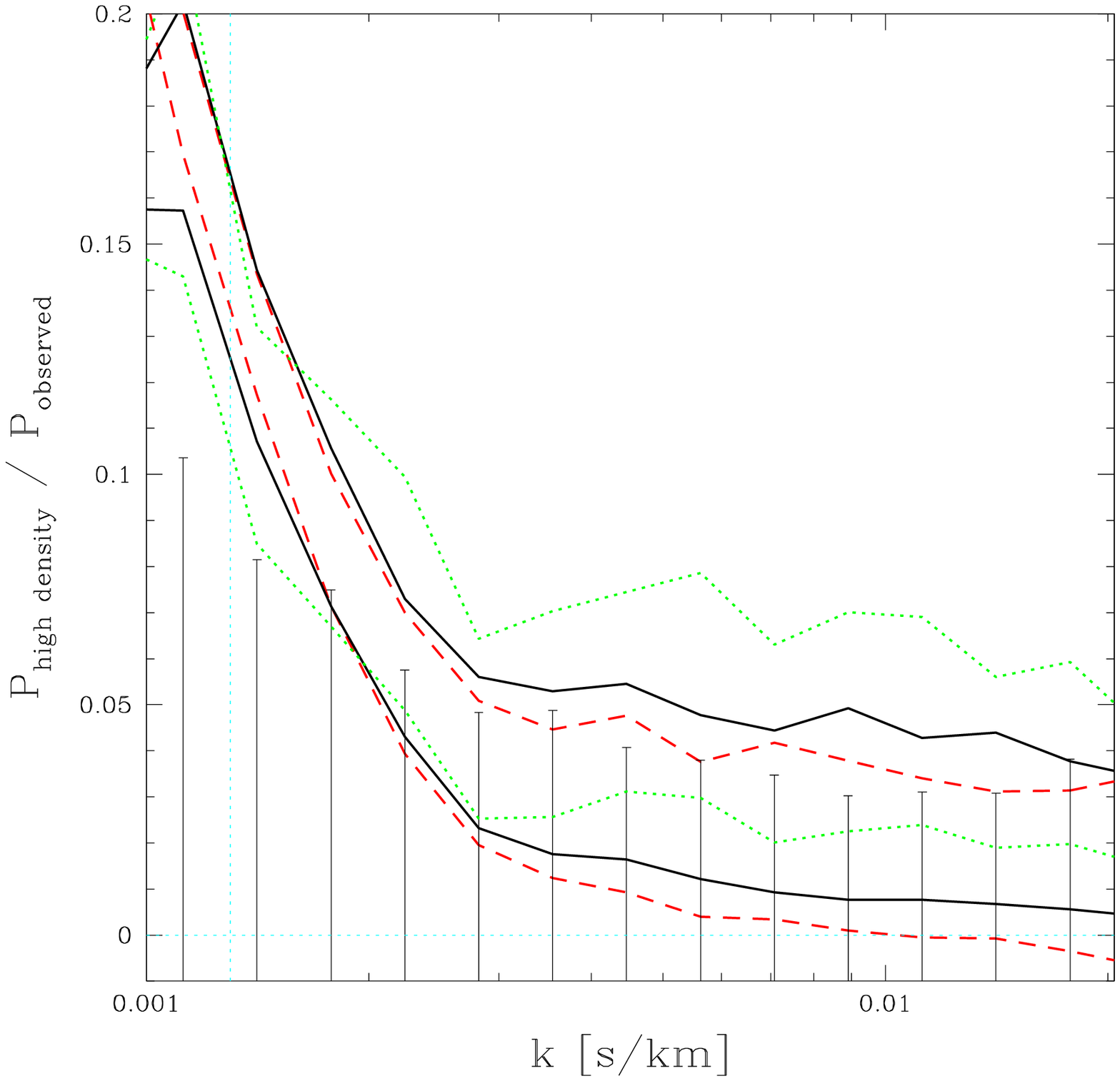}
\caption{
Change in $\PF$ when DLAs 
and LLSs (systems with $N(\hi)>1.6\times 10^{17} {\rm~atoms~cm^{-2}}$)
with the observed 
column density distribution are inserted into mock
spectra of the \lyaf, relative to the observed \lyaf\ $\PF$,
from \cite{2005MNRAS.360.1471M}.  
The upper curves show the case where the high density 
systems are inserted randomly, 
while for the lower
curves the LLSs and DLAs were inserted at the highest density maxima 
in mock \lyaf\ spectra.
Red (dashed), black (solid), and green (dotted) lines show $z=2.2$,
3.2, and 4.2.  The error bars indicate the fractional error on the
observed $\PF$ at $z=3.2$ (the errors at $z=2.2$ are very similar,
while the errors at $z=4.2$ are $\sim 2$ times bigger).
Note that a consistent systematic error that is $1\sigma$ for any
single point is very significant, because we have many $z$ and $k$
bins to average over.
}
\label{mockHD}
\end{figure}

Very high density systems are not necessarily well reproduced
by our hydrodynamic simulations 
\citep{2003ApJ...598..741C,1996ApJ...471..582M,
2001ApJ...559..131G,2004MNRAS.348..421N,2004MNRAS.349L..33V}.
\cite{2005MNRAS.360.1471M} 
investigate this issue in some detail, finding that the 
presence of damping wings is important, although much of the
effect comes from systems below
the traditional cutoff for damped \lya\ systems (neutral
column density 
$N(\hi)=2\times 10^{20} {\rm~atoms~cm^{-2}}$, \cite{1986ApJS...61..249W,
1986ApJ...310..583S}).  
\cite{2005MNRAS.360.1471M}
give templates for the contribution of high density systems
to $\PF$, constrained by the observed column density distribution
of these systems \citep{2003MNRAS.345..480P,2003MNRAS.346.1103P,
2004PASP..116..622P}.  
We reproduce examples from the two templates 
that we use in this paper in Figure \ref{mockHD}.
The differences between the two cases in the figure 
is that in one case the high density systems are located at
peaks in the mock density field, while in the other they are
located randomly.  Relative to the case when the systems 
are located randomly,
when the systems are located in high density regions 
there is little effect on the small-scale power, because the 
affected regions are already saturated (the relatively low 
equivalent width systems, which account for the
small scale power, produce little change when they are 
inserted). The randomly located
case is not realistic, but we include it to show that our 
fits are not sensitive to this kind of detail (see below).
Based on the discussion in \cite{2005MNRAS.360.1471M},
we will  
assign an overall 30\% error to the size of this effect in our
fits.  A more careful study could probably reduce this
error, but our results are not especially sensitive to it.

\begin{figure}
\plotone{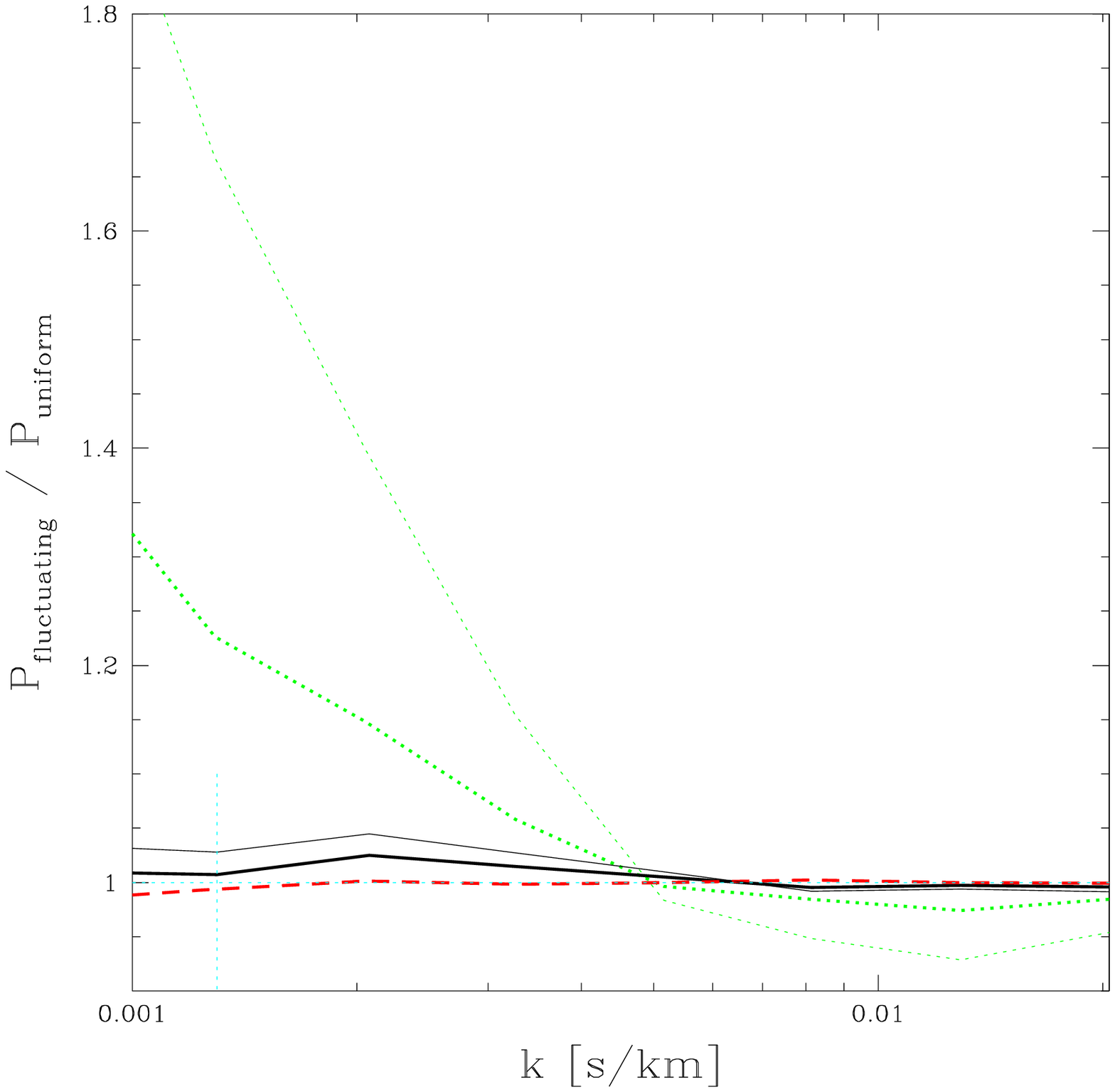}
\caption{
Effect of a fluctuating UV background,
from \cite{2005MNRAS.360.1471M}.  
Thick lines show the template we use in our standard fitting,
when we assume all of the UV background comes from quasars. 
Thin lines show a case where the mean free path for ionizing
photons in the IGM has been arbitrarily reduced by a factor of two 
and is meant to show the worst case scenario (the fluctuations
increase with decreasing mean free path).
Red (dashed), black (solid), and green (dotted) lines show $z=2.05$,
3.29, and 4.58.  The thick and thin dashed lines are 
indistinguishable.
}
\label{uvtemplate}
\end{figure}

\cite{2005MNRAS.360.1471M,2004ApJ...610..642C}, and \cite{2004MNRAS.350.1107M} 
investigate the potential
influence of a fluctuating UV background on $\PF$.
These papers find
an effect that increases dramatically as the mean free path
for an ionizing photon decreases with increasing redshift.
The effect only becomes significant at the high end of the
redshift range we
consider in this paper.  Figure \ref{uvtemplate} shows 
examples of the
templates we use to include this effect in our fitting, taken 
from \cite{2005MNRAS.360.1471M}. 
These correspond to the quasar luminosity function from
\cite{2002AJ....123.1247F}, with quasar lifetime of $10^7$ years, 
and include 
light-cone effects described by \cite{2004ApJ...610..642C}. 
The models in Figure \ref{uvtemplate} are the extreme 
(maximum fluctuation) cases.
More detailed analysis of other 
models is presented in \cite{2005MNRAS.360.1471M}.
In contrast to the case of damping wings, we have little direct
constraint on the redshift evolution of this effect.  We will
include nuisance parameters for both the amplitude and 
evolution of the effect in our fits, and find that including 
this freedom
increases the error on the linear power spectrum measurement,
but does not change the central value significantly. 
 
\subsection{Parameter Dependence of $P_F$}

\begin{figure}
\plotone{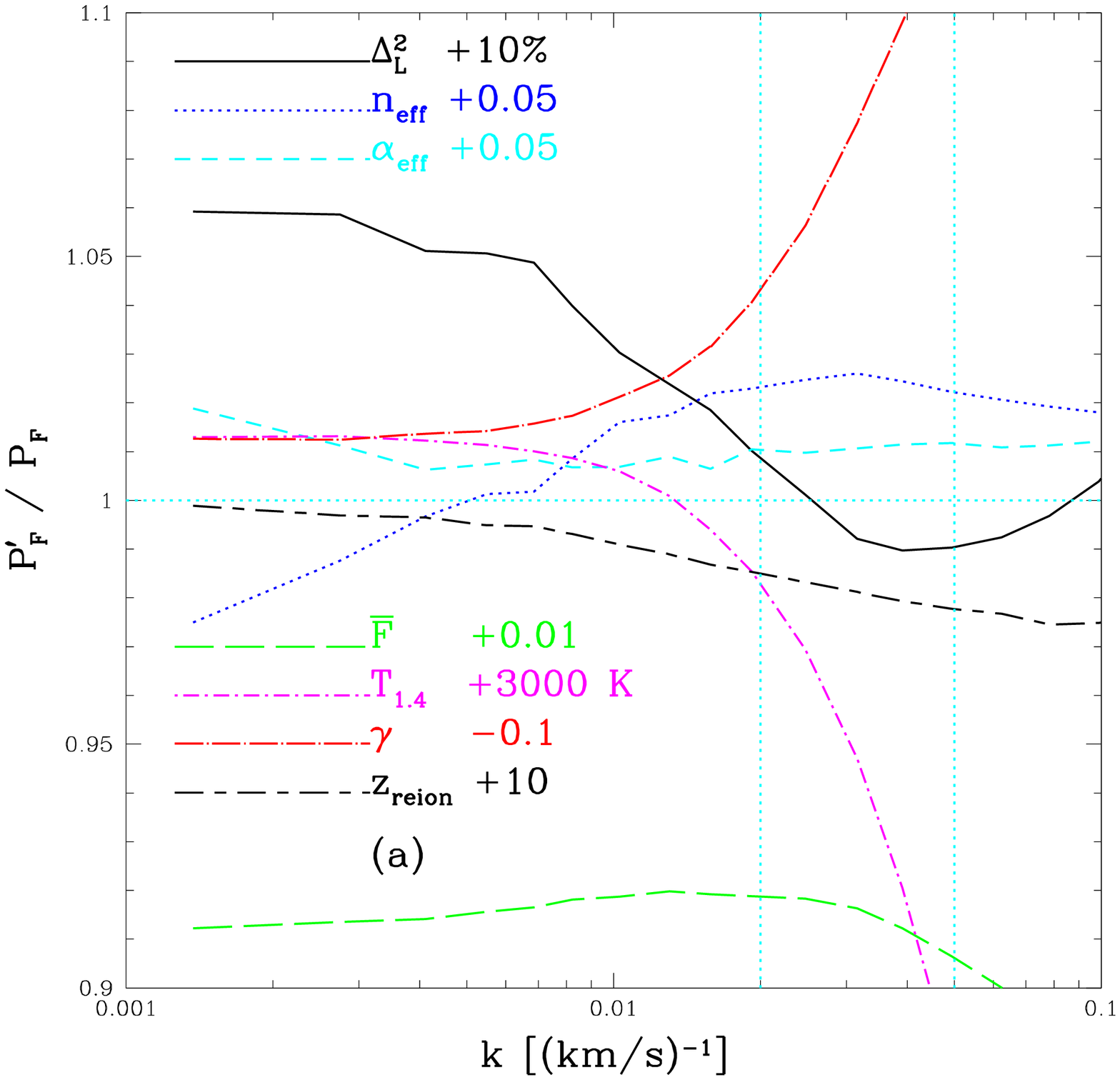}
\caption{
Parameter dependence of $\PF$ in $(40,512)$ simulations, 
as a ratio of $\PF$ after
one parameter, $p$, is changed by $\Delta p$ to $\PF$ for a 
central set of parameters.  
Solid (black) line:  $\DL$ increased by 10\%,  
dotted (blue) line:  $\Delta \neff = 0.05$,  
dashed (cyan) line:  $\Delta \aleff = 0.05$,  
long-dashed (green) line:  $\Delta \bF = 0.01$,  
dot-dashed (magenta) line:  $\Delta \Tp = 3000$ K,  
dot-long-dashed (red) line:  $\Delta \gamma = -0.1$,  
dashed-long-dashed (black) line:  $\Delta z_{\rm rei}  = 10$.
The panels show central values representing the middle and 
extremes of our redshift range:
(a) $\bF=0.85$, $a=0.32$, 
(b) $\bF=0.67$, $a=0.24$, 
(c) $\bF=0.4$, $a=0.2$.  Note that these figures are intended 
primarily as a qualitative demonstration, as detailed corrections  
have not been applied (e.g., the up-turn of the
$\DL$ dependence at very high $k$ is an effect of limited
resolution).
}
\label{pardep}
\end{figure}

We now discuss the parameter dependence of $\PF$ in 
our simulations.  Much of this has been shown already in 
the \cite{2003ApJ...585...34M} plots of the 
three-dimensional flux power spectrum, but it is
useful to see directly the effects on the 
one-dimensional $\PF$.
Figures \ref{pardep}(a-c) show 
examples of the fractional change in $\PF$ when 
$\DL$ is increased by 10\%, $\neff$ is increased
by 0.05, $\aleff$ is increased by 0.05, $\bF$ is
increased by 0.01, $\Tp$ is increased by 3000 K,
$\gmo$ is decreased by 0.1, or reionization is 
moved from $z=7$ to $z=17$.  
\begin{figure}
\plotone{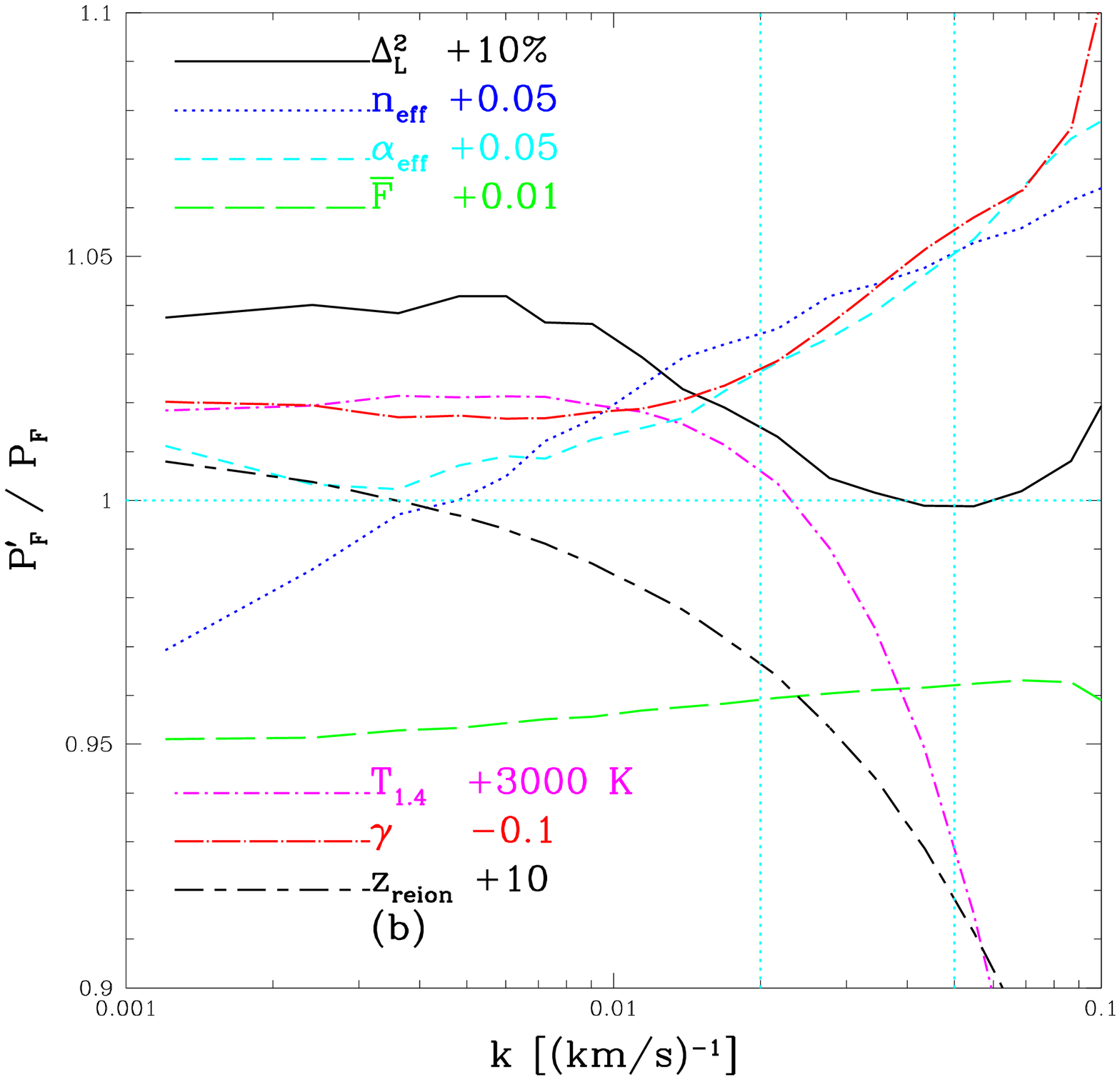}
\end{figure}
\begin{figure}
\plotone{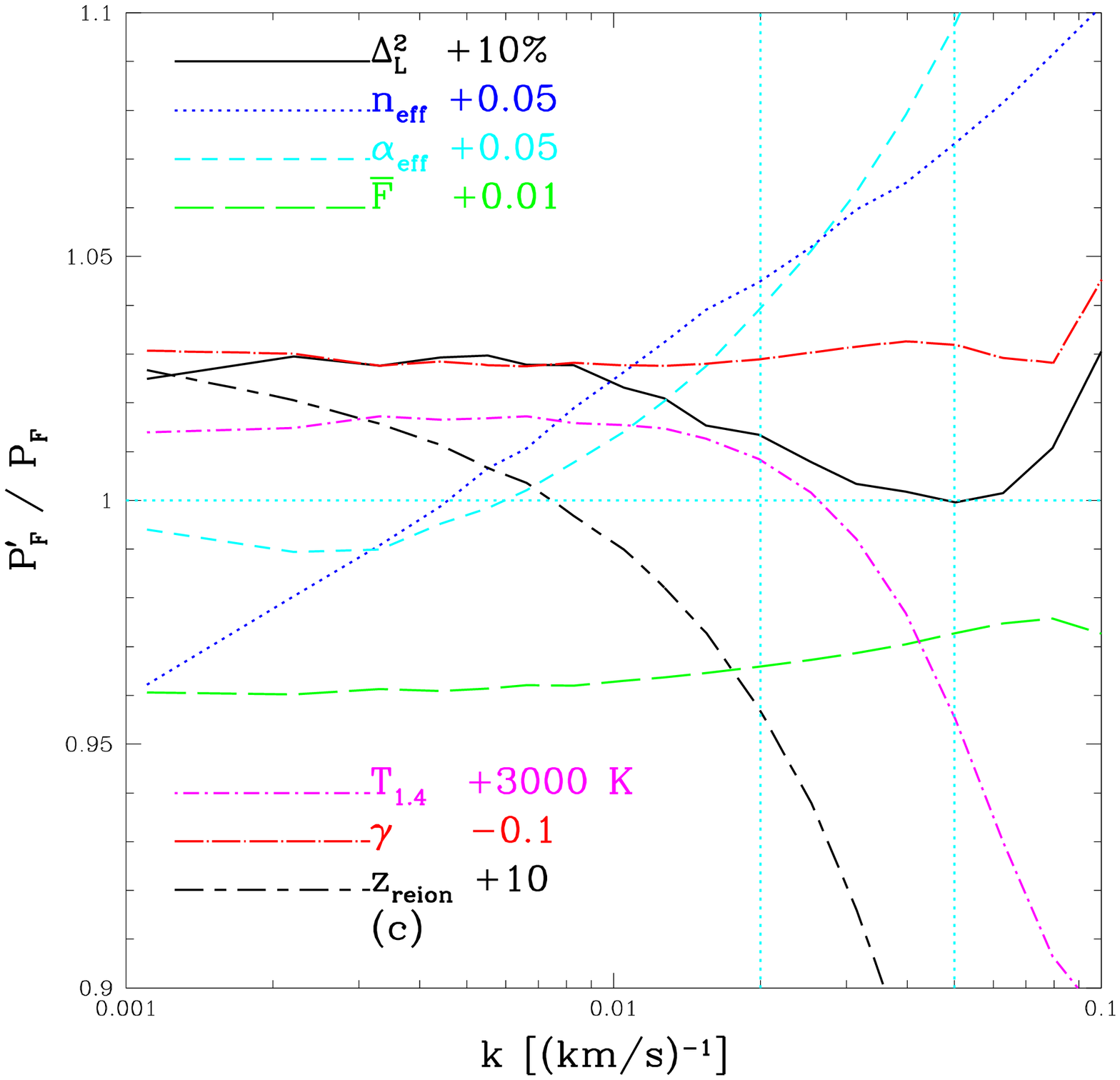}
\end{figure}
The starting values
are our simulation standard 
$\Delta^2(k_s=1~{\rm h/Mpc},a_s=0.24)=0.26$, 
$n_{\rm eff}(k_s,a_s)=-2.3$, 
$\alpha_{\rm eff}(k_s,a_s)=-0.2$, 
$\Tp=17000$ K, and $\gmo=0.6$, 
with $\bF=$(0.85, 0.67, 0.4) at $a=$(0.32, 0.24,
0.20).  We used (40,512) simulations for this 
figure.  $\PF$ for the central model (the denominator 
in the plot) is taken essentially directly from a 
simulation output, but the changes involve some 
interpolation to achieve the desired size of change.

The parameter dependences are generally non-trivial.  
Increasing $\DL$
enhances the power on large scales, but actually suppresses
the power on small scales. \cite{2003ApJ...585...34M} shows that this is 
a finger-of-god-like effect of peculiar velocities suppressing
power along the line of sight: if the amplitude is higher the velocities 
are higher, which leads to a suppression of power on small scales.  
Note that these dependences
can be affected slightly by limited resolution at high $k$, e.g., 
when $\DL$ is increased in 
(20,512) simulations the suppression of $\PF$ continues
to increase at $k>0.04\ikms$.  
Changing $\neff$ 
produces a fairly simple and expected change in the slope of $\PF$, except
at high $k$ and low $z$.
Changing $\aleff$ produces curvature in $\PF$, although the
effect almost disappears at low $z$.  $\bF$ produces a relatively
flat, large change, which is commonly assumed to be degenerate 
with $\DL$, although we see that the shapes are not the same, nor are
the relative effects at different redshifts: as a result, the data 
can break the degeneracy within the flux power spectrum analysis 
itself without the need to bring in external constraints.
Increasing
$\Tp$ primarily suppresses the power at high $k$, not surprisingly,
although it also produces a small change in large-scale bias.
Decreasing $\gmo$ produces an overall bias, but also a sharp increase
in power at high $k$, in the two lower redshift cases.  This is 
an indication that the power is sensitive to structures with
overdensity greater than 1.4, since their temperature is reduced
by a decrease in $\gmo$, leading to reduced thermal broadening
suppression of power.  At $z=4$ (and, more importantly, $\bF=0.4$),
$\Delta=1.4$ appears to be the most relevant overdensity. 
Finally, increasing the redshift of reionization allows more 
time for pressure to suppress small-scale structure (Jeans smoothing).
The power suppression extends to smaller $k$ than the thermal 
broadening
effect, because it acts on the three-dimensional field instead of
only along the line of sight.  This effect decreases rapidly with
decreasing redshift and increasing $\bF$, allowing us to constrain 
it in a full fit to the data.

\subsection{Combining and Interpolating Between 
Simulations\label{seccombinterpsim}}

\begin{figure}
\plotone{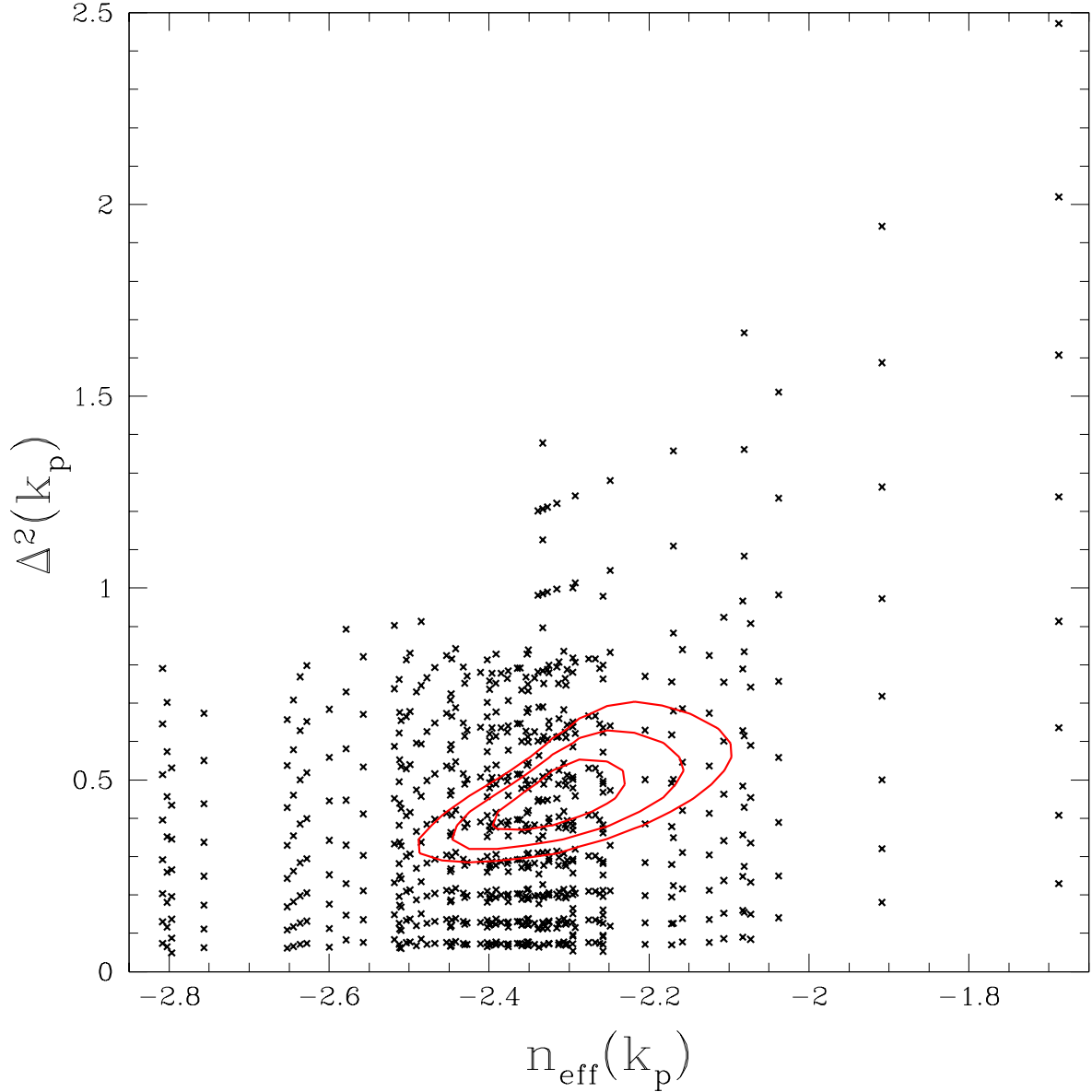}
\caption{
Each point shows the position of one of our (40,512) simulation outputs in 
the linear power spectrum amplitude-slope plane, at $k_p=0.009~\ikms$.  To 
distinguish between multiple runs using the same input power spectrum,
the points are plotted at the effective position computed from the 
power as realized in the randomly generated initial conditions.
For comparison, the red lines show the 1, 2, and 3 $\sigma$ 
error contours (at $z_p=3$) that we find later in the paper by fitting 
the observational data.
}
\label{showgrid}
\end{figure}

In this subsection we describe the procedure we use to turn 
hundreds of simulations (and more than 100,000 power spectrum 
calculations, after variations of $a$, $\bF$, $\Tp$, and $\gmo$) 
into a prediction for $\PF$ for any given 
set of input parameters.   
There are some subtleties in this process that we describe
in full, in preparation for releasing a code that can be
used as a black box calculator of the \lyaf\ $\chi^2$.  
Our simulation set can not be described as a simple grid in 
parameter space, so in Figure \ref{showgrid} we plot all of the 
points we cover in the $\DL$-$\neff$ plane, for (40,512).
Our grids of (20,512) and (20,256) simulations are almost identical 
to this (40,512) grid.

The adopted interpolation method is designed to deal with some of 
the special
features of our problem. The calculation of the \lyaf\ $\chi^2$ is 
an essential ingredient that goes into joint parameter estimation, 
together with similar $\chi^2$ calculations from the CMB, galaxy clustering, 
supernovae and other ingredients. The CMB and galaxy clustering 
depend on linear
theory calculations, so for each model we need to run a linear 
perturbation calculation like CMBFAST \citep{1996ApJ...469..437S}, 
which is relatively fast. To determine the error distributions in 
a parameter space of models one typically uses the Monte Carlo 
Markov Chain (MCMC) method, which requires $\chi^2$ calculations 
for tens of thousands of models. 

Each \lyaf\ simulation is relatively expensive
(compared to a CMBFAST run), while simultaneously only providing
a noisy estimate of the quantity of interest.  
To minimize the number of simulations needed, we take advantage of
the fact that $\PF$ in our simulations has a smooth $k$ dependence, 
and a smooth dependence on the input parameters.  We condense every
$P_F(k)$ calculation into a few numbers using a fitting 
formula, and then condense this information even further using another
fitting formula for the dependence of the parameters describing 
$P_F(k)$ on the more fundamental cosmological and \lyaf\ model 
parameters.  The idea is to use the minimum number of parameters
needed to describe the true (infinite simulation limit) power,
in contrast to a more standard local interpolation 
between $P_F(k)$ predictions binned by $k$.

The $P_F(k)$ fitting formula is simple, motivated by the 
smoothness of the power spectrum in the models we simulate.
\begin{equation}
\ln P_F(k) = \sum_{\alpha=0}^{N_{\rm log}} P_\alpha~\left[\ln\left(k/k_\star\right)
\right]^\alpha + 
\sum_{\alpha=1}^{N_{\rm lin}} P_{\alpha+N_{\rm log}}~k^\alpha~,
\end{equation}
where $k_\star=0.3\ihmpc$ and $N_{\rm log}$ and $N_{\rm lin}$ can be 
chosen to give the appropriate amount of freedom (we use 3 and 2
terms, respectively, as our standard).  For each simulated $P_F(k)$
we determine the parameters $P_\alpha$ by a $\chi^2$ fit weighted by 
statistical errors on $P_F(k)$ bands determined by measuring the
variance in a set of simulations of one model with many 
different seeds for the random initial condition generator. 

The general structure of our method for associating $P_\alpha$
with physical model parameters $p_i$ (e.g., linear power 
spectrum amplitude, $\bF$, etc.), is as follows:  
we define a conveniently transformed set of model parameters, $p_i$, 
and use them in a linear least-squares fit for the
coefficients $A_{\alpha \nu_1\nu_2\nu_3...\nu_{N_p}}$ in the formula
\begin{equation}
P_{\alpha s} = \left(\prod_{i=1}^{N_p}\sum_{\nu_i=0}^{N_i}p_{is}^{\nu_i}\right) 
A_{\alpha \nu_1\nu_2\nu_3...\nu_{N_p}}~,
\label{interpeq}
\end{equation}
where $s$ labels a simulation, $p_{is}$ means the value of the $i$th physical
parameter in the $s$th simulation, there are $N_p$ parameters, and the term
in parentheses should be thought
of as an operator acting on $A_{\alpha \nu_1\nu_2\nu_3...\nu_{N_p}}$.  
Equation \ref{interpeq}
is just a compact way of writing the formula one effectively uses for 
multi-polynomial interpolation, e.g., for $N_p=2$ parameters and $N_i=1$ we have
$P_{\alpha s}=A_{\alpha 0 0} + A_{\alpha 0 1}~p_{1 s} + A_{\alpha 1 0}~p_{2 s}+
A_{\alpha 1 1}~ p_{1 s}~ p_{2 s}$, i.e., the formula for bi-linear interpolation. 
Equation \ref{interpeq} has the important practical
advantage of being linear in all the
parameters, so it is easy to perform multiple fits to $\sim 10^5$ data 
points.  
After some experimentation, we chose $\ln[-\ln(\bF)]$,
$\ln\Tp$, and $\ln [2-0.7~(\gmo)]$ to be the parameters 
$p_i$ for the \lyaf\ model.
Reionization will be treated outside this formalism, as discussed below.  
All that remains is to define a way to turn a given linear power 
spectrum, say, from CMBFAST, into the rest of $p_i$.

In the infinite-dimensional space of possible input linear power spectra, 
we have many relatively smooth models,
from pure power laws with $-2.75\leq n \leq-2.1$, to $\Lambda$CDM transfer
function models
with $-2.8\leq \neff\leq-2.15$ and $-0.37\leq\aleff\leq-0.03$, to the 
primordial black hole models described in
\cite{2003ApJ...594L..71A} 
(these have extra white noise power that dominates at small scales),
to warm dark matter models where the small-scale power is erased 
\citep{2000ApJ...543L.103N}.
Nevertheless, it is easy to produce a model that cannot be
obtained exactly by interpolation between the models we have (e.g., we
do not include variations in the baryon density, because we do not expect
their effect to be independently measurable from $\PF$).  We deal with
this problem by defining a set of parameters to project any power spectrum
onto, akin to $\DL$, $\neff$, and $\aleff$.  
The basic formula for these parameters is
\begin{equation}
\Delta^2_l = \int_{\ln k_{\rm min}}^{\ln k_{\rm max}} 
d\ln k~\Delta_L^2(k)~P_l[x(k)] \exp[-(k R_c)^2]~,
\label{Deltaell}
\end{equation}
where $x(k) = \ln(k/k_0) / \ln (k_{\rm max}/k_0)$, 
$k_0 = (k_{\rm max} k_{\rm min})^{1/2}$, $k_{\rm min}=0.126\ihmpc$, 
$k_{\rm max}=15.8\ihmpc$, $R_c=0.2\hmpc$, and $P_l(x)$ is a Legendre 
polynomial of order $l$ (e.g., 1, $x$, $(3 x^2-1)/2$, ...).
This is nothing more than a convenient way of defining a measure of
the amplitude, slope, curvature, etc. of the power spectrum.  
There is nothing fundamental, or even decisively optimal, about
the choice of weighting $P_L(k)$ by $k^3$ in Equation \ref{Deltaell}
(we tried, and could almost just as well have used, other powers of $k$).
$k_{\rm min}$
was chosen to include the smallest $k$ in our simulations. 
The weighting term controlled by $R_c$ was introduced to reduce the influence 
of high-$k$ power on our interpolation parameters
($\Delta^2_l$), after we found 
by running simulations with spikes of power in relatively narrow bands of
$k$ that the very high $k$ linear power we are suppressing by this term
has diminishing effect on 
the \lyaf\ flux power, presumably because of some combination of pressure
smoothing and non-linear transfer of power from large to small scales
\citep{1991ApJ...374L...1H,2003ApJ...590....1Z}. 
The value $R_c=0.2 \hmpc$ was chosen to maximize the accuracy of the 
fit to the simulations for the number of Legendre polynomial terms we use 
(generally 4). 
$k_{\rm max}$ 
was chosen to center the Legendre polynomials 
near the wavenumber $k_s=1\ihmpc$ that
we used as the pivot point when setting the power spectra 
in our simulations. 
When applying Equation \ref{Deltaell} to our numerical 
simulations, we sum over the discrete set of mode amplitudes 
actually present in the simulation.   
Finally, for the parameters $p_i$ in Equation \ref{interpeq} 
we actually use $\ln \Delta^2_0$ 
and $\Delta^2_{i>0}/\Delta^2_0$, so that only the first
evolves with redshift and the rest are pure measures of power spectrum 
shape.

We apply the above formalism to each type of simulation separately.
When we need to extrapolate small-box simulations down to smaller $k$
than they contain directly, we assume the extrapolation should fall
somewhere between $P_F(k)={\rm constant}$ and 
$d \ln P_F/d\ln k={\rm constant}$, i.e., $P_F(k)$ in the \lyaf\
never decreases with decreasing $k$, and the second derivative is
generally negative.  We introduce a free parameter controlling 
our position between these limits.  This issue
is only significant when extrapolating the $L=10\hmpc$ hydrodynamic
simulations and their comparison HPM simulations to our largest 
scales.

\section{Fitting the Observed $\PF$ \label{secfitting}}

In this section we explain how we perform $\chi^2$ fits to the 
observational data to estimate the linear power spectrum.
We begin with the description of all the parameters that 
go into the fit. We then describe the data itself and present 
our main results next. The remainder of this section is devoted to 
the various consistency checks we performed, both using 
internal constraints from the data and modifying the standard 
fitting procedure. 

\subsection{Parameters \label{secparameters}}

We vary 34 parameters, 3 of which are fixed for our primary result, 
but varied for consistency checks.
We give a bulleted summary before defining each in detail.
In brackets we give the actual number of parameters for each type. 

\begin{itemize}

\item
$\DL$, $\neff$, $\aleff$ (3) \\
Standard linear power spectrum amplitude, slope, and curvature
on the scale of the \lyaf, assuming a typical
$\Lambda$CDM-like Universe.  $\aleff$ is fixed to -0.23 
for the main result.

\item
$\gprime$, $\sprime$ (2) \\
Modifiers of the evolution of the amplitude and slope with redshift, 
to test for
deviations from the expectation for $\Lambda$CDM.  Fixed for main result.

\item
$\bF(z_p)$, $\nu_F$ (2) \\
Mean transmitted flux normalization and redshift evolution.

\item
$T_{i=1..3}$, $\tgamma_{i=1..3}$ (6) \\
Temperature-density relation parameters, including redshift evolution.

\item
$\xrei$ (1)\\ 
Degree of Jeans smoothing, related to the redshift and temperature of
reionization.

\item
$f_{\rm SiIII}$, $\nu_{\rm SiIII}$ (2) \\
Normalization and redshift evolution of the SiIII-Ly$\alpha$ cross-correlation
term.

\item
$\epsilon_{n,i=1..11}$ (11) \\
Freedom in the noise amplitude in the data in each SDSS redshift bin.

\item
$\alpha_R$ (1)\\
Freedom in the resolution for the SDSS data.

\item
$A_{\rm damp}$ (1)\\
Normalization of the power contributed by high density systems.

\item
$a_{\rm NOSN}$, $a_{\rm NOMETAL}$ (2)\\
Admixture of corrections from the NOSN and NOMETAL hydrodynamic simulations.

\item
$A_{\rm UV}$, $\nu_{\rm UV}$ (2) \\
Normalization and redshift evolution of the correction for fluctuations in 
the ionizing background.

\item
$x_{\rm extrap}$ (1)\\
Freedom in the extrapolation of our small simulation results to low $k$.

\end{itemize}

The linear theory power spectrum, comprising the primary result of the
paper, is described by an
amplitude, $\DL=k_p^3 P_L(k_p,z_p)/2 \pi^2$, with normalization
convention such that $\sigma_L^2 = \int_{-\infty}^{\infty} d\ln k~\Delta_L^2(k)$,
where $\sigma_L^2$ is the variance of the linear theory density field; 
slope, $\neff=d\ln P_L / d\ln k \left. \right|_{z_p,k_p}$, and curvature,
$\aleff=d n_{\rm eff} / d\ln k \left. \right|_{z_p,k_p}$.  Together these describe an
approximate power spectrum:
\begin{equation}
\Delta^2_L(k,z) 
\simeq \left[\frac{D(z)}{D(z_p)}\right]^2 \DL 
\left[\frac{k}{k_{\star}(z)}\right]^{3+\neff+1/2~\aleff \ln\left[k/k_{\star}(z)\right]}~,
\end{equation}
where $k$ is measured in $\kms$, with $z_p=3.0$ and $k_p=0.009\ikms$.  
We preserve the linear theory prediction that
only the amplitude of the power spectrum evolves in comoving coordinates by defining
$k_\star(z) = k_p [H(z_p)/(1+z_p)] / [H(z)/(1+z)] \simeq k_p [(1+z_p)/(1+z)]^{1/2}$. 
We compute $D(z)$ and $H(z)$ for a typical $\Omega_m=0.3$ $\Lambda$CDM model, although
at the level of our error bars this is indistinguishable from an Einstein-de Sitter
model (i.e., $D(z)/D(z_p)\simeq a/a_p$). 
In practice we actually measure these power spectrum 
parameters as deviations from a CMBFAST power 
spectrum for a flat $\Lambda$CDM model with $\Omega_m=0.3$, $h=0.7$, and $\Omega_b=0.04$,
which has $\aleff=-0.23$ and $\neff=-2.26$ (the latter is 
for primordial power spectrum slope $n=1.0$). Note that $\aleff$
only weakly changes with cosmological parameters. i.e. $-0.25<
\aleff< -0.15$ over the 
range of interest. 

When we measure a growth factor we parameterize it by 
$\gprime$ in $D(z)/D(z_p)=(a/a_p)^{\gprime}$,
where $\gprime$ should not be measurably different from 1 for a standard cosmology.
Unexpected evolution of the slope is parameterized by 
$n_{\rm eff}[z,k_\star(z)]=\neff+\sprime (z-z_p)$. These parameters are included 
to test for deviations
from the expected Einstein-de Sitter Universe, but are fixed to their expected 
values for the standard fit.

We describe $\bF(z)$ by a power law in effective optical depth,
$\bF(z) = \exp(\ln[\bF(z_p)][(1+z)/(1+z_p)]^{\nu_F})$.  Even if the truth
is not quite consistent with this representation, 
we expect that the power spectrum parameters
will be mostly sensitive to the overall normalization, so this parameterization
should be sufficient, i.e., small wiggles or 
curvature might lead to a bad fit to the redshift evolution of $\PF$, but 
are not likely to cause significant bias in the extraction of $P_L(k)$.

We allow considerable 
freedom in the temperature-density relation, because it is possible
that its evolution is not monotonic 
\citep{2000MNRAS.318..817S,2000ApJ...534...41R,2001ApJ...562...52M,
2001ApJ...557..519Z}.  
$T_{1.4}(z)$ is
parameterized by quadratic interpolation between three points, 
$T_1=T_{1.4}(z=2.4)$, $T_2=T_{1.4}(z=3.0)$, $T_3=T_{1.4}(z=3.9)$. 
Similarly, 
$(\gmo)(z)$ is described by three parameters at the same redshifts as
$T_{1.4}$.  Because 
we have only weak observational constraints, but 
theoretical limits $0\lesssim \gmo \lesssim 0.6$ 
\citep{1997MNRAS.292...27H}, we use a 
parameterization that lends itself to enforcing an upper and
lower limit. 
The exact form 
is $(\gmo)(z) = 0.7 (\tanh[\tgamma(z)]+1)/2-0.05$, where 
$\tgamma(z)$ is defined by quadratic interpolation between  
$\tgamma_1=\tgamma(z=2.4)$, $\tgamma_2=\tgamma(z=3.0)$, 
$\tgamma_3=\tgamma(z=3.9)$.  This form naturally applies the constraint
$-0.05 \lesssim \gmo\lesssim 0.65$.  We 
add $(\tgamma_i/10)^2$ to $\chi^2$ to prevent the parameters from wandering
off to infinity.

Differing reionization histories are included by multiplying our standard
power spectrum prediction by $1+f(\xrei) [P_{{\rm high}~z}(k,z)/P_{\rm standard}(k,z)-1]$,
where $P_{{\rm high}~z}$ is an HPM simulation in which the temperature was
set to 50000 K at $z=17$ and evolved as a power law down to our usual values
at $z<4$, while $P_{\rm standard}$ was our standard case with $T=25000$ K at
$z=7$. We use $f(\xrei)= 1.6~(\tanh[\xrei]+1)/2-0.3$, and add 
$(\xrei/10)^2$ to $\chi^2$.  The lower limit $\xrei>-0.3$ was chosen to allow
for reionization at $z=7$ (this would be $\xrei=0$), minus 0.2 to allow for 
the hydrodynamic simulation resolution correction discussed 
in \S \ref{sechydrosims}, 
minus another 0.1 to allow for any residual small errors.  The upper limit
was chosen largely arbitrarily to allow for very early, hot reionization
(this limit has no effect in practice).

As discussed in \cite{2004astro.ph..5013M}, cross-correlation between SiIII and 
\lya\ absorption by the same gas leads to small wiggles in the observed
power spectrum.  As suggested in that paper, we use a linear bias model
to roughly describe this effect, with 
$\delta_{\rm SiIII}=a(z) \delta_{{\rm Ly}\alpha}$ and
$a(z) = f_{\rm SiIII} [(1+z)/3.2]^{\nu_{\rm SiIII}}/[1-\bF(z)]$.
$f_{\rm SiIII}$ and $\nu_{\rm SiIII}$ are the two free parameters in our
fit (we could constrain $f_{\rm SiIII}\geq 0$ but this is unnecessary because
the $\PF$ data completely rules this limit out). 
We refer the reader to \cite{2004astro.ph..5013M} for a discussion of the
parameters describing uncertainty in the noise determination in each 
SDSS $\PF$ bin, and the parameter describing the resolution uncertainty.

Following \cite{2005MNRAS.360.1471M},
the power contributed by high density absorbers is included by simply 
adding the template shown in Figure \ref{uvtemplate}, 
multiplied by the parameter 
$A_{\rm damp}$ to the simulation prediction, i.e., 
$P^\prime_F(k,z)=\PF+A_{\rm damp} P_{\rm damp}(k,z)$.  We add 
$[(A_{\rm damp}-1)/0.3]^2$ to $\chi^2$, constraining the contribution
to be near the prediction based on the observed column density 
distribution (see \cite{2005MNRAS.360.1471M}).

The difference between the three hydrodynamic simulations we studied
is allowed for by the following form for the calculation of
$P_{\rm hydro}$ that we use to calibrate the HPM simulations:  
$P_{\rm hydro}(k,z)= 
(1-x_1-x_2) P_{\rm FULL}(k,z) + x_1 P_{\rm NOSN}(k,z)+
x_2 P_{\rm NOMETAL}(k,z)$, where $x_1=[\tanh(a_{\rm NOSN})+1]/2$
and $x_2=[\tanh(a_{\rm NOMETAL})+1]/2$.  $a_{\rm NOMETAL}$ and
$a_{\rm NOSN}$ are the two parameters in our fit, with the usual
addition to $\chi^2$ of $(a/10)^2$.  We impose a hard constraint
$x_1+x_2<1$, but this is generally not activated because the 
fits prefer $P_{\rm FULL}$ to the alternatives.

The UV background fluctuation effect presented above should be present
at some level, but may be diluted by contributions to the background
from galaxies, and re-radiation by the IGM gas 
\citep{1996ApJ...461...20H}.  The relative
amount of radiation from different sources is expected to change with
redshift, so we do not feel comfortable using only a single normalization
parameter.  We implement the UV background fluctuation effect by 
multiplying the predicted $\PF$ by the factor $1+f(z) [U(k,z)-1]$,
where $U(k,z)$ is the ratio shown in Figure \ref{uvtemplate} and
$f(z)=(\tanh[A_{\rm UV}+\nu_{\rm UV} (z-4.2)]+1)/2$, i.e., we allow
somewhere between no effect and the full effect, and allow for 
a transition between the two extremes with redshift.  We
add $(A_{\rm UV}/10)^2$ and $(\nu_{\rm UV}/2)^2$ to $\chi^2$; the
former is the usual finiteness constraint, but the second is a 
non-trivial constraint on the rapidity with which the transition 
from domination by quasars to other sources can take place (our
constraint gives, for example, a $\chi^2$ penalty of 1 to a 
transition from 10\% of the full effect to 90\% if it occurs
over $\Delta z=1$).

Finally, we have a parameter controlling the extrapolation of
simulation predictions of $\PF$ to $k<k_L=2\pi/L$.  
We use $P(k)=x P(k_L)+(1-x) P(k_L) 
(k/k_L)^{n_F}$, where $n_F$ is the logarithmic derivative
of $P_F(k)$ at $k_L$.  We use our usual method to impose
$0<x<1$, $x=[\tanh(x_{\rm extrap})+1]/2$, where 
$x_{\rm extrap}$ is our final free parameter.  
This issue is only important for the hydrodynamic correction from 
$L=10\hmpc$ and not for the HPM resolution correction:  
our $40\hmpc$ simulations cover all of the observed points
we use, and the extrapolation from $L=20\hmpc$ is not long
enough to allow significant freedom in practice.

\subsection{Data}

The observational data constraints in our fit are
largely those described in \cite{2004astro.ph..5013M}.  
We fit to a total of 132 SDSS
$\PF$ points in the range $0.0013\ikms < k < 0.02\ikms$
(12 points each in 11 redshift bins from $z=2.2$ to $z=4.2$).  
We add 39 HIRES
$\PF$ points with $k<0.05\ikms$ from 
\cite{2000ApJ...543....1M}.  We do not include points
from \cite{2002ApJ...581...20C} and 
\cite{2004MNRAS.347..355K} in our standard analysis 
for reasons discussed in \S\ref{highresPF} and \cite{2004astro.ph..5013M}.
In \S\ref{highresPF}, we present an alternative 
analysis that does include
these measurements, finding similar
results to our standard analysis.

For $\bF$ we use the HIRES constraints $\bF=(0.458\pm0.034,
0.676\pm0.032,0.816\pm0.023)$ from 
\cite{2000ApJ...543....1M} (slightly modified 
to allow for systematic uncertainties, as discussed 
in \cite{2003MNRAS.342L..79S}).  
We do not use the tighter constraints in 
\cite{2003ApJ...596..768S} and \cite{2003AJ....125...32B}.
As we will see, the constraints we do use have essentially
no effect on the result, and we consider this to be a good
thing.  The $\PF$ fit itself constrains $\bF$ to better
than 0.01, with no external constraints.  Therefore, in order
for an external constraint to help much, it would have to be
accurate to this level -- not just have formal error bars
at this level, but actually deal with continuum fitting 
issues and metal absorption at this level.  Furthermore,
damping wings and UV fluctuations affect the predicted values
of $\bF$ in the simulations, and while our current analysis
can in principle account for this, we would not want to have
to do it very accurately.  The bottom line of this discussion 
is that it is advantageous that the power spectrum data constrain 
$\bF$ internally, rather than relying on external constraints on 
the mean flux, since those are controversial and do not account 
for all of the effects we have to worry about. We consider this a 
major improvement in the analysis of the \lyaf\ over previous 
analyses, where the data were not sufficiently precise to 
allow for this internal calibration of the mean flux. 

For the temperature-density relation we use 
$T_{1.4}=(20100\pm3400,20300\pm2400,20700\pm 2800)$K
and $\gamma-1=(0.43\pm0.45,0.29\pm0.3,0.52\pm 0.14)$ at 
$z=(3.9,3.0,2.4)$, in addition to the theoretical 
constraints $-0.05<\gamma-1<0.65$ \citep{1997MNRAS.292...27H}.  
These measurements are from \cite{2001ApJ...562...52M},
with 2000 K added in quadrature to the temperature errors
to allow for systematic errors.
\cite{2000MNRAS.318..817S} and \cite{2000ApJ...534...41R} present 
additional constraints which we do not use for reasons
similar to those discussed for $\bF$ -- we do not believe
any of these analyses have been done sufficiently carefully to
justify smaller errors than the ones we are using.
In fact, in this case we will see that the constraints we
are using do matter somewhat, and we do not consider 
them to be especially conservative, so assuming errors any smaller 
than this could lead to a reduction of statistical errors at the expense 
of introducing a systematic error.  
Nevertheless, it is informative to compare our results to
those using the temperature-density relation constraints in 
\cite{2000MNRAS.318..817S}.  For coding simplicity, we use
these measurements
as re-binned by \cite{2001ApJ...562...52M}:  for $z=$(2.46, 3.12, 3.58),
$T_{1.4}=(16000\pm 1300,~19600\pm1200,~14900\pm 1600)$K and 
$\gmo=(0.34\pm 0.07,~0.06\pm 0.07~,~0.22\pm 0.10)$.
Note that the 
power spectrum-based temperature determination of 
\cite{2001ApJ...557..519Z} is effectively part of 
our analysis (our analysis uses the
same basic approach as \cite{2001ApJ...557..519Z}
in many ways).  

\subsection{Basic $\PL$ Results}

\begin{figure}
\plotone{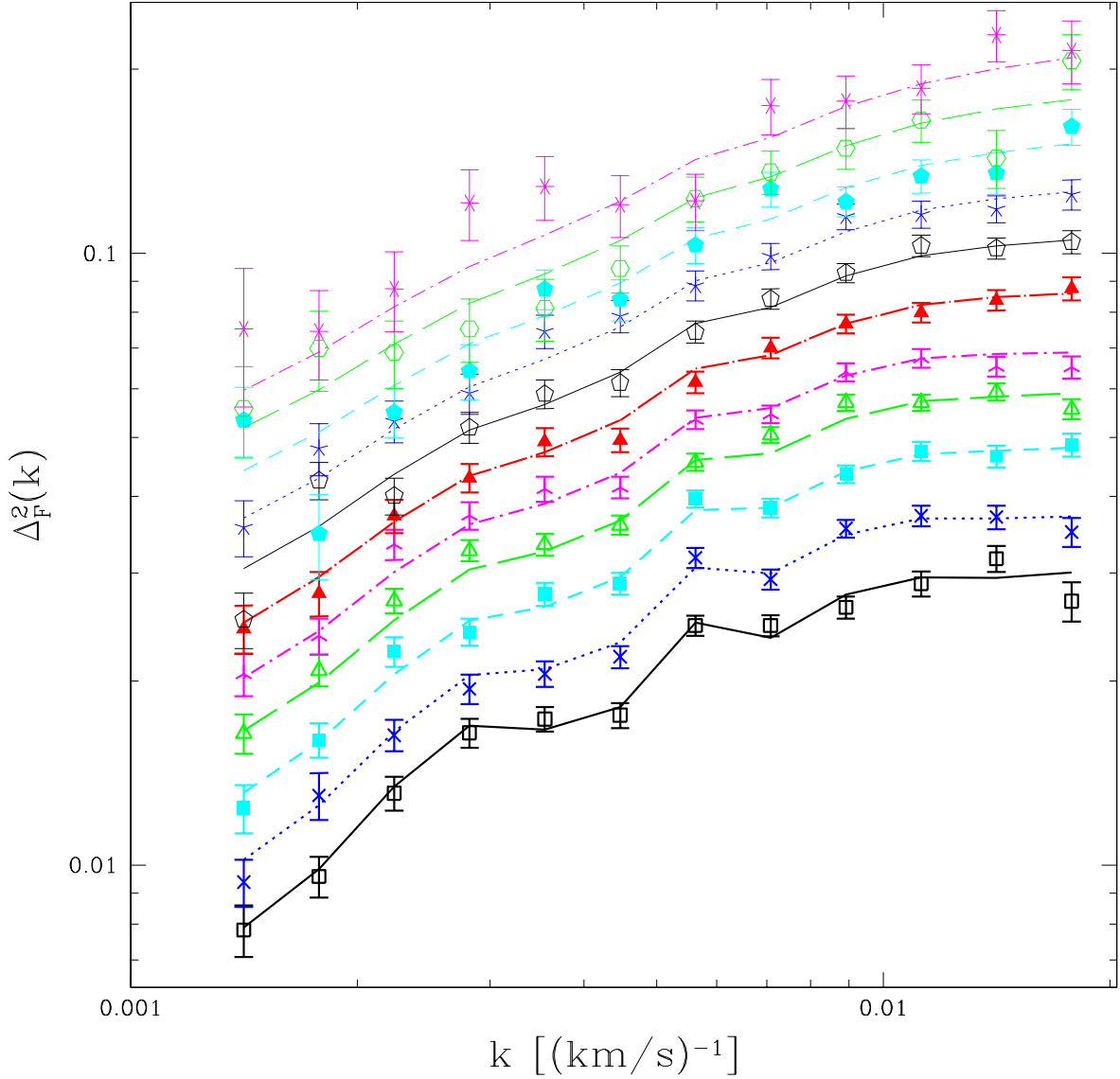}
\caption{
Points with error bars show the observed $\PF$ from SDSS.  
Lines show our best
fitting model.  From bottom to top ---
z=2.2:  black, solid line, open square;
z=2.4:  blue, dotted line, 4-point star (cross);
z=2.6:  cyan, dashed line, filled square;
z=2.8:  green, long-dashed line, open triangle;
z=3.0:  magenta, dot-dashed line, 3-point star;
z=3.2:  red, dot-long-dashed line, filled triangle;
z=3.4:  black, thin solid line, open pentagon;
z=3.6:  blue, thin dotted line, 5-point star;
z=3.8:  cyan, thin dashed line, filled pentagon;
z=4.0:  green, thin long-dashed line, open hexagon;
z=4.2:  magenta, thin dot-dashed line, 6-point star.
Note that the wiggles in the theory curve are caused by
SiIII-\lya\ cross-correlation.
}
\label{showfit}
\end{figure}

\begin{figure}
\plotone{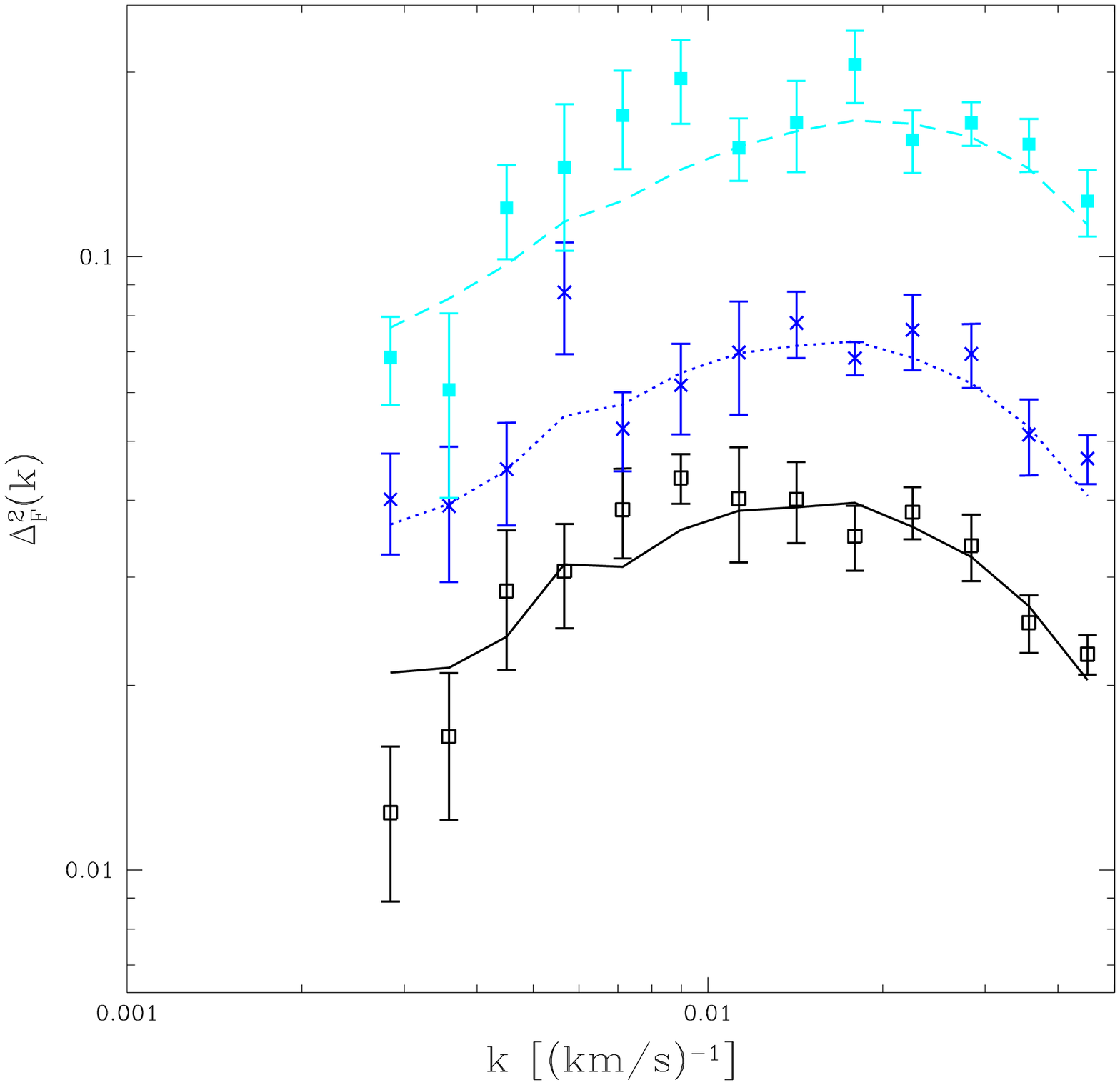}
\caption{
Points with error bars show the observed $\PF$ from HIRES
\citep{2000ApJ...543....1M}.  
Lines show our best
fitting model.  From bottom to top ---
z=2.4:  black, solid line, open square;
z=3.0:  blue, dotted line, 4-point star (cross);
z=3.9:  cyan, dashed line, filled square.
}
\label{showhiresfit}
\end{figure}

\begin{figure}
\plotone{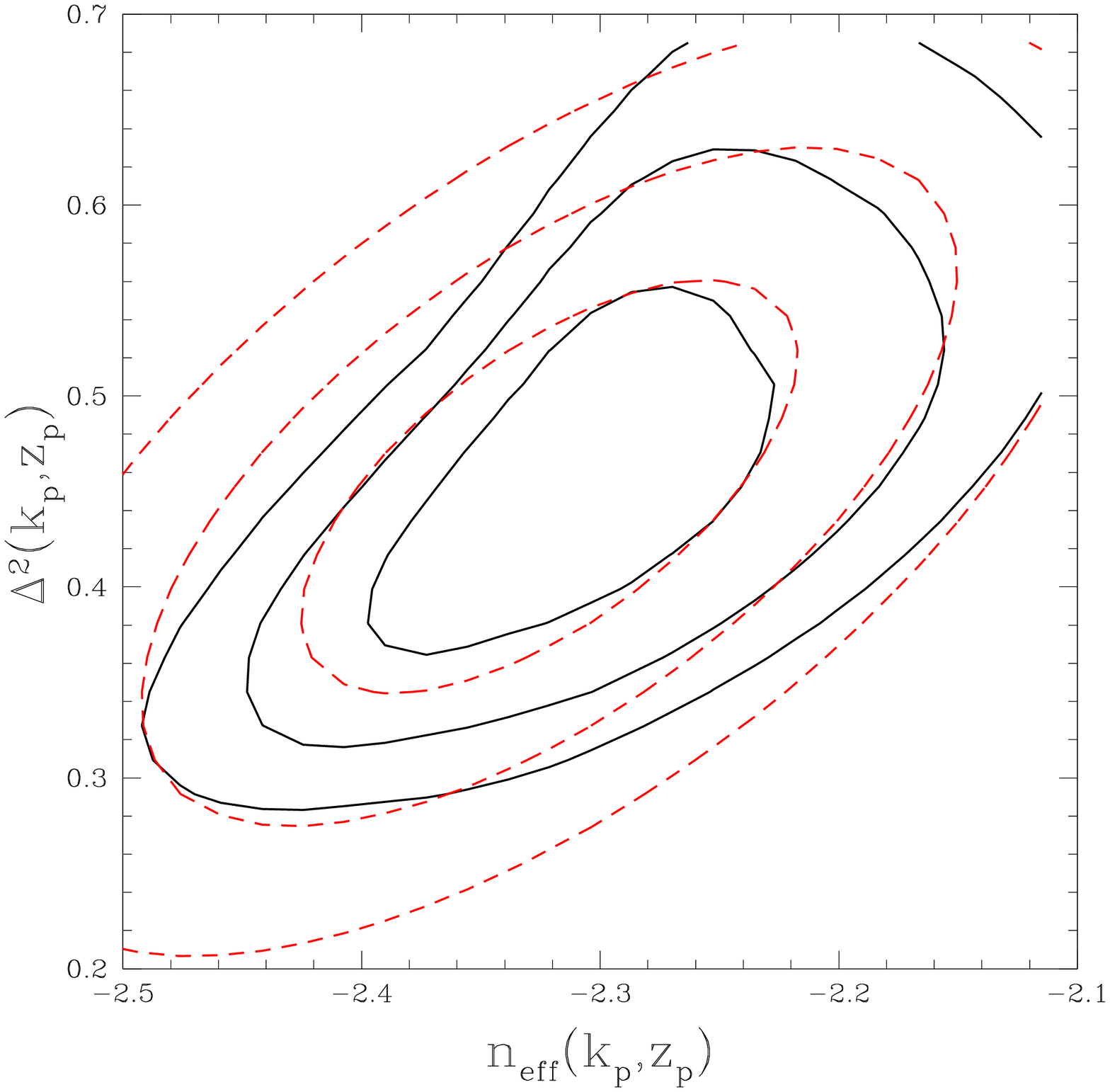}
\caption{
Contours of $\Delta \chi^2=2.3$, 6.2, and 11.8, minimized over the 
other parameters (solid black lines).  For comparison, we show the 
same contours implied using derivatives of $\chi^2$ with
respect to the parameters at the best fit point (red dashed lines).
}
\label{ennampcontour}
\end{figure}

\begin{figure}
\plotone{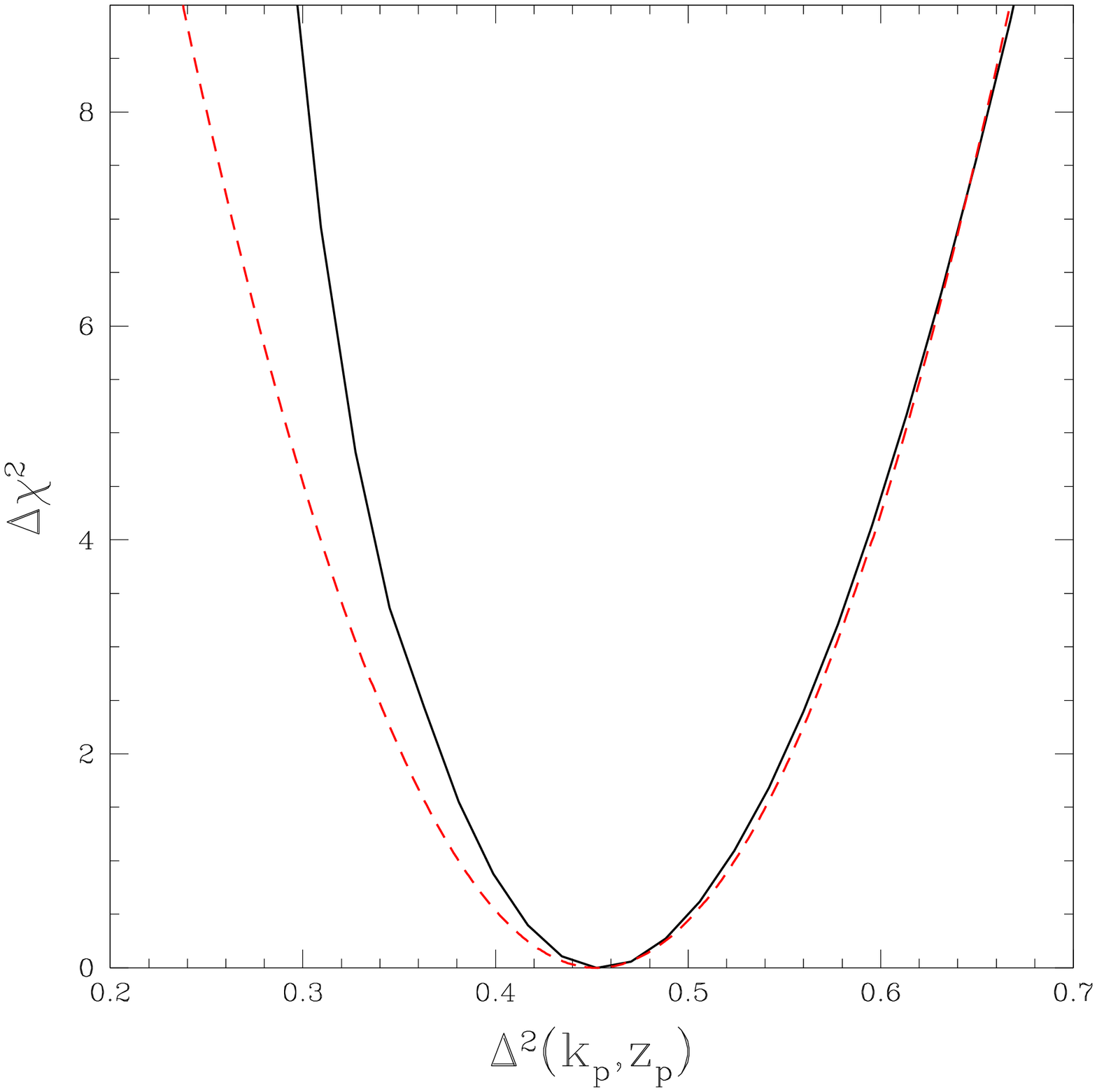}
\caption{
$\Delta \chi^2$ as a function of $\DL$ (a), and $\neff$ (b), 
minimized over the 
other parameters (solid black line), or
implied by derivatives of $\chi^2$ at the best fit point
(red dashed lines).
}
\label{chisq1D}
\end{figure}

We show the basic fit to the SDSS $\PF$ points in Figure \ref{showfit}, 
and the HIRES $\PF$ points in Figure \ref{showhiresfit}.
We find
$\chi^2=185.6$ for the fit, for $\sim 161$ degrees of freedom, which
is reasonable (a value
this high would occur 9\% of the time by chance). 
The best fit power spectrum parameters are 
$\Delta^2_L(k_p=0.009~{\rm s/km},z_p=3.0)=
0.452_{-0.057~-0.116}^{+0.069~+0.141}$ 
and slope $\neff=-2.321_{-0.047~-0.102}^{+0.055~+0.131}$, where the errors
are 1 and 2 $\sigma$ ($\Delta \chi^2=1$ and 4 as the parameter of interest
is varied while minimizing over the other parameters).  The formal (i.e.,
computed by derivatives of $\chi^2$ at the best fit point)
correlation coefficient of the errors is $r=0.63$, with 
$1\sigma$ errors $\pm 0.072$ and $\pm 0.069$ on $\DL$ and $\neff$,
respectively.  Figure \ref{ennampcontour} shows the contours of 
$\Delta \chi^2$ in the $\DL-\neff$ plane, compared to the contours one 
would estimate from derivatives at the best fit point.
We see that, while the local derivative errors are reasonably reflective
of the true errors, they are far from perfect.  This is not surprising,
both because we have various non-Gaussian priors on nuisance parameters,
and because the errors generally expand with increasing linear power
because of nonlinearities.  Fits combining the \lyaf\ with other probes
of cosmology should use the full contours for maximum accuracy.
Figures \ref{chisq1D}(a,b) show $\Delta \chi^2$ for
each parameter minimized over the other. 
\begin{figure}
\plotone{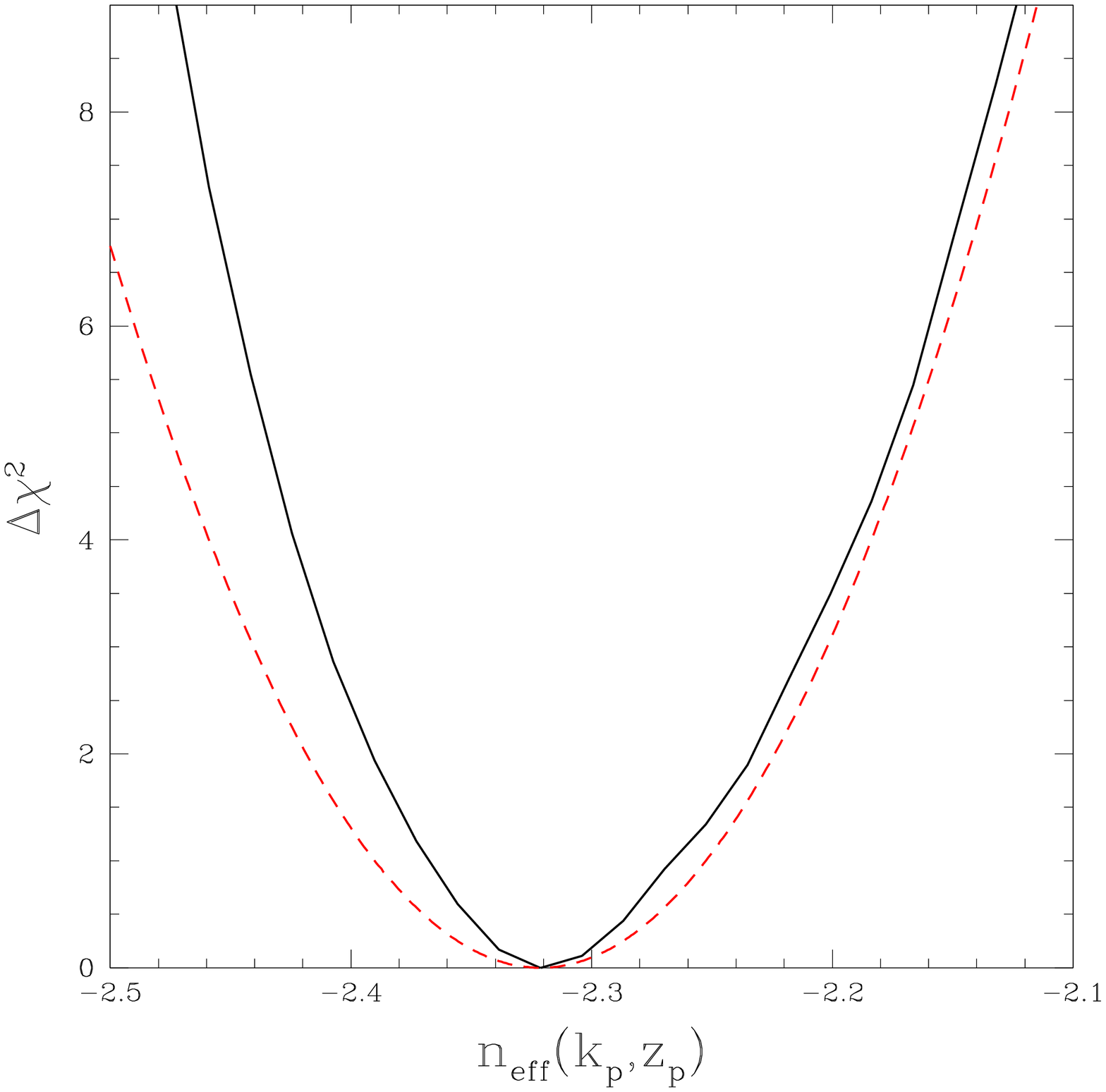}
\end{figure}
We use these curves to determine the asymmetric errors we quote 
on the standard result.

For the standard fit, we left $\aleff=-0.23$, the value in our 
$\Omega_m=0.3$ reference model (with primordial $\alpha=0$).  
If we include $\aleff$ as 
a free parameter, $\chi^2$ improves by 1.7, a change that would
occur 19\% of the time by chance.  
The best fit value is $\aleff=-0.135\pm0.094$.  Note that since
we have chosen the pivot point $k_p=0.009\ikms$ to make the
errors on $\neff$
and $\aleff$ approximately independent, the inferred value of
$\neff$ does not change significantly when $\aleff$ is varied.
In practice, the best fit value of $\DL$ does not change either.  

We provide an electronic table of 
$\chi^2[\Delta^2_L,n_{\rm eff},\alpha_{\rm eff}]$, a sample of which is 
shown as Table \ref{chisqtabsamp}.  The table is also available at:\\
{\it http://www.cita.utoronto.ca/$\sim$pmcdonal/LyaF/lyafchisq.txt }.\\
\begin{deluxetable}{lccc}\tablecolumns{4}\tablecaption{
$\chi^2[\Delta^2_L,n_{\rm eff},\alpha_{\rm eff}]$  
\label{chisqtabsamp}}
\tablehead{ \colhead{$\DL$} &\colhead{$\neff$}& \colhead{$\aleff$}& 
\colhead{$\chi^2$}}
\startdata
0.452491 & -2.35568 & -0.228985 & 186.798 \\
0.452491 & -2.33848 & -0.228985 & 185.872 \\
0.452491 & -2.32128 & -0.228985 & 185.595 \\
0.452491 & -2.30408 & -0.228985 & 185.821 \\
0.452491 & -2.28688 & -0.228985 & 186.22 \\
0.452491 & -2.26968 & -0.228985 & 186.723 \\
\enddata
\tablecomments{$z_p=3.0$, $k_p=0.009~{\rm s/km}$.  Points with $\chi^2<0$ are 
either outside the range where we have simulations, or an initial estimate 
indicated that $\chi^2$ would be very high there.  
The $\aleff$ dependence is included only to allow more accurate
computation of $\chi^2$ near $\aleff = -0.23$, the value for typical 
$\Lambda$CDM
models constrained by WMAP.  This table is not intended for 
models with power spectra
qualitatively different in shape from standard $\Lambda$CDM.
[The complete version of this table is in the electronic edition of
the Journal.  The printed edition contains only a sample.]}
\end{deluxetable}
The table covers the range $0.095<\DL<0.685$ in logarithmic
steps of 
$\delta \Delta_L/\Delta_L=0.06$, covers $-2.665<\neff<-1.977$ in steps of 0.017,
and $-0.33<\aleff<-0.13$ in steps of 0.1.
A computer code that takes the linear theory power spectrum at $z=3$ and 
produces $\chi^2$ can be found at 
{\it http://www.cita.utoronto.ca/$\sim$pmcdonal/code.html }\\
under the name ``LyaFChiSquared.''
This table (or code) will be suitable for joint analyses with
the CMB and other observations like those performed in 
\cite{2005PhRvD..71j3515S}.  It should not be trusted for models where
$P_L(k)$ is not effectively described by $\DL$, $\neff$, and
$\aleff$, or models where the values of these parameters deviate
substantially from those in typical $\Lambda$CDM-like models,
e.g., warm dark matter models \citep{2000ApJ...543L.103N} 
or primordial black hole models \citep{2003ApJ...594L..71A} 
(the code will produce a warning if a suspect power spectrum is 
input).
Our simulation database does contain these models.

\subsection{Consistency Checks:  Evolution of Slope and Amplitude}

If we believe the Universe is effectively Einstein-de Sitter (EdS)
in the redshift range we probe then the evolution of 
$P_L(k,z)$ is
completely specified (for typical $\Lambda$CDM-like models). 
Here we test this by measuring the growth factor and the
change in the slope of the power spectrum with redshift.

When we allow a power law modification of the growth factor,
we find a decrease of 2.8 in $\chi^2$, which would occur by
chance 9\% of the time.  The measured growth is 
$\gprime=1.46\pm 0.29$ (note that $\gprime>1$ means the growth is 
faster than EdS, the opposite of what one would expect if dark 
energy was present).  We consider this to be an ambiguous
result.  The deviation from the expectation is not very 
significant, and the constraint is not tight enough to call
this an important consistency check: it rules out
gross deviations, but not deviations at the level of the
statistical errors on our main result. Still, it would 
be interesting to explore this further, including additional
statistics like the bispectrum, as this method
can be one of the few ways to study the presence of dark 
energy at $z>2$ \citep{2003MNRAS.344..776M}. 

When we allow evolution in $n_{\rm eff}$ at fixed comoving $k$
through the parameter $\sprime$ in 
$n_{\rm eff}[z,k_\star(z)]=\neff+\sprime (z-z_p)$, $\chi^2$ 
improves by only 1.8 (probability 18\%).  
The measured value is $\sprime=0.051\pm0.041$.
The size of this error bar is a remarkable, 
and counterintuitive, result.  
The evolution of $n_{\rm eff}$ across the redshift range 
we probe is constrained more tightly than $\neff$
itself.  In retrospect, this result is not so hard to understand:
a substantial part of the error on $\neff$ comes from 
degeneracy with $\DL$, which causes the measured values 
of $n_{\rm eff}$ at different redshifts to move up or down
together, depending on the value of $\DL$ considered.

So far we have shown three consistency 
tests [$\aleff$, $\gprime$, $\sprime$], 
none of which show compellingly significant deviation 
from our expectation.  
Can these be combined to give a 
significant deviation? The answer is no: when we free
all three parameters at the same time, $\chi^2$ only 
decreases by 3.5 relative to the standard fit. This increase
occurs by chance 32\% of time with 3 free parameters. 
We can interpret this as a sign that 
the deviations are statistical in nature and are 
not consistent with each 
other in terms of being caused by 
a common source of systematic error. 

To summarize: in this subsection we have demonstrated that we
can make precise measurements of the slope of $\PL$ at
multiple redshifts.  In the model we use for the interpretation,
these values will be tightly correlated, so they can not be combined
to give an even better overall measurement, but they act as a 
stringent discriminator against any physical effect which changes the
inferred value of $n_{\rm eff}$ in a way that is not redshift
independent.  Remarkably, we could detect a redshift dependent 
effect even if its influence on $n_{\rm eff}$ was smaller than 
the size of our overall error on $\neff$.

\subsection{Consistency Checks: 
Modifications of the Fitting Procedure \label{secmodifications}}

\begin{deluxetable}{lccccccc}
\tablecolumns{8}
\tablecaption{Effect of modifications of
the fitting procedure on the inferred linear
power spectrum and its errors \label{modtab}}
\tablehead{ \colhead{Variant\tablenotemark{a}} &
\colhead{$\Delta_L^2$}
& \colhead{$n_{\rm eff}$}
& \colhead{$\chi^2$ \tablenotemark{b}}
& \colhead{$\Delta \chi^2$ \tablenotemark{c}}
}
\startdata
Standard fit & $ 0.452 \pm 0.072 $ & $ -2.321 \pm 0.069 $ & 185.6 & 0.0 \\
No hydrodynamic corrections & $ 0.377 \pm 0.041 $ & $ -2.284 \pm 0.046 $ & 191.8 & 4.0 \\
Fixed extrapolation & $ 0.456 \pm 0.071 $ & $ -2.303 \pm 0.058 $ & 185.9 &  0.2 \\
Fixed to FULL & $ 0.453 \pm 0.070 $ & $ -2.322 \pm 0.063 $ & 185.4 &  0.0 \\
Fixed to NOSN & $ 0.435 \pm 0.059 $ & $ -2.262 \pm 0.054 $ & 187.9 &  1.9 \\
Fixed to NOMETAL & $ 0.394 \pm 0.048 $ & $ -2.374 \pm 0.055 $ & 188.3 &  1.3 \\
No $L=40\hmpc$ simulations & $ 0.439 \pm 0.065 $ & $ -2.328 \pm 0.069 $ & 190.0 &  0.1 \\
$\Omega_m=0.4$, HS transfer func. & $ 0.454 \pm 0.074 $ & $ -2.307 \pm 0.067 $ & 187.6 &  0.1 \\
No damping wings (DW) & $ 0.366 \pm 0.042 $ & $ -2.398 \pm 0.050 $ & 188.7 &  1.8 \\
DW power known to 10\% & $ 0.452 \pm 0.071 $ & $ -2.321 \pm 0.067 $ & 185.6 & 0.0 \\
Randomly located DW & $ 0.435 \pm 0.070 $ & $ -2.333 \pm 0.067 $ & 186.8 &  0.1 \\
No UVBG fluctuations & $ 0.446 \pm 0.067 $ & $ -2.338 \pm 0.049 $ & 187.4 &  0.2 \\
Strong attenuation UVBG & $ 0.452 \pm 0.072 $ & $ -2.320 \pm 0.067 $ & 185.1 & 0.0 \\
Galaxy-based UVBG & $ 0.452 \pm 0.069 $ & $ -2.346 \pm 0.059 $ & 187.4 &  0.3 \\
$\bF$ errors $\times 2$ & $ 0.452 \pm 0.077 $ & $ -2.321 \pm 0.071 $ & 184.9 & 0.0 \\
$\bF$ errors $\times \frac{1}{2}$ & $ 0.455 \pm 0.062 $ & $ -2.320 \pm 0.066 $ & 188.2 & 0.0 \\
Fix $\bF$ to best & $ 0.452 \pm 0.030 $ & $ -2.321 \pm 0.048 $ & 185.6 & 0.0 \\
TDR errors $\times 2$ & $ 0.530 \pm 0.106 $ & $ -2.299 \pm 0.078 $ & 180.4 &  0.8 \\
TDR errors $\times \frac{1}{2}$ & $ 0.455 \pm 0.055 $ & $ -2.305 \pm 0.065 $ & 192.0 &  0.0 \\
Schaye TDR & $ 0.524 \pm 0.059 $ & $ -2.307 \pm 0.072 $ & 195.4 &  1.4 \\
HIRES $P_F$ errors $\times 2$ & $ 0.493 \pm 0.086 $ & $ -2.276 \pm 0.081 $ & 153.8 & 0.9 \\
HIRES $P_F$ errors $\times \frac{1}{2}$ & $ 0.442 \pm 0.070 $ & $ -2.335 \pm 0.053 $ & 292.1 & 0.1 \\
SDSS $P_F$ errors $\times \frac{1}{2}$ & $ 0.468 \pm 0.053 $ & $ -2.301 \pm 0.033 $ & 584.3 &  0.1 \\
Fix nuisance params. to best & $ 0.452 \pm 0.010 $ & $ -2.321 \pm 0.012 $ & 185.6 & 0.0 \\
Inc. Croft/Kim, no back. sub. & $ 0.355 \pm 0.051 $ & $ -2.366 \pm 0.054 $ & 313.3 &  2.9 \\
Include Croft \& Kim & $ 0.408 \pm 0.064 $ & $ -2.364 \pm 0.063 $ & 215.9 & 0.4 \\
Drop bad Croft $z$ & $ 0.411 \pm 0.064 $ & $ -2.366 \pm 0.064 $ & 206.1 &  0.3  \\
Add Kim only & $ 0.466 \pm 0.082 $ & $ -2.318 \pm 0.076 $ & 178.7 &  0.1 \\
standard w/HIRES back. sub. & $ 0.503 \pm 0.094 $ & $ -2.305 \pm 0.081 $ & 161.9 &  0.6 \\
\enddata
\tablecomments{$z_p=3.0$, $k_p=0.009~\ikms$.}
\tablenotetext{a}{The meaning of each variant is
explained in \S\ref{secmodifications}.}
\tablenotetext{b}{Standard $\chi^2$ for the fit, for $\sim 161$ degrees of
freedom, plus 20-24 for \cite{2004MNRAS.351.1471K}, plus 44-65 for \cite{2002ApJ...581...20C}
(see details in \S \ref{highresPF}).}
\tablenotetext{c}{$\Delta \chi^2$ between the variant best fit amplitude
and slope and the standard best fit values
(essentially unrelated to $\chi^2$ for the fit). }
\end{deluxetable}
Our plan in this subsubsection is to investigate the sensitivities of our 
measurement to various changes in our treatment, to look for potential
problems and 
identify the important areas for future improvement.
Table \ref{modtab} shows the effect of changes in various components of our 
fitting procedure.
For each modification of the procedure, we give the new best fits and errors 
for $\DL$ and $\neff$, and $\chi^2$ for the new fit, along with 
$\Delta \chi^2$ between the variant and standard best fit power spectrum
parameters.
We evaluate $\Delta \chi^2$ between the two pairs of parameters in the 
context of both 
the standard and modified fitting scenarios, 
and report the smaller change -- 
this method of comparison shows the significance of the modification 
in a more informative way 
than simply comparing the change in parameters to the error bars,
because it accounts for correlations and deviations from Gaussianity of the 
errors (we report the smaller $\Delta \chi^2$ because when we have two
measurements with different sized errors, we do not generally expect 
the measurement with larger errors to fall within the error contours
of the better measurement).
Note that, as discussed above, these $\pm 1\sigma$ errors are only 
intended to be indicative of the true errors, which will not be perfectly 
Gaussian (in fact, the Gaussian errors are sometimes so bad that we 
probably should not even report them, as we see, for example, in the 
standard fit). 

Our first modification is to remove the hydrodynamic correction to the HPM
prediction of $\PF$.  The change in the result, particularly the amplitude,
is significant, although not huge, and $\chi^2$ for the fit increases
significantly (indicating that the data prefers to have the correction).  
Note that the reduction in the error bars comes from three things:  
decreasing the amplitude of the power spectrum always reduces the errors,
removing the hydrodynamic correction effectively removes the freedom
to modify the large scale power prediction by modifying the form of 
extrapolation of the correction ($x_{\rm extrap}$ discussed above), and
we lose the freedom to choose between the three different forms of galaxy 
feedback in the hydrodynamic simulations.
Note that, as we see from the next line in the Table, the removal of this 
extrapolation uncertainty (we fix $x_{\rm extrap}=0$) 
is not what changes the best fit values or $\chi^2$,
since removing this alone does relatively little.

Next we try using each of the hydrodynamic simulations individually for the
correction, rather than letting the fit choose between them.  Using FULL has
no effect, except to reduce the error bars, because the fit prefers it
(the slight reduction in $\chi^2$ for FULL versus standard fit 
is an artifact of the way we impose the
boundaries on the simulation-type multipliers).  Using the NOSN and NOMETAL 
simulations leads to small but noticeable changes in the result, although
these are disfavored by the increase in $\chi^2$.

While our usual method is to use $(40,512)$ simulations for the main $\PF$
prediction, corrected for limited resolution by comparing (20,512) to (20,256)
simulations, we tried performing the fit simply using (20,512) simulations
(with the usual form of extrapolation to larger scales).  The results are
essentially unchanged, although $\chi^2$ increases somewhat.  This simple test
actually rules out a variety of potential problems with the details of our 
$\PF$ calculation.  One is the
possibility that we have statistical errors in the simulation predictions. 
We have a similar number of each size simulation, which means the 
(40,512) simulations have 8 times the total volume compared to 
(20,512). Thus, it would take an unlikely
fluke to make the (20,512)-based measurement agree with our (40,512)-based
measurement if even the larger simulations had significant statistical 
error [(20,512) would have even bigger errors].  Another is that 
substantial 
systematic errors from the limited size of the $L=40\hmpc$ boxes are
disfavored, because $L=20\hmpc$ should then give an even larger 
error.  Finally, the validity of the resolution 
correction is confirmed by this test.

Our standard fit is based on the CMBFAST transfer function
for the $\Omega_m=0.3$ model defined above, and uses this model for the
growth factor and Hubble parameter.  We tried basing the fit on a model
with $\Omega_m=0.4$, $\Omega_b=0.05$, $h=0.65$, and the 
\cite{1996ApJ...471..542H} (HS) transfer function (which is also 
the model used
in the simulations).  We expect that this should give results essentially
identical to our standard fit.  
There is no significant change in the fitted 
parameters, but there is a surprisingly large increase (2.0) in $\chi^2$.

Removing the power from high density systems with damping wings has a 
significant effect on the result, reducing the slope and amplitude
and their errors.  
This is not especially worrisome since 
the correction that we make can not be very wrong
because it is constrained by direct observations of these systems.  Reducing
our usually assumed 30\% error on the size of the effect to 10\% does
not change the fit results significantly.  Using the unrealistic 
template where the high density systems are randomly distributed in 
the IGM does not change our fit results although it does 
increase $\chi^2$ by 1.2.  

Removing the freedom to include UV background fluctuations in the fit
does not change the central values from the fit, but does significantly
reduce the error on $\neff$, at the cost of increasing $\chi^2$ by 1.8.  
Switching to the UV background fluctuation
template for the case where the mean free path of ionizing photons
has been arbitrarily halved (this allows a larger maximum effect) 
gives results very similar to the standard fit. 
We also tried using the template from \cite{2005MNRAS.360.1471M} 
where Lyman-break galaxies are the source of the ionizing radiation,
finding a modest reduction in the error on $\neff$, and a small 
increase in $\chi^2$, but ultimately no significant change in our
results.

Next we arbitrarily increase or decrease the errors on the observations 
we use.  This is intended to elucidate the importance of the different
constraints -- the central values that come out of the fits
when errors are arbitrarily reduced should not be taken seriously.

It may be surprising that the constraint on $\bF$ actually has little
effect on the fit, despite the well-known fact that $\PF$ is extremely
sensitive to $\bF$.  The effect of the constraint is so small because 
the observed power
spectrum itself constrains $\bF$ to about $\pm 0.01$, much better than
the constraint we have imposed.  As we mentioned above, this presents
a difficult target for direct measurements of $\bF$, which have to be
accurate to this level, including all systematic effects, to be useful.
To show that the inclusion of $\bF$ in the fit is important, just not
constrained by the external measurements, we 
repeat the fit with $\bF$ fixed to its best value, so that it doesn't
contribute to the errors on other parameters.
We find that the errors on the inferred power spectrum, especially
on the amplitude, are reduced dramatically, as one would expect.

The observational constraint we impose on the temperature-density 
relation does
have a noticeable effect.  Doubling the errors on the observations
of $T_{1.4}$ and $\gmo$ leads to a 13\% increase in the error on
$\neff$, and 47\% increase in the error on $\DL$.  Halving the 
errors reduces the errors on $\neff$ and $\DL$ by 6\% and 24\%,
respectively.  Reassuringly, the best fit values of the 
parameters do not change very much when the constraints are
modified.  For comparison, we tried fitting using the much tighter
temperature-density relation constraints from \cite{2000MNRAS.318..817S},
as re-binned by \cite{2001ApJ...562...52M}:  for $z=$(2.46, 3.12, 3.58),
$T_{1.4}=(16000\pm 1300,~19600\pm1200,~14900\pm 1600)$K and 
$\gmo=(0.34\pm 0.07,~0.06\pm 0.07~,~0.22\pm 0.10)$.
The fit is not especially good, with $P(>\chi^2)=3.2$\%.  The $\PL$
results change at the $1\sigma$ level, with the error on $\DL$
decreasing substantially.  The changes in parameter values are
consistent with our expectation for random changes based on the
change in error bar [e.g., a change in the $\DL$ error from $\pm 0.072$
to $\pm 0.053$ implies an expected change 
$\pm (0.072^2-0.053^2)^{1/2}=\pm 0.049$ in the measured value
of $\DL$].
There is clearly a lot of room for improvement in the 
temperature-density relation constraint, which we plan to 
address with future work. 

The HIRES measurement of $\PF$ that we include is fairly important
to the errors on our result, although, again, less important to the central
values.  Doubling the HIRES errors leads to a 17\% increase in 
the error on $\neff$, while halving them reduces this error by 
23\%.  The errors on $\DL$ increase by 19\% when the HIRES errors
are doubled, but remain essentially unchanged when they are halved.  
Finally, improving the errors on the SDSS $\PF$ measurement leads
to a 26\% improvement in the amplitude measurement and 52\% in the
slope measurement.  We are unable to perform a HIRES-only fit without
modifications of the procedure because the result is not well 
constrained to within the region where we have simulations.  An
SDSS-only fit is better constrained, but still has very 
large errors, i.e., both
high resolution data and SDSS are necessary for good results.

Finally, out of curiosity, we fix all the nuisance parameters to 
their best fit values, so the only free parameters are 
$\DL$ and $\neff$.  This tells us how well we could do if we
did not need to worry about uncertainties in the \lyaf\ model.  
The resulting errors are $\pm 0.010$ and $\pm 0.012$, respectively.

In summary:  While nothing that we have seen 
necessarily indicates a 
problem, the importance of some of the corrections indicates that they 
need to be dealt with carefully in the future, especially if the 
statistical errors can be reduced.  Reducing the statistical 
errors on the amplitude by more than
$\sim 30$\% will probably require improvements in more than one of 
the components of the measurement; however, the $\neff$ errors should
improve in proportion to the improvement in the SDSS $\PF$ statistics.  
The reader should keep in mind that,
to keep Table \ref{modtab} finite, we did not include combinations of
changes.  An improvement that does not seem useful alone, e.g., 
reducing the error on the power from damping wings, can 
become useful if another uncertainty that it is degenerate with is 
also removed. 

\subsection{Consistency Checks:  Alternative Treatment of High 
Resolution $\PF$ \label{highresPF}}

Finally, we consider the high resolution $\PF$ measurements we have
not included in our standard fit \citep{2002ApJ...581...20C,
2004MNRAS.347..355K,2004MNRAS.351.1471K}. 
As we found in \cite{2004astro.ph..5013M}, the fit is poor when these
measurements are included: $\chi^2=313.3$ for $\sim 250$ degrees of
freedom (our usual 161 plus 65 points from \cite{2002ApJ...581...20C} and 24 
from \cite{2004MNRAS.351.1471K}).  In Table \ref{modtab}, the line 
``Inc. Croft/Kim, no back. sub.'' shows this fit (the meaning of ``no
background subtraction'' will become clear shortly).
A value of
$\chi^2$ this high will only occur by chance 0.4\% of the time, and the increase
of 127.7 in $\chi^2$ for 89 additional degrees of freedom is similarly
unlikely.  Adding \cite{2002ApJ...581...20C} alone increases $\chi^2$
by 99.7 ($P(>\chi^2)=0.4$\%), while \cite{2004MNRAS.351.1471K} alone
increases $\chi^2$
by 40.0 ($P(>\chi^2)= 2.1$\%).  The fit using \cite{2004MNRAS.351.1471K} alone
is better than it was before the correction of the wavelengths
of the bins \citep{2004MNRAS.347..355K}, partially because the ($k=0.0010~{\rm s/km}$,
$z=2.58$)
point with the improbably small error bar is no longer within the $k$ range we
are using; however, the fit is still not good enough to be comfortable.

Because we would like to be able to use the additional statistical power
of \cite{2002ApJ...581...20C} and 
\cite{2004MNRAS.347..355K,2004MNRAS.351.1471K}, we investigate possible reasons
for the bad fits, starting with the statistical error bars.  
\cite{2004astro.ph..5013M}
pointed out that the \cite{2004MNRAS.351.1471K} point 
at $z=2.58$, $k=0.0010~{\rm s/km}$ is inconsistent with the SDSS data, and 
any reasonable extrapolation of the rest of the \cite{2004MNRAS.351.1471K} data
(the change in wavelength scale does not change this).  It seems likely that
the error bar is simply underestimated, possibly because there was not 
enough data to perform a robust jackknife error estimate.  While this point
is no longer included in our $k$ range, its existence suggests that the 
\cite{2004MNRAS.347..355K} errors are not fully reliable.  We attempt a
correction to the errors based on the following assumptions:  the 
true $\PF$ should
generally increase with decreasing $k$, and the fractional error should
also increase (because the data contain fewer modes per bin, and the noise
power is insignificant in these spectra).  Starting
from high $k$, we simply increase the error on each point as necessary to
guarantee monotonicity (7 of 24 of the points with 
$0.0013~\ikms < k < 0.05~\ikms$ have their
errors increased).  We apply the same adjustment to the
\cite{2002ApJ...581...20C} errors (13 of 65 increase), but leave
the \cite{2000ApJ...543....1M} errors unchanged, because 
\cite{2000ApJ...543....1M} performed tests of their bootstrap error 
computation on mock data and already applied a
correction based on the results.  These error corrections have only a
small effect on the results:  $\chi^2$ decreases to 306.4 (still a poor
fit) and the parameter values and error bars change by $<2$\% (not shown
in Table \ref{modtab}).  Unfortunately,
the potential problem of poorly determined jackknife error bars seems
unlikely to be the cause of our poor fits.

Next we investigate the possibility that the treatment of DLAs in the 
high resolution data leads to problems.  In each of the measurements, 
DLAs were removed, while our theoretical predictions in the fits assume
they are in the data.  Note that the error this causes will not be the
full size of our damping wing correction (see Figure \ref{mockHD}), 
because much of the correction
comes from systems with column density less than
$2\times 10^{20} {\rm cm}^{-2}$,
which were not necessarily completely removed from any of the high 
resolution data (some such systems were removed by \cite{2004MNRAS.347..355K},
but we do not know how complete this removal was).  There is no reason not to be
conservative in accounting for this possible error, because the high resolution
data is not important to the low $k$ constraints (where damping wings are 
important), so we simply add a component corresponding to uncertainty at the
level of the full amplitude 
of the damping wing correction to the error covariance matrices of all of
the high
resolution data (i.e., $C_{ij}^\prime = C_{ij}+
P_{\rm damp}(k_i,z) P_{\rm damp}(k_j,z)$, where $C_{ij}$
is the covariance matrix for redshift bin $z$).  
The only effect is to decrease $\chi^2$ by 3.1 to 303.3
(the power spectrum parameter values and errors change by less than 2\% --
not shown in Table \ref{modtab}).
The treatment of DLAs in the high resolution data does not seem to be 
important to our goodness of fit.       

One of the advances of \cite{2004astro.ph..5013M} was a careful measurement
and subtraction of background power (e.g., from metal absorption), using 
the $1268 < \lr < 1380$ \AA\ region in the quasar rest frame.
Further investigating possible reasons for the bad fits when we include
\cite{2002ApJ...581...20C}
and \cite{2004MNRAS.347..355K,2004MNRAS.351.1471K}, we discovered that the 
background
power measured by \cite{2004MNRAS.347..355K} in their quasar spectra in the 
restframe wavelength range $1265.67<\lambda<1393.67$ (derivable from 
their Figure 2) is quite significant for our fits.   \cite{2004MNRAS.347..355K}
made no correction for this background, but we can make a very rough 
correction in the following way:  First, we read the fractional background from their
figure, using the power spectra computed without continuum fitting (F3 in their 
notation), and 
finding the numbers: ($k$, $P_{1266,1394}/P_{1026,1203}$) = 
(0.0011,  0.778),
(0.0015,  0.470),
(0.0021,  0.233),
(0.0030,  0.221),
(0.0042,  0.141),
(0.0060,  0.090),
(0.0085,  0.081),
(0.0120,  0.073),
(0.0169,  0.090),
(0.0239,  0.049),
(0.0338,  0.060), and
(0.0478,  0.090), where $P_{\lambda_1,\lambda_2}$ means power measured in the
rest frame wavelength range $\lambda_1<\lambda<\lambda_2$ (wavelength in \AA,
$k$ in s/km).
Then, for each $\PF$ point used in our fit we estimate the absolute background 
to subtract by multiplying the
measured $\PF$ by this fraction.    
This background is similar to or larger than that found by 
\cite{2004astro.ph..5013M} in SDSS data at the same redshift, depending on 
the $k$ value
and SDSS noise level considered.  
In particular, on large scales it is much larger.  This means that
some of the background must be related to the observing and data reduction 
process rather than metal absorption or quasar continuum power 
(\cite{2004astro.ph..5013M} removed only a very small amount of power by 
dividing
the spectra by the mean quasar continuum).  For this reason, we will 
henceforth exclude \cite{2004MNRAS.347..355K} points with $k<0.003~\ikms$,
where the differences between continuum fitted and not-continuum fitted
spectra becomes important, and the disagreement with SDSS data
on the background level
becomes severe (this same cutoff was suggested by \cite{2004MNRAS.347..355K},
although we do not agree that the reason for it is likely to be simple
continuum fluctuations).
When we perform the background 
subtraction, we propagate an independent 35\% statistical
error on the 
fractional background power for each point (roughly estimated from the error 
bars on the 
power measurements in \cite{2004MNRAS.347..355K}), along with the error implied
by the uncertainty in the measured power itself. 
Our estimate of the fractional background using \cite{2004MNRAS.347..355K}
in 
principle applies only to the redshift range in their figure, while the 
true ratio 
of background to \lyaf\ power will inevitably change with redshift, 
therefore we allow 
substantial freedom in the overall amplitude of the subtracted component, equal
to 50\% of the amplitude at $1\sigma$ (i.e., at 95\% confidence, anything from 
no background to twice the fiducial background is allowed).  
\cite{2000ApJ...543....1M} and \cite{2002ApJ...581...20C} present no measurement
of the background in their data, although it must be present at some level. 
We perform the same background subtraction procedure just described on their
measurements of $\PF$, and apply the same $k$ limit, although we note
that this procedure is highly dubious because the background may depend on the
details of the observations (e.g., the details of sky subtraction).

\begin{figure}
\plotone{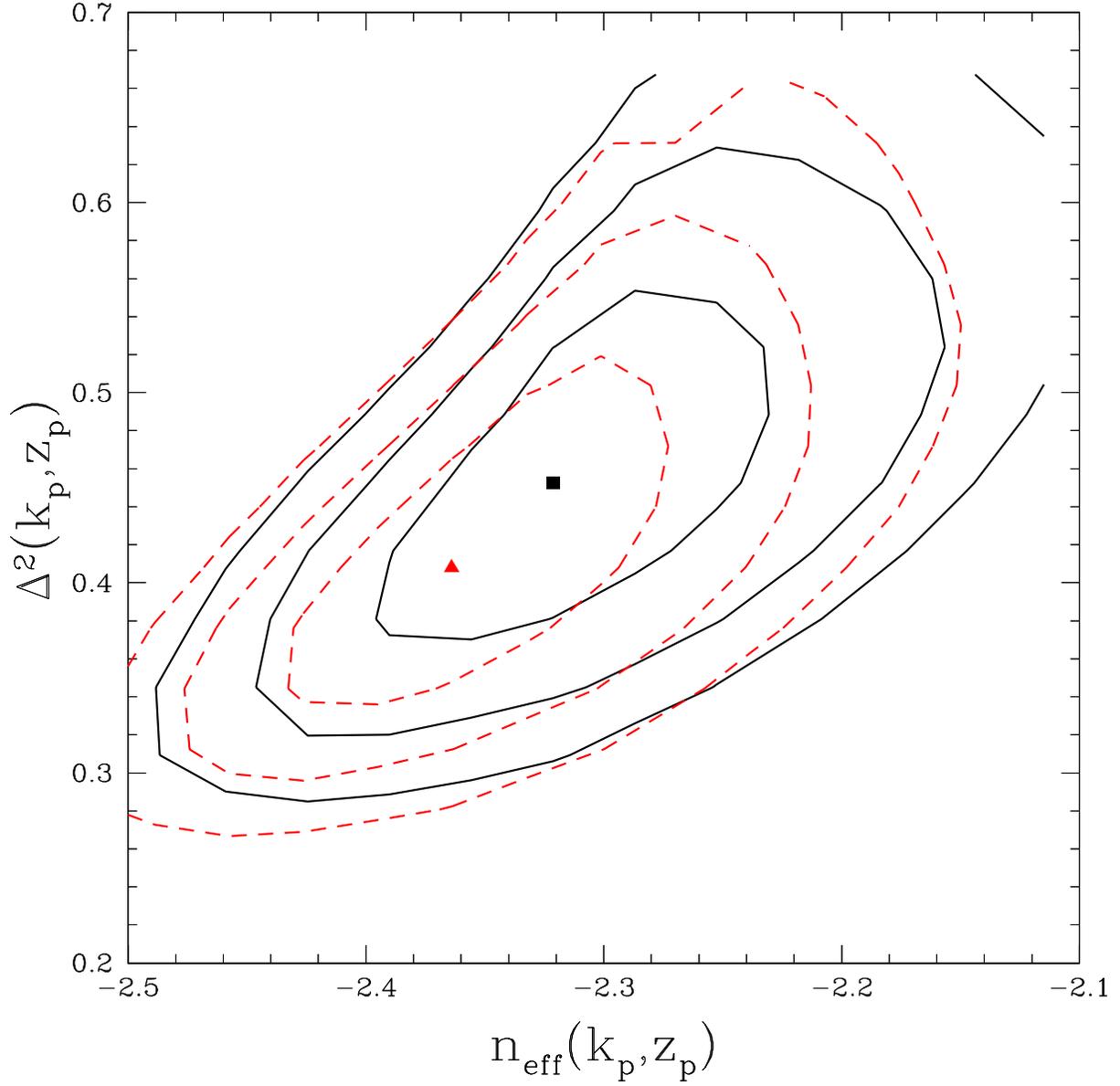}
\caption{
Comparison of standard fit results (black, solid lines and square) to 
a fit including the \cite{2002ApJ...581...20C} and 
\cite{2004MNRAS.347..355K,2004MNRAS.351.1471K} high resolution 
$\PF$ measurements, with a background correction to all the high 
resolution measurements (red, dashed lines and triangle).
The points show the minimum $\chi^2$ while
contours show $\Delta \chi^2=2.3$, 6.2, and 11.8, minimized over the 
other parameters.  
}
\label{wcroftkimcontours}
\end{figure}

Using this prescription for the background subtraction from high resolution
data, along with the previously described monotonicity constraint on the
error bars, and DLA uncertainty, we find $\chi^2=215.9$ for $\sim 224$ 
degrees of freedom, a perfectly good fit.  The fitted $\PL$ parameters
change very little relative to our standard fit, 
and the errors on amplitude and slope both decrease by $\sim 10$\% 
(see the ``Include Croft \& Kim'' entry in Table \ref{modtab}).  
Figure \ref{wcroftkimcontours} shows the change of the constraint in the
$\DL-\neff$ plane.
The change in parameter values falls along the degeneracy direction,
making the combined change even less significant than the individual
changes might appear to be.
Note that changes in parameter values of the size we do see are 
not a sign of even small systematic
disagreement between the data sets -- they are perfectly consistent with  
the expected change whenever
independent data is added and extra freedom is allowed in the fit.
Table \ref{modtab} also shows the case where we remove the redshift bin
from \cite{2002ApJ...581...20C} that \cite{2004astro.ph..5013M} identified
as suspect.  The results do not change much, and $\chi^2$ only decreases
by 9.7, indicating that we have included enough uncertainty in the 
covariance matrix to allow this bin to match the other data (this is not to
say that we believe simple background is responsible for the strange 
results in this bin).  To explore the relative importance of the different
$\PF$ measurements, we perform the fit without \cite{2002ApJ...581...20C},
and using only \cite{2000ApJ...543....1M} as in our standard fit (but
including the DLA error, and the background subtraction -- this is ``standard
w/HIRES back. sub.'' in Table \ref{modtab}).  The fit with 
only \cite{2000ApJ...543....1M} gives a 30\% larger amplitude error than
our standard fit (and 10\% larger $\neff$ error), while adding
\cite{2004MNRAS.347..355K,2004MNRAS.351.1471K} brings us about half way
back to the standard fit (note that one of the significant advances of
\cite{2004MNRAS.347..355K} is a $\PF$ measurement at $z<2.1$, which we
are ignoring because it is outside the SDSS range).  Adding the 
\cite{2002ApJ...581...20C} measurement accounts for the rest of the 
improvement in the errors.  None of these variant fits give significantly
different central values for the $\PL$ parameters.

This treatment of the high resolution $\PF$ measurements is admittedly
ad hoc, but nonetheless informative.  Some 
improvement could be made in the $\PL$ error bars if all of the data
could be used.  It is clear, however, that an investigation of the 
background in the spectra actually used in these papers is needed.
Ultimately, the results we obtain when all the measurements are included
are very similar to our standard results, reassuring us that the 
cosmological results are not sensitive to the details of the high
resolution data.

\subsection{Consistency Checks:  Nuisance parameter results 
\label{nuisancevalues}}

Many of the nuisance parameters are interesting in 
themselves; however, we are hesitant to present their values
and error bars from the fits in this paper. 
Unlike the case of the power spectrum
slope and amplitude, we have not checked carefully that 
the resulting measurements of the other parameters are
reliable at the level of precision we could quote.
We hope to present complete results for other parameters in the
future, but for now their role in our fits should be  
seen as simply to be descriptors of various forms of 
uncertainty in the power spectrum extraction.
To reassure the reader that the values are reasonable, we give
some central values from the fit; however, the errors should be
considered to be unknown (which is of course practically 
equivalent to infinite errors).

\begin{figure}
\plotone{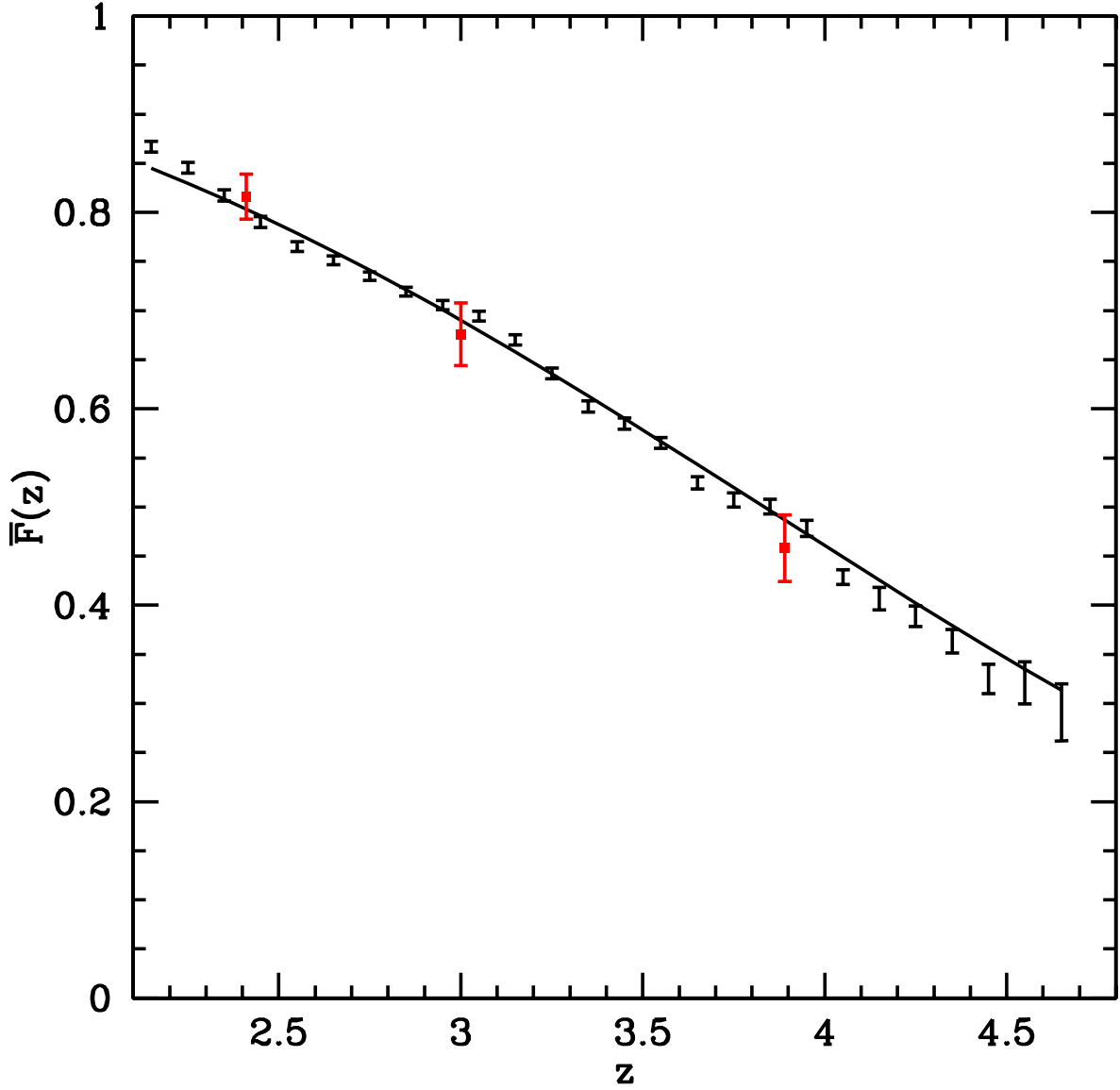}
\caption{Evolution of the mean transmitted flux fraction, $\bF(z)$.  
The curve is our indirect measurement 
from the fit to the power spectrum.  The black bars show a direct 
estimate from the SDSS spectra using a PCA determination of their continua
(the overall normalization of these points is arbitrary).  The red points with 
large error bars are from HIRES spectra \citep{2000ApJ...543....1M}.  
Intended for qualitative use only -- we are not certain that
the SDSS measurements are reliable at the level of their error bars, and the
errors are correlated. 
}
\label{meanFevolution}
\end{figure}

Our $\bF(z)$ results are probably the most interesting, because the measurement
is quite precise, and the method is completely different from the usual 
direct measurement.  Recall that we 
parameterized $\bF(z)$  by 
$\bF(z) = \exp(\ln[\bF(z_p)][(1+z)/(1+z_p)]^{\nu_F})$.  
The fitting results are:  $\bF(z_p=3)=0.69$,  
$\nu_F=3.3$.  Unfortunately, we do not know that these $\bF$ numbers are 
robust estimates of
the value of $\left<\exp(-\tau)\right>$ that we should expect
to observe directly.
We plot this result, along with direct estimates from 
\cite{2000ApJ...543....1M} and from the SDSS spectra based on the PCA 
continuum
method described in \S 2.5 of \cite{2004astro.ph..5013M}, in 
Figure \ref{meanFevolution}. 
The PCA measurement is also still preliminary, because we have not been 
able to rigorously demonstrate 
convergence to the level of the statistical error bars when increasing the 
number of eigenvectors used to determine the continuum.
Each of these measurements alone may not be perfectly reliable, but together 
they present a clear, consistent
picture of the evolution of the mean absorption (although note that the
PCA measurement only constrains the evolution up to a single overall 
normalizing factor, which has been adjusted to match the other measurements).  

The other nuisance parameter results are as follows:
The temperature-density relation from the fit is 
$T_{1.4}=(20000,~20000,~15000)$K, $\gmo=(0.5,~0.5,~0.1)$  at 
$z=(2.4,~3.0,~3.9)$.  The reionization/filtering length parameter is at the 
lower limit of
the range we allow (note that this parameter plays a dual role as the final 
insurance against error caused by limited simulation resolution, as we showed 
in Figure \ref{hydrorestest}).  The SiIII normalization is 
$f_{\rm SiIII}=0.013$ with redshift evolution poorly constrained.  
The damping wing power 
normalization is $A_{\rm damp}=1.0$.  The full-physics (FULL) 
hydrodynamic simulation is favored over the NOMETAL and NOSN simulations.
Finally, the presence of significant power from UV background fluctuations
is disfavored, but the constraint is weak.

\section{Conclusions \label{conclusions}}

Our primary result is the measurement of the amplitude and
slope of the linear theory power spectrum at $z\sim 3$ on 
$\sim 1 \hmpc$ scales:
$\Delta^2_L(k_p=0.009~{\rm s/km},z_p=3.0)=
0.452_{-0.057~-0.116}^{+0.069~+0.141}$ 
and $\neff=-2.321_{-0.047~-0.102}^{+0.055~+0.131}$ (these are
1 and 2 $\sigma$ errors, with correlation $r\simeq0.63$).  
These were measured as the amplitude and tilt of a CMBFAST
power spectrum for a flat $\Lambda$CDM model with $\Omega_m=0.3$, 
$\Omega_b=0.04$, and $h=0.7$, and correspond to $\sigma_8=0.85$,
$n=0.94$ for this model; however, we emphasize that these $\sigma_8$
and $n$ numbers aren't
especially meaningful because they are model dependent.  The
real power of the \lyaf\ measurement is achieved when it is 
combined with the CMB measurements on larger scales
\citep{2005PhRvD..71j3515S}.
If we additionally
allow variation in the curvature of the power spectrum, we find 
$\aleff=-0.135\pm 0.094$ (the expected value for $\Lambda$CDM
models with zero primordial running is $\aleff\sim -0.23$). 
As a consistency check, we estimated a power
law growth factor, $D(a)\propto a^{\gprime}$, finding 
$\gprime=1.46\pm0.29$, consistent with the expectation 
$\gprime=1.0$. 
We also estimated the evolution of the inferred slope 
at a fixed comoving $k$ with redshift, which is expected to be zero, 
and found $\sprime=d n_{\rm eff}/dz=0.051\pm0.041$.

The use of the SDSS $\PF$ measurement \citep{2004astro.ph..5013M} 
represents an improvement over past work.  In addition
to more data, we have improved the analysis method in several ways:
Our method is different from others
(with exception of \cite{2001ApJ...557..519Z}), in that we assume nothing
about the dependence of $\PF$ on $P_L(k)$ and other parameters (other
than the smoothness assumptions implicit in our interpolation 
procedure).  As a result, our errors on the power spectrum parameters
properly incorporate partial degeneracies and correlations 
with each other and with
nuisance parameters such as the mean absorption level $\bF$. 
We calibrate our HPM simulations using fully hydrodynamic simulations 
and include non-negligible uncertainty in the calibration, found by comparing
simulations with three different versions of the physics.
Our most important addition to the \lyaf\ model is power contributed
by high density systems with damping wings (many of them below the
traditional column density of DLAs), as investigated by 
\cite{2005MNRAS.360.1471M}.
This increases the measured slope and amplitude, and their
error bars.  We also include the possibility of UV background fluctuations,
which turn out to be easy to constrain because their effect changes
rapidly with redshift.
Note that the systematic error tests
in \cite{2004astro.ph..5013M} show smaller errors than we present 
here because most of these effects were not included in that analysis.

There is plenty of room for improvement in every aspect of the measurement.
The accumulation of SDSS spectra will improve the large-scale $\PF$ 
measurement and in turn the errors on $P_L$. 
We showed that improved measurements of $\PF$ from high resolution 
data will also help significantly.  We only used the HIRES-based 
measurement of \cite{2000ApJ...543....1M}, because the other existing
measurements 
\citep{2002ApJ...581...20C,2004MNRAS.347..355K,2004MNRAS.351.1471K} 
show signs of problems --
they would produce bad $\chi^2$ fits
if we included them.  We investigated the reason for the disagreement
and concluded that it is probably the presence of significant unsubtracted
background power in the high resolution measurements.  After accounting for
this in a very rough way, we obtain results consistent with our standard
results.
For future measurements of the power spectrum 
(or any \lyaf\ statistic), we suggest two steps that can help diagnose
problems:  (1) The measurement, including the error estimation, should be 
performed on mock spectra, where the correct result is known, ideally 
constructed in a format that allows
exactly the same analysis code to be applied from end to end.  In addition
to allowing high precision tests for any bias in the measurement, the 
ability to produce many complete sets of mock spectra allows a test
of the commonly used jackknife or bootstrap errors, which, in particular,
may underestimate the errors for small samples of data.
(2) The measurement should be performed on the red side of the
\lya\ emission line and if there is any detection there it should 
be accounted for (keeping in mind that systematic errors that look
small relative to the statistical error on a single data point can
still be very significant when they affect many points).   

On the theory side, the main improvements to be made are  
in the size and number of hydrodynamic simulations.  The
errors on the linear power spectrum 
$P_L$ could be reduced if we did not have to extrapolate
as far beyond the scale of the simulation boxes.  It might be possible
to further reduce the errors if we better understood the causes for
the differences between simulations with and without supernova energy
feedback and metal cooling.
Improving  and understanding better the 
hydrodynamic simulations should be the top priority for the near 
future.
We have
not shown that improving the accuracy of the damping wing and 
UV background fluctuation calculations \citep{2005MNRAS.360.1471M} can 
improve the $P_L$ measurement, but the accuracy can certainly be
improved and we suspect that this will become important if other
errors can be reduced.

Finally, the measurement of $P_L$ can be improved by additional
\lyaf\ statistics like the bispectrum \citep{2003MNRAS.344..776M}.
While we have not found any indication that fundamental issues 
stand in the way of
an even more 
precise measurement of $P_L$ from the \lyaf, 
our current analysis does not take advantage of the full statistical 
power in the data and 
the challenge of constructing a 
sufficiently
accurate calculational procedure should not be taken lightly. 

\acknowledgements

    Funding for the creation and distribution of the SDSS Archive has
been provided by the Alfred P. Sloan Foundation, the Participating
Institutions, the National Aeronautics and Space Administration, the
National Science Foundation, the U.S. Department of Energy, the
Japanese Monbukagakusho, and the Max Planck Society. The SDSS Web site
is http://www.sdss.org/. 

    The SDSS is managed by the Astrophysical Research Consortium (ARC)
for the Participating Institutions. The Participating Institutions are
The University of Chicago, Fermilab, the Institute for Advanced Study,
the Japan Participation Group, The Johns Hopkins University, Korean
Scientist Group, Los Alamos National Laboratory, the
Max-Planck-Institute for Astronomy (MPIA), the Max-Planck-Institute
for Astrophysics (MPA), New Mexico State University, University of
Pittsburgh, Princeton University, the United States Naval Observatory,
and the University of Washington. 

Some of the computations used facilities at Princeton provided in part 
by NSF grant AST-0216105, and some computations were performed at NCSA. 
RC acknowledges grants AST-0206299 and NAG5-13381.
US is supported by a fellowship from the
David and Lucile Packard Foundation,
NASA grants NAG5-1993, NASA NAG5-11489 and NSF grant CAREER-0132953.
DPS is supported by NSF grant AST03-07582.  SB is supported by NSF AST-0307705.

We thank Nick Gnedin for the HPM code.
We thank Joop Schaye for helpful discussions and comments on 
the manuscript.

\newpage
\bibliography{apjmnemonic,cosmo,cosmo_preprints}
\bibliographystyle{apj} 

\end{document}